\documentclass[11pt]{article}
\usepackage{geometry}
\usepackage{times}
\usepackage{graphics}
\usepackage{tensor}
\usepackage{natbib}
\usepackage{makeidx}
\usepackage{latexsym}
\usepackage{bm}
\usepackage{amsmath}
\usepackage{amsthm}
\usepackage{amssymb}
\usepackage{setspace}
\usepackage{url}
\usepackage[toc,page]{appendix}
\makeindex
\newcommand{\sm}{\bm}
\newcommand{\beq}{\begin{equation}}
\newcommand{\eeq}{\end{equation}}
\newcommand{\dif}[2]{\frac{{\rm d} #1}{{\rm d} #2}}
\newcommand{\ddif}[3]{\frac{{\rm d}^2 #1}{{\rm d} #2 {\rm d} #3}}

\newcommand{\ildif}[2]{{\rm d} #1/{{\rm d} #2 }}
\newcommand{\ilddif}[3]{{\rm d}^2 #1/{{\rm d} #2 {\rm d} #3}}
\newcommand{\ilpdif}[2]{\partial #1/{\partial #2 }}
\newcommand{\pdif}[2]{\frac{\partial #1}{\partial #2}}
\newcommand{\pddif}[3]{\frac{\partial^2 #1}{\partial #2 \partial #3}}
\newcommand{\ptdif}[4]{\frac{\partial^3 #1}{\partial #2 \partial #3 \partial #4}}
\newcommand{\ilpddif}[3]{\partial^2 #1/{\partial #2 \partial #3}}
\newcommand{\defn}{\begin{quote}{\bf Definition. }}
\newcommand{\edefn}{\end{quote}}
\newcommand{\thm}{\begin{theorem}}
\newcommand{\ethm}{\end{theorem}}

\newcommand{\bmat}[1]{\left ( \begin{array}{#1}}
\newcommand{\emat}{\end{array}\right )}
\newcommand{\E}{\mathbb{E}}

\newcommand{\ts}{^{\sf T}} %transposition
\newcommand{\its}{^{\sf -T}}

\newcommand{\X}{{\bf X}}

\newcommand{\bp}{{\bm \beta}}

\newcommand{\tr}[1]{\text{tr}\left ( {#1} \right )}

\theoremstyle{definition}

\theoremstyle{plain}
\newtheorem{theorem}{Theorem}

\newcommand{\eps}[3]
{{\begin{center}
 \rotatebox{#1}{\scalebox{#2}{\includegraphics{#3}}}
 \end{center}}
}

\newcommand{\dsp}{1}
\newcommand{\matsize}{\small} %%{\footnotesize}
\renewcommand{\baselinestretch}{\dsp}

\begin{document}
%\title{On modelling with reduced rank smoothing splines}
%\title{On doubly generalized additive models}
\title{Smoothing parameter and model selection for general smooth models}
%\author{}
\author{ Simon N. Wood$^0$, Natalya Pya$^1$ and Benjamin S\"afken$^2$\\$~^0$ School of Mathematics, University of Bristol, Bristol BS8 1TW U.K.  \\$~^1$Mathematical Sciences, University of Bath, Bath BA2 7AY U.K.\\ $~^2$Georg-August-Universit\"at G\"ottingen, Germany\\{\tt simon.wood@bath.edu}}

\maketitle

\begin{abstract}
This paper discusses a general framework for smoothing parameter estimation for models with regular likelihoods constructed in terms of unknown smooth functions of covariates. Gaussian random effects and parametric terms may also be present. By construction the method is numerically stable and convergent, and enables smoothing parameter uncertainty to be quantified. The latter enables us to fix a well known problem with AIC for such models, thereby improving the range of model selection tools available. The smooth functions are represented by reduced rank spline like smoothers, with associated quadratic penalties measuring function smoothness. Model estimation is by penalized likelihood maximization, where the smoothing parameters controlling the extent of penalization are estimated by Laplace approximate marginal likelihood. The methods cover, for example, generalized additive models for non-exponential family responses (for example beta, ordered categorical, scaled t distribution, negative binomial and Tweedie distributions), generalized additive models for location scale and shape (for example two stage zero inflation models, and Gaussian location-scale models), Cox proportional hazards models and multivariate additive models. The framework reduces the implementation of new model classes to the coding of some standard derivatives of the log likelihood.  
 
\end{abstract}

\section{Introduction}

This paper is about smoothing parameter estimation and model selection in statistical models with a smooth regular likelihood, where the likelihood depends on smooth functions of covariates and these smooth functions are the targets of inference. Simple Gaussian random effects and parametric dependencies may also be present. When the likelihood (or a quasi-likelihood) decomposes into a sum of independent terms each contributed by a response variable from a single parameter exponential family distribution, then such a model is a generalized additive model \citep[GAM,][]{h&t86,h&t90}. GAMs are widely used in practice \citep[see e.g.,][]{ruppert.wand.carroll,fahrmeir2013}, with their popularity resting in part on the availability of statistically well founded smoothing parameter estimation methods, that are numerically efficient and robust \citep{wood2000,wood2011} and perform the important task of estimating how smooth the component smooth functions of a model should be. 

The purpose of this paper is to provide a general method for smoothing parameter estimation when the model likelihood does not have the convenient exponential family (or quasi-likelihood) form. For the most part we have in mind regression models of some sort, but the proposed methods are not limited to this setting. The simplest examples of the extension are generalized additive models where the response distribution is not in the single parameter exponential family. For example, when the response has a Tweedie, negative binomial, beta, scaled t or some sort of ordered categorical or zero inflated distribution. Examples of models with a less GAM like likelihood structure are Cox proportional hazard and Cox process models, scale-location models, such as the GAMLSS class of \cite{rigby2005}, and multivariate additive models \citep[e.g.][]{yee1996}. Smooth function estimation  for such models is not new: what is new here is the general approach to smoothing parameter estimation, and the wide variety of smooth model components that it admits.   

\renewcommand{\baselinestretch}{1}
\begin{figure}
\vspace*{-.6cm }

\eps{-90}{.5}{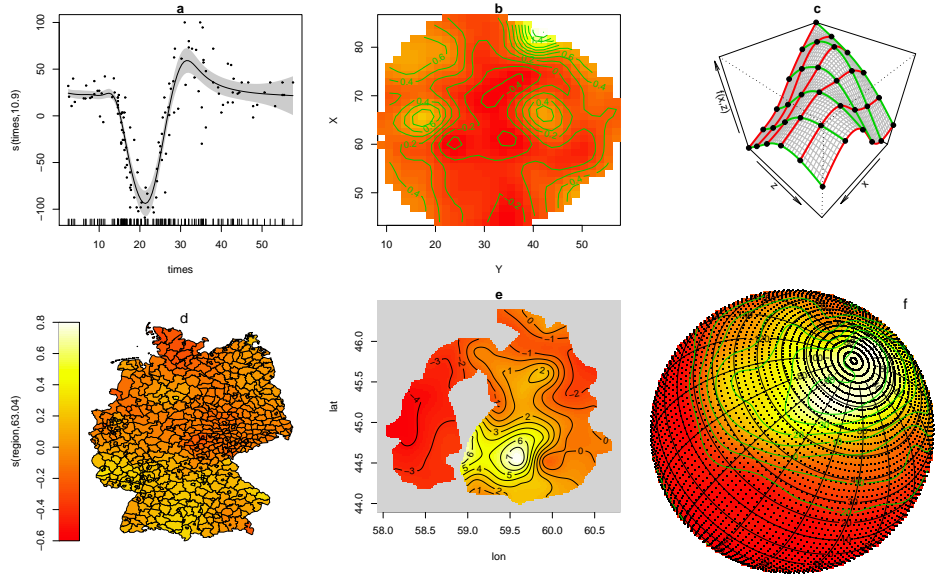}
\vspace*{-.5cm}

\caption{\small Examples of the rich variety of smooth model components that can be represented as reduced rank basis smoothers, with quadratic penalties and therefore can routinely be incorporated as components of a GAM. This paper develops methods to allow their routine use in a much wider class of models. a) one dimensional smooths such as cubic, P- and adaptive splines. b) isotropic smooths of several variables, such as thin plate splines and Duchon splines. c) Non-isotropic tensor product splines used to model smooth interactions. d) Gaussian Markov random fields for data on discrete geographies. e) Finite area smoothers, such as soap film smoothers. f) Splines on the sphere. Another important class are simple Gaussian random effects.\label{smooth.fig}}
\end{figure}
\renewcommand{\baselinestretch}{\dsp}

The proposed method broadly follows the strategy of \cite{wood2011} that has proved successful for the GAM class. The smooth functions will be represented using reduced rank spline bases with associated smoothing penalties that are quadratic in the spline coefficients. There is now a substantial literature showing that the reduce rank approach is well-founded, and the basic issues are covered in Supplementary Appendix A (henceforth `SA A'). More importantly, from an applied perspective, a wide range of spline and Gaussian process terms can be included as model components by adopting this approach (figure \ref{smooth.fig}). We propose to estimate smoothing parameters by Newton optimization of a Laplace approximate marginal likelihood criterion, with each Newton step requiring an inner Newton iteration to find maximum penalized likelihood estimates of the model coefficients. Implicit differentiation is used to obtain derivatives of the coefficients with respect to the smoothing parameters. This basic strategy works well in the GAM setting, but is substantially more complex when the simplifications of a GLM type likelihood no longer apply.

Our aim is to provide a general method that is as numerically efficient and robust as the GAM methods, such that (i) implementation of a model class requires only the coding of some standard derivatives of the log likelihood for that class and (ii) much of the inferential machinery for working with such models can re-use GAM methods (for example interval estimation or p-value computations). An important consequence of our approach is that we are able to compute a simple correction to the conditional AIC for the models considered, which corrects for smoothing parameter estimation uncertainty and the consequent deficiencies in a conventionally computed  conditional AIC \citep[see][]{greven.kneib2010}. This facilitates the part of model selection distinct from smoothing parameter estimation. 

The paper is structured as follows. Section \ref{gen.fram} introduces the general modelling framework. Section \ref{smooth.sec} then covers smoothness selection methods for this framework, with section \ref{general.model} developing a general method, section \ref{gamlss.sec} illustrating its use for the special case of distributional regression and section \ref{egam} covering the simplified methods that can be used in the even more restricted case of models with a similar structure to generalized additive models. Section \ref{dist.sec} then develops approximate distributional results accounting for smoothing parameter uncertainty which are applied in section \ref{new.aic} to propose a corrected AIC suitable for the general model class. The remaining sections present simulation results and examples, while extensive further background, and details for particular models, are given in the supplementary appendices (referred to as `SA A', `SA B' etc., below).

\section{The general framework \label{gen.fram}}

Consider a model for an $n$-vector of data, $\bf y$, constructed in terms of unknown parameters, $\bm \theta$, and some unknown functions, $g_j$, of covariates, $x_j$. Suppose that the log likelihood for this model satisfies the Fisher regularity conditions, has 4 continuous derivatives and can be written $l({\bm \theta}, g_1,g_2,\ldots,g_M) = \log f({\bf y}|{\bm \theta}, g_1,g_2,\ldots,g_M)$. In contrast to the usual GAM case the likelihood need not be based on a single parameter exponential family distribution, and we do not assume that the log likelihood can be written in terms of a single additive linear predictor.  Now let the $g_j(x_j)$ be represented via basis expansions of modest rank ($k_j$),
$$
g_j(x) = \sum_{i=1}^{k_j} \beta_{ji} b_{ji}(x)
$$
where the $\beta_{ji}$ are unknown coefficients and the $b_{ji}(x)$ are known basis functions such as splines, usually chosen to have good approximation theoretic properties. With each $g_j$ is associated a smoothing penalty, which is quadratic in the basis coefficients and measures the complexity of $g_j$.
 Writing all the basis coefficients and $\bm \theta$ in one $p$-vector $\bm \beta$, then the $j^{\text{th}}$ smoothing penalty can be written as $\bp \ts {\bf S}^j \bp$, where ${\bf S}^j$ is a matrix of known coefficients, but generally has only a small non zero block. The estimated model coefficients are then
\beq
\hat \bp = \underset{\beta}{\text{argmax}}%\argmax_{\beta} 
\left \{l(\bp) - \frac{1}{2} \sum_j^M \lambda_j \bp \ts {\bf S}^j \bp\right \} \label{plig}
\eeq
given $M$ smoothing parameters, $\lambda_j$, controlling the extent of penalization. A slight extension is that the smoothing penalties may be such that several $\lambda_i \bp \ts {\bf S}^i \bp$ are associated with one $g_j$, for example when $g_j$ is a non-isotropic function of several variables. Note also that the framework can incorporate Gaussian random effects, provided the corresponding precision matrices can be written as $\sum \lambda_i \bp \ts {\bf S}^i \bp$ (where the ${\bf S}^i$ are known).

From a Bayesian viewpoint $\hat \bp$ is a posterior mode for $\bp$. The Bayesian approach views the smooth functions as intrinsic Gaussian random fields with prior $f_\lambda$ given by $N({\bf 0},{\bf S}^{\lambda -})$ where ${\bf S}^{\lambda -}$ is a Moore-Penrose (or other suitable) pseudoinverse of $ \sum_j \lambda_j {\bf S}^j$. Then the posterior modes are $\hat \bp $ from (\ref{plig}), and in the large sample limit, assuming fixed smoothing parameter vector, $\bm \lambda$, we have $\bp|{\bf y} \sim N(\hat \bp, (\bm{{\cal I}}+{\bf S}^\lambda)^{-1})$, where $\bm{{\cal I}}$ is the expected negative Hessian of the log-likelihood (or its observed version) at $\hat \bp$. An empirical Bayesian approach is appealing here as it gives well calibrated inference for the $g_j$ \citep{wahba83,silverman85,nychka88,marra.wood2012} in a GAM context. Appropriate summations of the elements of $\text{diag}\{(\bm{{\cal I}}+{\bf S}^\lambda)^{-1}\bm{{\cal I}}\}$ provide estimates of the `effective degrees of freedom' of the whole model, or of individual smooths. 

Under this Bayesian view, smoothing parameters can be estimated to maximize the log marginal likelihood 
\beq
{\cal V}_r({\bm \lambda}) = \log \int f({\bf y}|\bp)f_\lambda(\bp) d \bp, \label{ML}
\eeq
or a Laplace approximate version of this \citep[e.g.][]{wood2011}. In practice optimization is with respect to $\bm \rho$ where $\rho_i = \log \lambda_i$. Marginal likelihood estimation of smoothing parameters in a Gaussian context goes back to \cite{anderssen1974} and \cite{wahba85}, while \citep{shun1995laplace} show that Laplace approximation of more general likelihoods is theoretically well founded. That marginal likelihood is equivalent to REML \cite[in the sense of][]{laird.ware} supports its use when the model contains Gaussian random effects.  Theoretical work by \cite{reiss.ogden2009} also suggests practical advantages at finite sample sizes, in that marginal likelihood is less prone to multiple local minima than GCV (or AIC). Supplementary Appendix B (SA B) also demonstrates how Laplace approximate marginal likelihood (LAML) estimation of smoothing parameters maintains statistical consistency of reduced rank spline estimates. The use of Laplace approximation and demonstration of statistical consistency requires the assumption that $\text{dim}(\bp) = O(n^\alpha)$ where $\alpha < 1/3$.

\section{Smoothness selection methods \label{smooth.sec}}

This section describes the general smoothness selection method, and a simplified method for the special case in which the likelihood is a simple sum of terms for each observation of a univariate response, and there is a single GAM like linear predictor.

The nonlinear dependencies implied by employing a general smooth likelihood result in unwieldy expressions unless some care is taken to establish  a compact notation. In the rest of this paper, Greek subscripts denote partial differentiation with respect to the given variable, while Roman superscripts are indices associated with the derivatives. Hence $D\indices*{*^i_\beta^j_\theta} = \ilpddif{D}{\beta_i}{\theta_j}$. Similarly $D\indices*{*^i_{\hat \beta}^j_\theta} = \ilpddif{D}{\beta_i}{\theta_j} |_{\hat \beta}$. Roman subscripts denote vector or array element indices. For matrices the first Roman sub- or super-script denotes rows, the second columns. Roman superscripts without a corresponding Greek subscripts are labels, for example ${\bm \beta}^1$ and ${\bm \beta}^2$ denote two separate vectors ${\bm \beta}$. For Hessian matrices only, $D\indices*{*_i^\beta_j^\theta} $ is element $i,j$ of the inverse of the matrix with elements $D\indices*{*^i_\beta^j_\theta}$.
If any Roman index appears in two or more multiplied terms, but the index is absent on the other side of the equation, then a summation over the product of the corresponding terms is indicated (the usual Einstein summation convention being somewhat unwieldy in this context). To aid readability, in this paper summation indices will be highlighted in bold. For example the equation $a_{{\sm i}j}b_{{\sm i}k}c^{{\sm i}l} + d_{jkl} = 0$ is equivalent to $\sum_i a_{ij}b_{ik}c^{il} + d_{jkl} = 0 $. An indexed expression not in an equation is treated like an equation with no indices on the other side (so $a_{i{\sm j}}b_{\sm j}$ is interpreted as $\sum_j a_{ij} b_j$).

\subsection{General model estimation \label{general.model}}

Consider the general case in which the log likelihood depends on several smooth functions of predictor variables, each represented via a basis expansion and each with one or more associate penalties.  The likelihood may also depend on some strictly parametric model components. The log likelihood is assumed to satisfy the Fisher regularity conditions and in addition we usually assume that it has 4 bounded continuous derivatives with respect to the parameters (with respect to $g_j(x)$ for any relevant fixed $x$ in the case of a smooth, $g_j$). Let the model coefficients be $\bm \beta$ (recalling that this includes the vector $\bm \theta$ of parametric coefficients and nuisance parameters). The penalized log likelihood is then
$$
{\cal L}({\bm \beta}) = l({\bm \beta}) - \frac{1}{2} \lambda_{\sm j} {\bm \beta} \ts {\bf S}^{\sm j} {\bm \beta},
$$
and we assume that the model is well enough posed that this has a positive definite maximum (at least after dealing with any parameter redundancy issues that can be addressed by linear constraint). Let $\hat {\bm \beta}$ be the maximizer of $\cal L$ and let $\bm{{\cal H}}$ be the negative Hessian, with elements $- {\cal L}\indices*{*^i_{\hat \beta}^j_{\hat \beta}}$. The log LAML (see SA C) is 
$$
{\cal V}({\bm \lambda}) = {\cal L}(\hat {\bm \beta}) + \frac{1}{2} \log{|{\bf S}^\lambda|_+} - \frac{1}{2} \log{|\bm{{\cal H}}|} + \frac{M_p}{2} \log(2 \pi),
$$  
where ${\bf S}^\lambda = \lambda_{\sm j} {\bf S}^{\sm j}$ and $|{\bf S}^\lambda|_+$ is the product of the positive eigenvalues of ${\bf S}^\lambda$. $M_p$ is the number of zero eigenvalues of ${\bf S}^\lambda$, when all $\lambda_j$ are strictly positive. The basic strategy is to optimize $\cal V$ with respect to ${\bm \rho} = \log ({\bm \lambda})$ via Newton's method. This requires $\hat {\bm \beta}$
to be obtained for each trial $\bm \rho$ via an inner Newton iteration, and derivatives of $\hat {\bm \beta}$ must be obtained by implicit differentiation. The log determinant computations have the potential to be computationally unstable, and reparameterization is needed to deal with this. The full Newton method based on computationally exact derivatives has the substantial practical advantage that it can readily be detected when ${\cal V}$ is indefinite with respect to a particular $\rho_i$, since then $\ilpdif{{\cal V}}{\rho_i} = \ilpdif{^2 {\cal{V}}}{\rho_i^2} \simeq 0$. Such indefiniteness occurs when a smoothing parameter, $\lambda_i, \to \infty$ or a variance component tends to zero, both of which are perfectly legitimate. Dropping a $\rho_i$ from Newton update when such indefiniteness is detected ensures that it takes a value which can be treated as `working infinity' without overflowing. Methods which use an approximate Hessian, or none, do not have this advantage.  

The proposed general method consists of outer and inner iterations, as follows.

\bigskip
\noindent {\bf Outer algorithm for $\bm \rho$}
\begin{enumerate}
\item Obtain initial values for ${\bm \rho} = \log (\bm \lambda)$, to ensure that the effective degrees of freedom of each smooth lies away from its maximum or minimum possible values.
\item Find initial $\hat \bp$ guesstimates (model specific).
\item Perform the initial reparameterizations required in section \ref{ldet.fix} to facilitate stable computation of $\log |{\bf S}^\lambda|_+$.
\item Repeat the following standard Newton iteration until convergence is detected at step (c)\ldots
\begin{enumerate}
\item Find $\hat \bp$, ${\cal V}_\rho^i$ and  ${\cal V}_{\rho\rho}^{ij}$ by the inner algorithm.
\item Drop any ${\cal V}_\rho^i$, ${\cal V}_{\rho\rho}^{ij}$ and ${\cal V}_{\rho\rho}^{ji}$ for which ${\cal V}_\rho^i \simeq {\cal V}_{\rho\rho}^{ii}\simeq 0$. Let ${\mathbb I}$ denote the indices of the retained terms.
\item Test for convergence. i.e. all ${\cal V}_\rho^i \simeq 0 $ and the Hessian (elements $-{\cal V}_{\rho\rho}^{ji}$) is positive semi-definite.
\item If necessary perturb the Hessian (elements $-{\cal V}_{\rho\rho}^{ji}$) to make it positive definite (guaranteeing that the Newton step will be a descent direction).
\item Define ${\bm \Delta}_{\mathbb I}$ as the subvector of ${\bm \Delta}$ indexed by $\mathbb I$, with elements $ - {\cal V}^{\rho\rho}_{i{\sm j}}{\cal V}^{\sm j}_\rho$, and set $\Delta_j = 0 $ $\forall$ $j \notin {\mathbb I}$.
\item While ${\cal V}({\bm \rho} + {\bm \Delta}) < {\cal V}({\bm \rho})$ set ${\bm \Delta} \leftarrow {\bm \Delta}/2$.
\item Set ${\bm \rho}\leftarrow{\bm \rho} + {\bm \Delta}$.
\end{enumerate}
\item Reverse the step 3 reparameterization.
\end{enumerate}

The method for evaluating $\cal V$ and its gradient and Hessian with respect to $\bm \rho$ is as follows, where ${\cal L}\indices*{*^{\hat \beta}_k^{\hat \beta}_j}$ denotes the inverse of ${\cal L}\indices*{*_{\hat \beta}^k_{\hat \beta}^j}$. 

\bigskip
\noindent {\bf Inner algorithm for $\bm \beta$}
\begin{enumerate}
\item Reparameterize to deal with any `type 3' penalty blocks as described in section \ref{ldet.fix}, so that computation of $\log |{\bf S}^\lambda|_+$ is stable, and evaluate the derivatives of $\log |{\bf S}^\lambda|_+$.
\item Use Newton's method to find $\hat {\bm\beta}$, regularizing the Hessian, and applying step length control, to ensure convergence even when the Hessian is indefinite and/or $\hat {\bm \beta} $ is not identifiable as described in section \ref{newton.sec}. 
\item Test for identifiability of $\hat {\bm \beta} $ at convergence by examining the rank of the $\bm{{\cal H}}$ as described in section \ref{newton.sec}. Drop unidentifiable coefficients. 
\item If coefficients were dropped, find the reduced $\hat {\bm\beta}$ by further steps of Newton's method (section \ref{newton.sec}). 
\item Compute $ \ildif{{\hat \beta}_i}{\rho_k} = {\cal L}\indices*{*^{\hat \beta}_i^{\hat \beta}_{\sm j}} \lambda_k S^k_{{\sm j}{\sm l}} {\hat \beta}_{\sm l}$ 
and hence $
l\indices*{*_{\hat \beta}^i_{\hat \beta}^j_\rho^l} = l\indices*{*_{\hat \beta}^i_{\hat \beta}^j_{\hat \beta}^{\sm k}}
\ildif{\hat \beta_{\sm k}}{\rho_l} 
$ (section \ref{imp.sec}).
\item Compute $
\ilddif{\hat \beta_i}{\rho_k}{\rho_l} = 
{\cal L}\indices*{*^{\hat \beta}_i^{\hat \beta}_{\sm j}} \left \{
\left ( -l\indices*{*_{\hat \beta}^{\sm j}_{\hat \beta}^{\sm p}_\rho^l} + \lambda_l S^l_{{\sm j}{\sm p}} 
\right ) \ildif{{\hat \beta}_{\sm p}}{\rho_k}
+ \lambda_k S^k_{{\sm j}{\sm p}} \ildif{{\hat \beta}_{\sm p}}{\rho_l}
\right \} + \delta_{k}^l \ildif{{\hat \beta}_i}{\rho_k},
$ (section \ref{imp.sec}).
\item Compute ${\cal L}\indices*{*^{\hat \beta}_{\sm k}^{\hat \beta}_{\sm j}} l\indices*{*^{\sm j}_{{\hat\beta}}^{\sm k}_{{\hat \beta}}_\rho^p_\rho^v}$ (model specific). (\ref{imp.sec})
\item The derivatives of $\cal V$ can now be computed according to section \ref{dlaml.sec}.
\item For each parameter dropped from $\hat \bp$ during fitting, zeroes must be inserted in $\hat \bp$, $\ilpdif{\hat \bp}{\rho_j}$ and the corresponding rows and columns of ${\cal L}\indices*{*^{\hat \beta}_{k}^{\hat \beta}_{ j}}$. The step 1 reparameterization is then reversed.
\end{enumerate}
The following subsections fill in the method details, but note that in order to implement a particular model in this class it is necessary to be able to compute, $l$, $l_{\beta}^i$ and $l\indices*{*^i_\beta^j_\beta}$, given $\bm \beta$, along with $l\indices*{*^i_{\hat \beta}^j_{\hat \beta}^k_\rho}$ given $\ildif{\hat {\bm \beta}}{\rho_k}$, and ${\cal L}\indices*{*^{\hat \beta}_{\sm k}^{\hat \beta}_{\sm j}} l\indices*{*^{\sm j}_{{\hat\beta}}^{\sm k}_{{\hat \beta}}_\rho^p_\rho^v}$ given $\ilddif{\hat {\bm\beta}}{\rho_k}{\rho_l}$. The last of these is usually computable much more efficiently than if $l\indices*{*^j_{{\hat\beta}}^k_{{\hat \beta}}_\rho^p_\rho^v}$ was computed explicitly.

\subsubsection{Derivatives and stable evaluation of $\log |{\bf S}^\lambda|_+$\label{ldet.fix}}

This section covers the details for outer step 3 and inner step 1. Stable evaluation of the log determinant terms is the key to stable computation with the LAML. SA C explains the issue. \cite{wood2011} proposes a solution which involves orthogonal transformation of the whole parameter vector $\bm \beta$, but in the general case the likelihood may depend on each smooth function separately and such a transformation is therefore untenable. It is necessary to develop a reparameterization strategy which does not combine coefficients from different smooths. This is possible if we recognise that ${\bf S}^\lambda$ is block diagonal, with different blocks relating to different smooths. For example, if $\mathbb{S}^j$ denotes the non-zero sub-block of ${\bf S}^j$,

\singlespacing \matsize 
\vspace*{-.8cm}
$$
{\bf S}^\lambda = \bmat{ccccc}
\lambda_1 \mathbb{S}^1 & . & . & . & . \\
. & \lambda_2 \mathbb{ S}^2 & . & . & . \\
. & . &  \lambda_{\sm j} \mathbb{ S}^{\sm j} & . & . \\
. & . & . & . & . \\
. & . & . & . & . \emat.
$$
\normalsize %\doublespacing
That is there are some blocks with single smoothing parameters, and others with a more complicated additive structure. There are usually also some zero blocks on the diagonal. The block structure means that the generalized determinant, its derivatives with respect to $\rho_k = \log \lambda_k$ and the matrix square root of ${\bf S}^\lambda$ can all be computed blockwise. So for the above example,
$$
\log |{\bf S}^\lambda|_+ = {\rm rank}(\mathbb{S}^1)\log (\lambda_1) +  \log |\mathbb{ S}^1|_+ + {\rm rank}(\mathbb{ S}^2)\log (\lambda_2) +  \log |\mathbb{ S}^2|_+ +
\log  | \lambda_{\bf j} \mathbb{S}^{\bf j} |_+ + \cdots
$$
For any $\rho_k$ relating to a single parameter block we have
$$
\pdif{\log |{\bf S}^\lambda |_+}{\rho_k} = \text{rank}(\mathbb{ S}^k)
$$
and zero second derivatives. For multi-$\lambda$ blocks there will generally be first and second derivatives to compute. There are no second derivatives `between-blocks'.

In general, there are three block types, each requiring different pre-processing.
\begin{enumerate}
\item Single parameter diagonal blocks. A reparameterization can be used so that all non-zero elements are one, and the rank pre-computed.
\item Single parameter dense blocks. An orthogonal reparameterization, based on the eigenvectors of the symmetric eigen-decomposition of the block, can be used to make these blocks look like the previous type (by similarity transform). Again the rank is computed.
\item Multi-$\lambda$ blocks will require the reparameterization method of \cite{wood2011} appendix B to be applied for each new $\bm \rho$ proposal, since the numerical problem that the re-parameterization avoids is $\bm \rho$ dependent (see SA C). Initially, before the smoothing parameter selection iteration, it is necessary to reparameterize in order to separate the parameters corresponding to the block into penalized and unpenalized sub-vectors. This initial re-parameterization can be based on the eigenvectors of the symmetric eigen decomposition of the `balanced' version of the block penalty matrix, $\sum_j \mathbb{ S}^j/\|\mathbb{ S}^j\|_F$, where $\| \cdot \|_F$ is the Frobenious norm. The balanced penalty is used for maximal numerical stability, and is usable because formally the spaces for the penalized and unpenalized components do not change with the smoothing parameters.    
\end{enumerate}
The reparameterizations from each block type are applied to the model, usually to the model matrices $\X^j$ of the individual smooth terms. The reparameterization information must be stored so that we can return to the original parameterization at the end. 

After the one off initial reparameterization just described, then step one of the inner algorithm requires only that the reparameterization method of \cite{wood2011} Appendix B be applied to the parameters corresponding to type 3 blocks, for each new set of smoothing parameters. 

\subsubsection{Newton iteration for $\hat {\bm \beta}$\label{newton.sec}}

This section provides details for inner steps 2-4. Newton iteration for $\hat {\bm \beta}$ requires the gradient vector, $\bm{{\cal G}}$, with elements 
$
{\cal L}\indices*{*^i_{\beta}} = l\indices*{*^i_{\beta}} - \lambda_{\sm k} S^{\sm k}_{i{\sm j}} { \beta}_{\sm j}
$
and negative Hessian matrix $\bm{{\cal H}}$ with elements 
$
-{\cal L}\indices*{*^i_{\beta}^j_{\beta}} = -l\indices*{*^i_{\beta}^j_{\beta}} + \lambda_{\sm k} S^{\sm k}_{ij}
$ 
(we will also use $\bf H$ to denote the Hessian of the negative unpenalized log likelihood with elements $-l\indices*{*^i_{\beta}^j_{\beta}}$). In principle Newton iteration proceeds by repeatedly setting $\bm \beta$ to ${\bm \beta} + {\bm \Delta} $, where ${\bm \Delta } = \bm{{\cal H}}^{-1}\bm{{\cal G}}$. In practice, Newton's method is only guaranteed to converge to a maximum of $\cal L$, provided (i) that the Hessian is perturbed to be positive definite if it is not, guaranteeing that the Newton direction is an ascent direction, (ii) that step reduction is used to ensure that the step taken actually increases $\cal L$ and (iii) that the computation of the step is numerically stable \citep[see][]{nocedal.wright}. 

$\cal L$ may be indefinite away from a maximum, but even near the maximum there are two basic impediments to stability and positive definiteness. Firstly, some elements of $\bm \beta$ may be unidentifiable. This issue will be dealt with by dropping parameters at convergence, as described shortly. The second issue is that some smoothing parameters may legitimately become very large during fitting, resulting in very large $\lambda_j {\bf S}^j$ components, poor scaling, poor conditioning and hence computational singularity. However, given the initial and step 1 reparameterizations such large elements can be dealt with by diagonal pre-conditioning of $\bm{{\cal H}}$. That is define diagonal matrix $\bf D$ such that $D_{ii} = |{\cal H}_{ii}|^{-1/2}$, and preconditioned Hessian $\bm{{\cal H}}^\prime ={\bf D}\bm{{\cal H}}{\bf D} $. Then $\bm{{\cal H}}^{-1} =  {\bf D}\bm{{\cal H}}^{\prime-1}{\bf D} $, with the right hand side resulting in much better scaled computation. %This step is only applied if all diagonal elements of $\bm{{\cal H}}$ are positive (formally set ${\bf D}_{ii}=1$), since otherwise the Hessian is anyway indefinite.  
In the work reported here the pivoted Cholesky decomposition of the perturbed Hessian ${\bf R}\ts {\bf R} = \bm{{\cal H}}^\prime + \epsilon {\bf I}$ is repeated with increasing $\epsilon$, starting from zero, until positive definiteness is obtained. The Newton step is then computed as
$
{\bm \Delta} = {\bf D}{\bf R}^{-1} {\bf R}\its {\bf D}\bm{{\cal G}}.
$
If the step to ${\bm \beta} + {\bm \Delta}$ fails to increase the likelihood then ${\bm \Delta}$ is repeatedly halved until it does. Note that the perturbation of the Hessian does not change the converged state of a Newton algorithm (although varying the perturbation strength can change the algorithm convergence rate). 

At convergence $\bm{{\cal H}}$ can at worst be positive semi- definite, but it is necessary to test for the possibility that some parameters are unidentifiable. The test should not depend on the particular values of the smoothing parameters. This can be achieved by constructing the balanced penalty $ {\bf S} = \sum_j {\bf S}^j/\|{\bf S}^j\|_F$ ($\|\cdot \|_F$ is the Frobenius norm, but another norm could equally well be used), and then forming the pivoted Cholesky decomposition ${\bf P}\ts{\bf P} = {\bf H}/\|{\bf H}\|_F + {\bf S}/\|{\bf S}\|_F $.  The rank of $\bf P$ can then be estimated by making use of \cite{cline1979}. If this reveals rank deficiency of order $q$ then the coefficients corresponding to the matrix rows and columns pivoted to the last $q$ positions should be dropped from the analysis. The balanced penalty is used to avoid dropping parameters simply because some smoothing parameters are very large. Given the non-linear setting it is necessary to repeat the Newton iteration to convergence with the reduced parameter set, in order that the remaining parameters adjust to the omission of those dropped. 

\subsubsection{Implicit differentiation \label{imp.sec}}

This section provides the details for inner steps 5-7. We obtain the derivatives of the identifiable elements of $\hat {\bm \beta}$ with respect to $\bm \rho$. All computations here are in the reduced parameter space, if parameters were dropped. At the maximum penalized likelihood estimate we have
$
{\cal L}_{\hat \beta}^i = l_{\hat \beta}^i - \lambda_{\sm k} S^{\sm k}_{i{\sm j}} {\hat \beta}_{\sm j} = 0
$
and differentiating with respect to $\rho_k = \log \lambda_k$ yields
$$
{\cal L}\indices*{^i_{\hat \beta}^k_\rho} = l\indices*{^i_{\hat \beta}^{\sm j}_{\hat \beta}} \dif{{\hat \beta}_{\sm j}}{\rho_k} - \lambda_k S^k_{i{\sm j}} {\hat \beta}_{\sm j} - \lambda_{\sm l} S^{\sm l}_{i{\sm j}} \dif{{\hat \beta}_{\sm j}}{\rho_k} = 0
\text{ and re-arranging, } 
\dif{{\hat \beta}_i}{\rho_k} = {\cal L}\indices*{*^{\hat \beta}_i^{\hat \beta}_{\sm j}} \lambda_k S^k_{{\sm j}{\sm l}} {\hat \beta}_{\sm l},
$$
given which we can compute $
l\indices*{*_{\hat \beta}^i_{\hat \beta}^j_\rho^l} = l\indices*{*_{\hat \beta}^i_{\hat \beta}^j_{\hat \beta}^{\sm k}}
\ildif{\hat \beta_{\sm k}}{\rho_l} 
$ 
from the model specification.  $-l\indices*{*_{\hat \beta}^i_{\hat \beta}^j_\rho^l} +\delta_{\sm k}^l \lambda_{\sm k} S^{\sm k}_{ij} $ are the elements of $\ilpdif{\bm{{\cal H}}}{\rho_l}$, required in the next section ($\delta^l_k$ is 1 for $l=k$ and 0 otherwise).  Then 
$$
\ddif{\hat \beta_i}{\rho_k}{\rho_l} = 
{\cal L}\indices*{*^{\hat \beta}_i^{\hat \beta}_{\sm j}} \left \{
\left ( -l\indices*{*_{\hat \beta}^{\sm j}_{\hat \beta}^{\sm p}_\rho^l} + \lambda_l S^l_{{\sm j}{\sm p}} 
\right ) \dif{{\hat \beta}_{\sm p}}{\rho_k}
+ \lambda_k S^k_{{\sm j}{\sm p}} \dif{{\hat \beta}_{\sm p}}{\rho_l}
\right \} + \delta_{k}^l \dif{{\hat \beta}_i}{\rho_k},
$$
which enables computations involving $\ilpddif{\bm{{\cal H}}}{\rho_k}{\rho_l}$, with elements $
- l\indices*{*^i_{{\hat\beta}}^j_{{\hat \beta}}_\rho^k_\rho^l} + \delta_{\sm k}^l \lambda_{\sm k} S^{\sm k}_{ij}$, and 
$$
l\indices*{*^i_{{\hat\beta}}^j_{{\hat \beta}}_\rho^k_\rho^l} = 
l\indices*{*^i_{{\hat \beta}}^j_{{\hat \beta}}^{\sm r}_{{\hat \beta}}^{\sm t}_{\hat \beta}} \dif{\hat \beta_{\sm r}}{\rho_k} \dif{\hat \beta_{\sm t}}{\rho_l} + 
l\indices*{*^i_{{\hat \beta}}^j_{{\hat \beta}}^{\sm r}_{{\hat \beta}}} \ddif{\hat \beta_{\sm r}}{\rho_k}{\rho_l}.
$$
As mentioned in section \ref{general.model}, it will generally be inefficient to form this last quantity explicitly, as it occurs only in the summations involved in computing the final trace in (\ref{deriv2}).

\subsubsection{The remaining derivatives\label{dlaml.sec}}

Recalling that $\bm{{\cal H}}$ is the matrix with elements 
$
-{\cal L}\indices*{*^i_{\beta}^j_{\beta}} = -l\indices*{*^i_{\beta}^j_{\beta}} + \lambda_{\sm k} S^{\sm k}_{ij}
$, we require (inner step 8)
$$
\pdif{{\cal V}}{\rho_k} = - \frac{\lambda_k}{2} \hat \bp \ts {\bf S}^k \hat \bp +
%%{\cal L}_{\hat \beta}^j\dif{\hat \beta_j}{\rho_k} + 
\frac{1}{2} \pdif{\log|{\bf S}^\lambda|_+}{\rho_k} - \frac{1}{2} \pdif{\log|\bm{{\cal H}}|}{\rho_k}
$$
and
$$
\pddif{{\cal V}}{\rho_k}{\rho_l} = - \delta_k^l  \frac{\lambda_k}{2} \hat \bp \ts {\bf S}^k \hat \bp
-  \dif{\hat {\bm \beta}\ts}{\rho_l}\bm{{\cal H}} \dif{\hat {\bm \beta}}{\rho_k} 
%% +{\cal L}^j_{\hat \beta} \ddif{\hat \beta_j}{\rho_k}{\rho_l} 
+ \frac{1}{2} \pddif{\log|{\bf S}^\lambda|_+}{\rho_k}{\rho_l}
 - \frac{1}{2} \pddif{\log|\bm{{\cal H}}|}{\rho_k}{\rho_l},
$$
where components involving ${\cal L}_{\hat \beta}^j$ are zero by definition of $\hat \bp$.
The components not covered so far are 
\beq
\pdif{\log|\bm{{\cal H}}|}{\rho_k} = {\rm tr} \left ( \bm{{\cal H}}^{-1} \pdif{\bm{{\cal H}}}{\rho_k}  \right )
\text{ and }
\pddif{\log|\bm{{\cal H}}|}{\rho_k}{\rho_l} = -{\rm tr} \left ( \bm{{\cal H}}^{-1} \pdif{\bm{{\cal H}}}{\rho_k} 
\bm{{\cal H}}^{-1} \pdif{\bm{{\cal H}}}{\rho_l}  \right ) +
{\rm tr} \left ( \bm{{\cal H}}^{-1} \pddif{\bm{{\cal H}}}{\rho_k}{\rho_l}  \right ). \label{deriv2}
\eeq
The final term above is expensive if computed naively by explicitly computing each term $\ilpddif{\bm{{\cal H}}}{\rho_k}{\rho_l}$, but this is unnecessary and the computation of ${\rm tr} \left ( \bm{{\cal H}}^{-1} \ilpddif{\bm{{\cal H}}}{\rho_k}{\rho_l}  \right ) $ can usually be performed efficiently as the final part of the model specification, keeping the total cost to $O(Mnp^2)$: see SA G and section \ref{gamlss.sec} for illustrative examples. 

The \cite{cox1972} proportional hazards model provides a straightforward application of the general method, and the requisite computations are set out in SA G in a manner that maintains  $O(Mnp^2)$ computational cost. Another example is the multivariate additive model, in which the means of a multivariate Gaussian response are given by separate linear predictors, which may optionally share terms. This model is covered in SA H and section \ref{mvn.mpg}. Section \ref{gamlss.sec} considers how another class of models falls into the general framework.

\subsection{A special case: GAMLSS models\label{gamlss.sec}}

\renewcommand{\baselinestretch}{1}
\begin{figure}
\eps{-90}{.55}{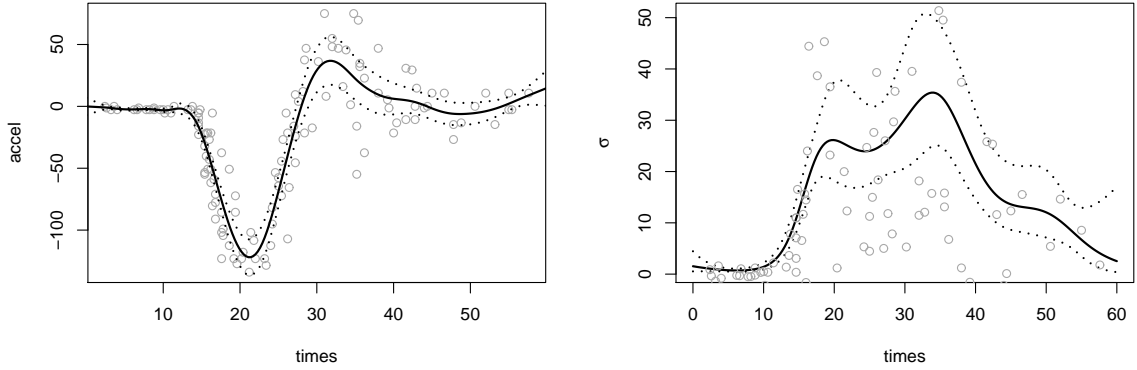}\vspace*{-.5cm}
\caption{\small A smooth Gaussian location scale model fit to the motorcycle data from \cite{silverman85}, using the methods developed in section \ref{gamlss.sec}. The left plot shows the raw data as open circles and an adaptive p-spline smoother for the mean overlaid. The right plot shows the simultaneous estimate of the standard deviation in the acceleration measurements, with the absolute values of the residuals as circles. Dotted curves are approximate 95\% confidence intervals.   The effective degrees of freedom of the smooths are 12.5 and 7.3 respectively.
 \label{mcycle.fig}}
\end{figure}
\renewcommand{\baselinestretch}{\dsp}

The GAMLSS (or `distributional regression') models discussed by \cite{rigby2005} \citep[and also][]{yee1996,klein2014dr,klein2015dr} fall within the scope of the general method. The idea is that we have independent univariate response observations, $y_i$, whose distributions depend on several unknown parameters, each of which is determined by its own linear predictor. The log likelihood is a straightforward sum of contributions from each $y_i$ (unlike the Cox models, for example), and the special structure can be exploited so that implementation of new models in this class requires only the supply of some derivatives of the log likelihood terms with respect to the distribution parameters.  Given the notational conventions established previously, the expressions facilitating this are rather compact (without such a notation they can easily become intractably complex). 

Let the log likelihood for the $i^{\rm th}$ observation be $l(y_i,\eta^1_i,\eta^2_i,\ldots)$ where the $\eta^k = {\bf X}^k {\bm \beta}^k$ are $K$ linear predictors. The Newton iteration for estimating ${\bm \beta} = ({\bm \beta}^{1\sf T},{\bm \beta}^{2\sf T},\ldots)\ts$ requires
$
l_{\beta^l}^j = l\indices*{*^{\sm i}_{\eta^l}} X^l_{{\sm i}j} $ 
and
$l\indices*{*^j_{\beta^l}^k_{\beta^m}} = l\indices*{*^{\sm i}_{\eta^l}^{\sm i}_{\eta^m}} X^l_{{\sm i}j} X^m_{{\sm i}k},
$
which are also sufficient for first order implicit differentiation.

LAML optimization also requires
%\begin{eqnarray*}
$$
l\indices*{*^j_{{\hat\beta}^l}^k_{{\hat \beta}^m}_\rho^p} =
l\indices*{*^j_{{\hat \beta}^l}^k_{{\hat \beta}^m}^{\sm r}_{{\hat \beta}^{\sm q}}} \dif{\hat \beta^{\sm q}_{\sm r}}{\rho_p} 
= 
l\indices*{*^{\sm i}_{{\hat \eta}^l}^{\sm i}_{{\hat \eta}^m}^{\sm i}_{{\hat \eta}^{\sm q}}}
 X_{{\sm i}j}^l X_{{\sm i}k}^m X_{{\sm i}{\sm r}}^{\sm q} \dif{\hat \beta^{\sm q}_{\sm r}}{\rho_p} = %\\ &=& 
l\indices*{*^{\sm i}_{{\hat \eta}^l}^{\sm i}_{{\hat \eta}^m}^{\sm i}_{{\hat \eta}^{\sm q}}} X_{{\sm i}j}^l X_{{\sm i}k}^m \dif{\hat \eta^{\sm q}_{\sm i}}{\rho_p}.
$$
%\end{eqnarray*}
Notice how this is just an inner product ${\bf X}\ts {\bf V} {\bf X}$, where the diagonal matrix ${\bf V}$ is the sum over $q$ of some diagonal matrices.
At this stage the second derivatives of $\hat {\bm \beta}$ with respect to $\bm \rho$ can be computed, after which we require only
\begin{eqnarray*}
l\indices*{*^j_{{\hat\beta}^l}^k_{{\hat \beta}^m}_\rho^p_\rho^v} &=& 
l\indices*{*^j_{{\hat \beta}^l}^k_{{\hat \beta}^m}^{\sm r}_{{\hat \beta}^{\sm q}}^{\sm t}_{\hat \beta^{\sm s}}} \dif{\hat \beta^{\sm q}_{\sm r}}{\rho_p} \dif{\hat \beta^{\sm s}_{\sm t}}{\rho_v} + 
l\indices*{*^j_{{\hat \beta}^l}^k_{{\hat \beta}^m}^{\sm r}_{{\hat \beta}^{\sm q}}} \ddif{\hat \beta^{\sm q}_{\sm r}}{\rho_p}{\rho_v}\\ &=& 
l\indices*{*^{\sm i}_{{\hat \eta}^l}^{\sm i}_{{\hat \eta}^m}^{\sm i}_{{\hat \eta}^{\sm q}}^{\sm i}_{\hat \eta^{\sm s}}}X_{{\sm i}j}^l X_{{\sm i}k}^m \dif{\hat \eta^{\sm q}_{\sm i}}{\rho_p} \dif{\hat \eta^{\sm s}_{\sm i}}{\rho_v} + 
l\indices*{*^{\sm i}_{{\hat \eta}^l}^{\sm i}_{{\hat \eta}^m}^{\sm i}_{{\hat \eta}^{\sm q}}}X_{{\sm i}j}^l X_{{\sm i}k}^m  \ddif{\hat \eta^{\sm q}_{\sm i}}{\rho_p}{\rho_v}.
\end{eqnarray*}
So to implement a new family for GAMLSS estimation requires mixed derivatives up to fourth order with respect to the parameters of the likelihood. In most cases what would be conveniently available is e.g. $l\indices*{*^i_{{\hat \mu}^l}^i_{{\hat \mu}^m}^i_{{\hat \mu}^q}^i_{\hat \mu^s}}$ rather than 
$l\indices*{*^i_{{\hat \eta}^l}^i_{{\hat \eta}^m}^i_{{\hat \eta}^q}^i_{\hat \eta^s}}$, where $\mu^k$ is the $k^\text{th}$ parameter of the likelihood and is given by $h^k(\mu^k) = \eta^k$, $h^k$ being a link function. 

To get from the $\mu$ derivatives to the $\eta$ derivatives, the rules (\ref{eta.mu1}) - (\ref{eta.mu4}) from Appendix \ref{egam.imd} are used. This is straightforward for any derivative that is not mixed. For mixed derivatives containing at least one first order derivative the transformation rule applying to the highest order derivative is applied first, followed by the transformations for the first order derivatives. This leaves only the transformation of $l\indices*{*^i_{{\hat \mu}^j}^i_{{\hat \mu}^j}^i_{{\hat \mu}^k}^i_{\hat \mu^k}}$ as at all awkward, but we have
$$
l\indices*{*^i_{{\hat \eta}^j}^i_{{\hat \eta}^j}^i_{{\hat \eta}^k}^i_{\hat \eta^k}} = 
(l\indices*{*^i_{{\hat \mu}^j}^i_{{\hat \mu}^j}^i_{{\hat \mu}^k}^i_{\hat \mu^k}}/h^{j\prime 2}_i - 
l\indices*{*^i_{{\hat \mu}^j}^i_{{\hat \mu}^k}^i_{\hat \mu^k}} h^{j\prime\prime}_i/h^{j\prime 3}_i )/ h^{k\prime 2}_i- 
(l\indices*{*^i_{{\hat \mu}^j}^i_{{\hat \mu}^j}^i_{{\hat \mu}^k}}/h^{j\prime 2}_i - 
l\indices*{*^i_{{\hat \mu}^j}^i_{{\hat \mu}^k}} h^{j\prime\prime}_i/h^{j\prime 3}_i
)h^{k\prime\prime}_i/h^{k\prime 3}_i.
$$
 
The general method requires ${\cal L}\indices*{*^{\hat \beta}_{\sm k}^{\hat \beta}_{\sm j}} l\indices*{*^{\sm j}_{{\hat\beta}}^{\sm k}_{{\hat \beta}}_\rho^p_\rho^v}$ to be computed, which would have $O\{M(M+1) n {\cal{P} }^2/2\}$ cost if the terms $l\indices*{*^j_{{\hat\beta}}^k_{{\hat \beta}}_\rho^p_\rho^v}$ were computed explicitly for this purpose (where $\cal{P}$ is the dimension of combined $\bp$). However this can be reduced to $O(n {\cal{P}}^2)$ using a trick most easily explained by switching to a matrix representation. For simplicity of presentation assume $K=2$, and define matrix ${\bf B}$ to be the inverse of the penalized Hessian, so that  $B_{ij} = {\cal L}\indices*{*^{\hat \beta}_i^{\hat \beta}_j}$. Defining 
$$
v^{lm}_i = l\indices*{*^i_{{\hat \eta}^l}^i_{{\hat \eta}^m}^i_{{\hat \eta}^{\sm q}}^i_{\hat \eta^{\sm s}}} \dif{\hat \eta^{\sm q}_i}{\rho_p} \dif{\hat \eta^{\sm s}_i}{\rho_v} + 
l\indices*{*^i_{{\hat \eta}^l}^i_{{\hat \eta}^m}^i_{{\hat \eta}^{\sm q}}} \ddif{\hat \eta^{\sm q}_i}{\rho_p}{\rho_v} 
\text { and } {\bf V}^{lm} = \text{diag}(v^{lm}_i) \text{ we have } 
$$

\vspace*{-1cm}

\singlespace
\beq
{\cal L}\indices*{*^{\hat \beta}_{\sm k}^{\hat \beta}_{\sm j}} l\indices*{*^{\sm j}_{{\hat\beta}}^{\sm k}_{{\hat \beta}}_\rho^p_\rho^v} =
\text{tr} \left \{ 
{\bf B} \bmat{cc} {\bf X}^{1\sf T} {\bf V}^{11} {\bf X}^1 & {\bf X}^{1\sf T} {\bf V}^{12} {\bf X}^2\\
{\bf X}^{2\sf T} {\bf V}^{12} {\bf X}^1 & 
{\bf X}^{2\sf T} {\bf V}^{22} {\bf X}^2\emat 
\right \} = \text{tr} \left \{ {\bf B} \bmat{cc}
{\bf X}^1 & {\bf 0}\\
{\bf 0} & {\bf X}^2
 \emat \ts
\bmat{cc}
{\bf V}^{11} {\bf X}^1 &  {\bf V}^{12} {\bf X}^2\\
{\bf V}^{12} {\bf X}^1 & {\bf V}^{22} {\bf X}^2
\emat
\right \}. \label{trick}
\eeq
Hence following the one off formation of ${\bf B} \bmat{cc}
{\bf X}^1 & {\bf 0}\\
{\bf 0} & {\bf X}^2
 \emat \ts$ (which need only have $O(n{\cal{P}}^2)$ cost), each trace computation has $O(Mn{\cal{P}})$ cost (since $\text{tr}({\bf C}\ts {\bf D}) = D_{{\sm i}{\sm j}}C_{{\sm i}{\sm j}}$).  %\doublespace
 
See SA I where a zero inflated Poisson model provides an example of the details.  Figure \ref{mcycle.fig} shows estimates for the model ${\tt accel}_i \sim N(f_1(t_i),\sigma^2_i)$ where $\log \sigma_i = f_2(t_i)$, $f_1$ is an adaptive P-spline and $f_2$ a cubic regression spline, while SA F.2 provides another application. Package {\tt mgcv} also includes multinomial logistic regression implemented this way and further examples are under development. An interesting possibility with any model which has multiple linear predictors is that one or more of those predictors should depend on some of the same terms, and SA H shows how this can be handled.

\subsection{A more special case: extended generalized additive models\label{egam}} 
 
For models with a single linear predictor in which the log likelihood is a sum of contributions per $y_i$, it is possible to perform fitting by iterative weighted least squares, enabling profitable re-use of some components of standard GAM fitting methods, including the exploitation of very stable orthogonal methods for solving least squares problems. Specifically, consider observations $y_i$, and let the corresponding log likelihood be of the form
$$
l = \sum_i l_i(y_i,\mu_i,{\bm \theta}, \phi)
$$
where the terms in the summation may also be written as $l_i$ for short, and $\mu_i$  is often $ \E(y_i) $, but may also be a latent variable (as in the ordered categorical model of SA K). Given $h$, a known link function, $h(\mu_i) = \eta_i$ where ${\bm \eta} = {\bf X}{\bm \beta} + {\bf o}$, $\bf X$ is a model matrix, $\bm \beta$ is a parameter vector and $\bf o$ is an offset (often simply 0). $\bm \theta$ is a parameter vector, containing the extra parameters of the likelihood, such as the $p$ parameter of a Tweedie density (see SA J), or the cut points of an ordered categorical model (see SA K). Notice that in this case $\bm \theta$ is not treated as part of $\bp$, since $\bm \theta$ can not always be estimated by straightforward iterative regression. Instead $\bm \theta$ will be estimated alongside the smoothing parameters. $\phi$ is a scale parameter, often fixed at one. Let $\tilde l_i = \max_{\mu_i} l_i(y_i,\mu_i,{\bm \theta}, \phi)$ denote the saturated log likelihood. Define the {\em deviance} corresponding to $y_i$ as $D_i = 2 (\tilde l_i - l_i) \phi$, where $\phi $ is the scale parameter on which $D_i$ does not depend. Working in terms of the deviance is convenient in a regression setting, where deviance residuals are a preferred method for model checking and the proportion deviance explained is a natural substitute for the $r^2$ statistic as a measure of goodness of fit (but see the final comment in SA I).  

In general the estimates of $\bm \beta$ will depend on some log smoothing parameter  $\rho_j = \log \lambda_j$, and it is notationally expedient to consider these to be part of the vector $\bm \theta$, although it is to be understood that $l$ does not actually depend on these elements of $\bm \theta$. Given $\bm \theta$, estimation of $\bm \beta$ is by minimization of the penalized deviance
$
{\cal D}(\bp, {\bm \theta}) = \sum_i D_i(\bp, {\bm \theta}) + \sum_j \lambda_j {\bm \beta} \ts {\bf S}^j {\bm \beta},
$
with respect to $\bp$. This can be achieved by penalized iteratively re-weighted least squares (PIRLS), which consists of iterative minimization of 
$
\sum_i w_i (z_i - {\bf X}_i{\bm \beta})^2 
+ \sum_j \lambda_j {\bm \beta} \ts {\bf S}^j {\bm \beta}
$
where the pseudo data and weights are given by 
$$
z_i = \eta_i - o_i - \frac{1}{2 w_i} \pdif{D_i}{\eta_i}, ~~~ w_i = \frac{1}{2} \dif{^2 D_i}{\eta_i^2}.
$$
Note that if $w_i=0$ (or $w_i$ is too close to 0), the penalized least squares estimate can be computed using only $w_i z_i$, which is then well defined and finite when $z_i$ is not.

Estimation of $\bm \theta$, and possibly $\phi$, is by  LAML. Writing $\bf W$ as the diagonal matrix of $w_i$ values, the log LAML is given by
$$
{\cal V}({\bm \theta}, \phi) = - \frac{{\cal D}(\hat \bp,{\bm \theta})}{2 \phi} + \tilde l({\bm \theta},\phi) - \frac{\log|{\bf X}\ts {\bf WX} + {\bf S}^\lambda| - \log|{\bf S}^\lambda|_+}{2} + \frac{M_p}{2} \log(2 \pi \phi).
$$
where $\bf W$ is evaluated at the $\hat {\bm \beta}$ implied by $\bm \theta$. To compute the derivatives of $\cal V$ with respect to $\bm \theta$ the derivatives of $\hat \bp$ with respect to $\bm \theta $ are required.  Note that $\cal V$ is a full Laplace approximation, rather than the `approximate' Laplace approximation used to justify PQL \citep{breslow.clayton}, so that PQL's well known problems with binary and low count data are much reduced. In particular: i) most PQL implementations estimate $\phi$ when fitting the working linear mixed model, even in the binomial and Poisson cases, where it is fixed at 1. For binary and low count data this can give very poor results. ii) PQL uses the expected Hessian rather than the Hessian, and these only coincide for the canonical link case. iii) PQL is justified by an assumption that the iterative fitting weights only vary slowly with the smoothing parameters, an assumption that is not needed here. 

The parameters $\bm \theta$ and $\phi$ can be estimated by maximizing ${\cal V}$ using Newton's method, or a quasi-Newton method. Notice that $\cal V$ depends directly on the elements of $\bm \theta$ via ${\cal D}$, $\tilde l$ and ${\bf S}^\lambda$, but also indirectly via the dependence of $\hat {\bm \mu}$ and $\bf W$ on $\hat {\bm \beta}$ and hence on $\bm \theta$. Hence each trial $\bm \theta, \phi$ requires a PIRLS iteration to find the corresponding $\hat {\bm \beta}$, followed by implicit differentiation to find the derivatives of $\hat {\bm \beta}$ with respect to $\bm \theta$. Once these are obtained the chain rule can be applied to find the derivatives of $\cal V$ with respect to $\bm \theta$ and $\phi$. 

As illustrated in SA C, there is scope for serious numerical instability in the evaluation of the determinant terms in $\cal V$, but for this case we can re-use the stabilization strategy from \cite{wood2011}, namely for each trial  $\bm \theta$ and $\phi$:
\begin{enumerate}
\item Use the orthogonal re-parameterization from Appendix B of \cite{wood2011} to ensure that $\log|{\bf S}^\lambda|_+$ can be computed in a stable manner.
\item Estimate $\hat \bp $ by PIRLS using the stable least squares method for negatively weighted problems from section 3.3 of \cite{wood2011}, setting structurally unidentifiable coefficients to zero.
\item Using implicit differentiation, obtain the derivatives of $\cal V$ required for a Newton update. 
\end{enumerate}
Step 3 is substantially more complicated than in \cite{wood2011}, and is covered in Appendix \ref{egam.imd}.

\subsubsection{Extended GAM new model implementation}

The general formulation above assumes that various standard information is available for each distribution and link. What is needed depends on whether quasi-Newton or full Newton is used to find $\hat {\bm \theta}$. Here is a summary of what is needed for each distribution
\begin{enumerate}
\item For finding $\hat {\bm \beta}$. $D^i_\mu$, $D\indices*{*^i_\mu^i_\mu}$, $h^\prime$ and $h^{\prime\prime}$. 
\item For $\hat {\bm \rho}$ via quasi-Newton. $h^{\prime\prime\prime}$, $D\indices*{*^i_\mu^j_\theta}$, $D_\theta^i$,
$D\indices*{*^i_\mu^i_\mu^i_\mu}$ and $D\indices*{*^i_\mu^i_\mu^j_\theta}$.
\item For $\hat {\bm \rho}$ via full Newton. $h^{\prime\prime\prime\prime}$, $D\indices*{*^i_\theta^j_\theta}$, $D\indices*{*^i_\mu^j_\theta^k_\theta}$, $D\indices*{*^i_\mu^i_\mu^i_\mu^i_\mu}$, $D\indices*{*^i_\mu^i_\mu^i_\mu^j_\theta}$ and $D\indices*{*^i_\mu^i_\mu^j_\theta^k_\theta}$.
\end{enumerate}
In addition first and second derivatives of $\tilde l$ with respect to its arguments are needed. All of these quantities can be obtained automatically using a computer algebra package. $\E D\indices*{^i_\mu^i_\mu}$ is also useful for further inference. If it is not readily computed then we can substitute $ D\indices*{^i_\mu^i_\mu}$, but a complication of penalized modelling is that $ D\indices*{^i_\mu^i_\mu}$ can fail to be positive definite at $\hat \bp$. When this happens $\E D\indices*{^i_\mu^i_\mu}$ can be estimated as the nearest positive definite matrix to $D\indices*{^i_\mu^i_\mu}$.

We have implemented beta, negative binomial,  scaled t models for heavy tailed data, simple zero inflated Poisson, ordered categorical and  Tweedie additive models in this way. The first three were essentially automatic: the derivatives were computed by a symbolic algebra package and coded from the results. Some care is required in doing this, to avoid excessive cancellation error, underflow or overflow in the computations. Overly naive coding of derivatives can often lead to numerical problems: SA I on the zero inflated Poisson provides an example of the sort of issues that can be encountered. The ordered categorical and Tweedie models are slightly more complicated and details are therefore provided in SA J and K (including further examples of the need to avoid cancellation error). 

\section{Smoothing parameter uncertainty \label{dist.sec}}

Conventionally in a GAM context smoothing parameters have been treated as fixed when computing interval estimates for functions, or for other inferential tasks.  In reality smoothing parameters must be estimated, and the uncertainty associated with this has generally been ignored except in fully Bayesian simulation approaches.  \cite{kass1989} proposed a simple first order correction for this sort of uncertainty in the context of i.i.d. Gaussian random effects in a one way ANOVA type design. Some extra work is required to understand how their method works when applied to smooths. It turns out that the estimation methods described above provide the quantities required to correct for smoothing parameter uncertainty. 

Assume we have several smooth model components, let $\rho_i = \log \lambda_i$ and ${\bf S}^\lambda = \sum_j \lambda_j {\bf S}^j$. Writing $\hat \bp_\rho$ for $\hat \bp$, to emphasise the dependence of $\hat \bp $ on the smoothing parameters, we use the Bayesian large sample approximation (see SB.4) 
\beq
{\bm \beta} | {\bf y}, {\bm \rho} \sim N(\hat {\bm \beta}_\rho, {\bf V}_\beta) \text{~~where~~}{\bf V}_\beta = (\hat {\bm{{\cal I}}}+{\bf S}^\lambda)^{-1} \label{post.smooth}
\eeq
which is exact in the Gaussian case, along with the large sample approximation 
\beq
{\bm \rho}|{\bf y} \sim N(\hat {\bm \rho}, {\bf V}_\rho) \label{post.rho}
\eeq
where ${\bf V}_\rho$ is the inverse of the Hessian of the negative log marginal likelihood with respect to $\bm \rho$. Since the approximation (\ref{post.rho}) applies in the interior of the parameter space, it is necessary to substitute a Moore-Penrose pseudoinverse of the Hessian if a smoothing parameter is effectively infinite, or otherwise to regularize the inversion (which is equivalent to placing a Gaussian prior on $\bm \rho$). Conventionally (\ref{post.smooth}) is used with $\hat {\bm \rho}$ plugged in and the uncertainty in $\bm \rho$ neglected. To improve on this note that if (\ref{post.smooth}) and (\ref{post.rho}) are correct, while $\bf z \sim N({\bf 0},{\bf I})$ and independently ${\bm \rho}^* \sim N(\hat {\bm \rho}, {\bf V}_\rho)$, then
$
{\bm \beta} | {\bf y} \overset{d}{=} \hat {\bm \beta}_{\rho^*} + {\bf R}_{\rho^*}\ts {\bm z} 
$
where ${\bf R}_{\rho^*}\ts {\bf R}_{\rho^*} = {\bf V}_\beta$ (and ${\bf V}_\beta$ depends on ${\bm \rho}^*$).  This provides a way of simulating from ${\bm \beta} | {\bf y}$, but it is computationally expensive as  $\hat {\bm \beta}_{\rho^*}$ and ${\bf R}_{\rho^*}$ must be computed afresh for each sample. (The conventional approximation would simply set ${\bm \rho}^* = \hat {\bm \rho}$.) Alternatively consider a first order Taylor expansion 
$$
{\bm \beta} | {\bf y} \overset{d}{=}\hat {\bm \beta}_{\hat \rho} + {\bf J}({\bm \rho} - \hat {\bm \rho}) + {\bf R}_{\hat \rho}\ts {\bm z} + 
\sum_k \left . \pdif{{\bf R}_{\rho}\ts {\bm z}}{{\rho}_k}\right |_{\hat {\bm \rho}}({ \rho}_k - \hat {\rho}_k) + r
$$
where $r$ is a lower order remainder term and ${\bf J} = \ildif{\hat {\bm \beta}}{{\bm \rho}} |_{\hat {\bm \rho}}$. Dropping $r$, the expectation of the right hand side is $\hat {\bm \beta}_{\hat \rho}$. Denoting the elements of ${\bf R}_{\rho}$ by $R_{ij}$, tedious but routine calculation shows that the 3 remaining random terms are uncorrelated with covariance matrix 
\beq
{\bf V}_\beta^\prime = {\bf V}_\beta + {\bf V}^\prime + {\bf V}^{\prime\prime} \text{~~where~~} {\bf V}^\prime = {\bf JV}_\rho {\bf J}\ts \text{~~and~~} 
 V^{\prime\prime}_{jm} = \sum_i^p \sum_l^M \sum_k^M \pdif{R_{ij}}{\rho_k} V_{\rho,kl} \pdif{R_{im}}{\rho_l}, \label{cov.cor}
\eeq
which is computable at $O(Mp^3)$ cost (see SA D). % spu.append 
Dropping ${\bf V}^{\prime\prime}$ we have the 
\cite{kass1989} approximation
$
{\bm \beta}|{\bf y} \sim N(\hat {\bm \beta}_{\hat \rho}, {\bf V}_\beta^*)
\text{ where }
{\bf V}_\beta^* = {\bf V}_\beta + {\bf J} {\bf V}_\rho {\bf J}\ts.
$
(A first order Taylor expansion of $\hat {\bm \beta}$ about $\bm \rho$ yields a similar 
correction for the frequentist covariance matrix of $\hat {\bm \beta}$:
$
{\bf V}_{\hat \beta}^* =(\hat { \bm{{\cal I}}} + {\bf S}^\lambda)^{-1} \hat {\bm{{\cal I}}}(\hat {\bm{{\cal I}} }+ {\bf S}^\lambda)^{-1} +  {\bf J} {\bf V}_\rho {\bf J}\ts
$, where $\hat { \bm{{\cal I}}}$ is the negative Hessian of the log likelihood).

SA D %spu.append 
shows that in a Demmler-Reinsch like parameterization,
for any penalized parameter $\beta_i$ with posterior standard deviation $\sigma_{\beta_i}$,
$$
\frac{\ildif{\hat {\beta_i}}{\rho_j}}{\ildif{({\bf R}\ts {\bf z})_i}{\rho_j}} \simeq 
\frac{\hat \beta_i}{z_i\sigma_{\beta_i}}.
$$
So the ${\bf J}({\bm \rho} - \hat {\bm \rho})$ correction is dominant for components that are strongly non-zero. This offers some justification for using the \cite{kass1989} approximation, but not in a model selection context, where near zero model components are those of most interest: hence in what follows we will use (\ref{cov.cor}) without dropping 
${\bf V}^{\prime\prime}$.

\section{An information criterion for smooth model selection \label{new.aic}}

When viewing smoothing from a Bayesian perspective the smooths have improper priors (or alternatively vague priors of convenience) corresponding to the null space of the smoothing penalties.  This invalidates model selection via marginal likelihood comparison. An alternative is a frequentist AIC \citep{akaike}, based on the conditional likelihood of the model coefficients, rather than the marginal likelihood. In the exponential family GAM context, \citet[][\S 6.8.3]{h&t90} proposed a widely used version of this {\em conditional} AIC in which the effective degrees of freedom of the model, $\tau_0$, is used in place of the number of model parameters (in the general setting  $\tau_0=\text{tr}\{{\bf V}_\beta\hat {\bm{{\cal I}}}\}$ is equivalent to the \citet{h&t90} proposal). But \cite{greven.kneib2010} show that this is overly likely to select complex models, especially when the model contains random effects: the difficulty arises because $\tau_0$ neglects the fact that the smoothing parameters have been estimated, and are therefore uncertain (a marginal AIC based on the frequentist marginal likelihood, in which unpenalized effects are not integrated out, is equally problematic, partly because of underestimation of variance components and consequent bias towards simple models). A heuristic alternative is to use $\tau_1 =\text{tr}(2 \hat {\bm{{\cal I}}} {\bf V}_{\beta} - \hat{\bm{{\cal I}}} {\bf V}_{\beta}\hat {\bm{{\cal I}}} {\bf V}_{\beta})$ as the effective degrees of freedom, motivated by considering the number of unpenalized parameters required to optimally approximate a bias corrected version of the model, but the resulting AIC is too conservative (see section \ref{sim.sec}, for example). \cite{greven.kneib2010} show how to exactly compute an effective modified AIC for the Gaussian additive model case based on defining the effective degrees of freedom as $ \sum_i \ilpdif{\hat y_i}{y_i}$ \citep[as proposed by][]{liang2008AIC}. \cite{yu.yau2012} and \cite{saefken2014} consider extensions to generalized linear mixed models. The novel contribution of this section is to use the results of the previous section to avoid the problematic neglect of smoothing parameter uncertainty in the conditional AIC computation in a manner that is easily computed and applicable to the general model class considered in this paper.

The derivation of AIC \citep[see e.g.][\S 4.7]{Davison} with the MLE replaced by the penalized MLE is identical up to the point at which  the AIC score is represented as 
\begin{eqnarray}
{\rm AIC} &=& - 2 l(\hat {\bm \beta}) + 2\E \left \{ 
(\hat {\bm \beta} - {\bm \beta}_d)\ts \bm{{\cal I}}_d(\hat {\bm \beta} - {\bm \beta}_d)
\right \} \\ &=& - 2 l(\hat {\bm \beta}) + 2 \text{tr} \left [  \E  
\{(\hat {\bm \beta} - {\bm \beta}_d)(\hat {\bm \beta} - {\bm \beta}_d)\ts \}\bm{{\cal I}}_d
\right ]
\end{eqnarray}
where $\bp_d$ is the coefficient vector minimizing the KL divergence and $\bm{{\cal I}}_d$ is the corresponding expected negative Hessian of the log likelihood. In an unpenalized setting $\E  
\{(\hat {\bm \beta} - {\bm \beta}_d)(\hat {\bm \beta} - {\bm \beta}_d)\ts
 \}$ is estimated as the observed inverse information matrix $ \hat{\bm{{\cal I}}}^{-1}$ 
 and $\tau^\prime = \text{tr} \left \{  \E  
(\hat {\bm \beta} - {\bm \beta}_d)(\hat {\bm \beta} - {\bm \beta}_d)\ts\bm{{\cal I}}_d
\right \}$ is estimated as $\text{tr} (\hat{\bm{{\cal I}}}^{-1}\hat{\bm{{\cal I}}}) = k$. 
Penalization means that the expected inverse covariance matrix of $\hat \bp $ is no longer well approximated by $\hat{\bm{{\cal I}}}$, and there are then two ways of proceeding. 

The first is to view $\bm \beta$ as a frequentist random effect, with predicted values $\hat \bp$. In that case the covariance matrix for the predictions, $\hat {\bm \beta}$, corresponds to the posterior covariance matrix obtained  when taking the Bayesian view of the smoothing process, so we have the conventional estimate $\tau = \text{tr} \{{\bf V}_\beta\hat{\bm{{\cal I}}}\}$ if we neglect smoothing parameter uncertainty, or $\tau = \text{tr}({\bf V}^\prime_\beta \hat{\bm{{\cal I}}})$ accounting for it using (\ref{cov.cor}). 

The frequentist random effects formulation is not a completely natural way to view smooths, since we do not usually expect the smooth components of a model to be re-sampled from the prior with each replication of the data. However in the smoothing context ${\bf V}_\beta$ has the interpretation of being the frequentist covariance matrix for $\hat \bp$ plus an estimate of the prior expectation of the squared smoothing bias (matrix), which offers some justification for using the same $\tau$ estimate as in the strict random effects case. To see this consider the decomposition 
$$
\E  \{ (\hat {\bm \beta} - {\bm \beta}_d)(\hat {\bm \beta} - {\bm \beta}_d)\ts  \} = \E  \{ (\hat {\bm \beta} - \E \hat{\bm \beta})(\hat {\bm \beta} - \E \hat {\bm \beta})\ts  \} + {\bm \Delta}_\beta {\bm \Delta}_\beta\ts
$$ 
where ${\bm \Delta}_\beta$ is the smoothing bias in $\hat {\bm \beta}$. The first term on the right hand side, above, can be replaced by the standard frequentist estimate ${\bf V}_{\hat \beta} = (\hat { \bm{{\cal I}}} + {\bf S}^\lambda)^{-1} \hat {\bm{{\cal I}}}(\hat {\bm{{\cal I}} }+ {\bf S}^\lambda)^{-1}$. Now expand the penalized log likelihood around $\bp_d$:
$$
l_p({\bm \beta}^\prime) \simeq l({\bm \beta}_d) + \pdif{l}{\bp\ts} ({\bm \beta}^\prime - {\bm \beta}_d) -
\frac{1}{2}  ({\bm \beta}^\prime -{\bm \beta}_d)\ts\bm{{\cal I}}_d({\bm \beta}^\prime -{\bm \beta}_d) -
\frac{1}{2}{\bm \beta}^{\prime {\sf T}} {\bf S}^\lambda  {\bm \beta}^\prime.
$$
Differentiating with respect to $\bp^\prime$ and equating to zero we obtain the approximation
$$
\hat \bp \simeq (\bm{{\cal I}}_d+{\bf S}^\lambda)^{-1} \left (\bm{{\cal I}}_d {\bm \beta}_d 
+ \left . \pdif{l}{\bp}\right |_{\beta_d} \right ).
$$
$\E \ildif{l}{\bp} |_{\beta_d} = 0$ by definition of $\bp_d$, so taking expectations of both sides we have $\E(\hat \bp) \simeq (\bm{{\cal I}}_d+{\bf S}^\lambda)^{-1}\bm{{\cal I}}_d {\bm \beta}_d$. Hence estimating $\bm{{\cal I}}_d$ by $ \hat{\bm{{\cal I}}}$ we have $\tilde {\bm \Delta}_\beta \simeq \{ ( \hat{\bm{{\cal I}}}+{\bf S}^\lambda)^{-1} \hat{\bm{{\cal I}}} - {\bf I}  \}{\bm \beta}_d$. Considering the expected value of $\tilde {\bm \Delta}_\beta \tilde {\bm \Delta}_\beta\ts$ according to the prior mean and variance assumptions of the model, we have the following.

\theoremstyle{definition}
\newtheorem{lemma}{Lemma}
\begin{lemma}
Let the setup be as above and let $\E_\pi$ denote expectation assuming the prior mean and covariance for $\bp$. Treating $ \hat{\bm{{\cal I}}}$ as fixed, then ${\bf V}_{\hat \beta} + \E_\pi (\tilde {\bm \Delta}_\beta \tilde {\bm \Delta}_\beta\ts) = {\bf V}_{\beta}.$
\end{lemma}
\noindent For proof see SA D. %spu.append}
This offers some justification for again using $\tau = \text{tr} \{{\bf V}_\beta\hat{\bm{{\cal I}}}\}$, or $\tau = \text{tr}({\bf V}^\prime_\beta \hat{\bm{{\cal I}}})$ accounting for $\bm \rho$ uncertainty. So both the frequentist random effects perspective and the prior expected smoothing bias approach result in 
\beq
{\rm AIC} = - 2 l(\hat {\bm \beta}) + 2 \text{tr}(\hat {\bm{{\cal I}}} {\bf V}^\prime_\beta).
\label{caic}
\eeq
This is the conventional \cite{h&t90} conditional AIC with an additive correction $2 \text{tr}\{\hat{ \bm{{\cal I}}}({\bf V}^\prime + {\bf V}^{\prime\prime})\}$, accounting for smoothing parameter uncertainty. The correction is readily computed for any model considered here, provided only that the derivatives of $\hat \bp$ and ${\bf V}_{\beta}$ can be computed: the methods of section \ref{smooth.sec} provide these. Section \ref{sim.sec} provides an illustration of the efficacy of (\ref{caic}).
\renewcommand{\baselinestretch}{1}

\begin{figure}

\vspace*{-1cm}

\eps{-90}{.6}{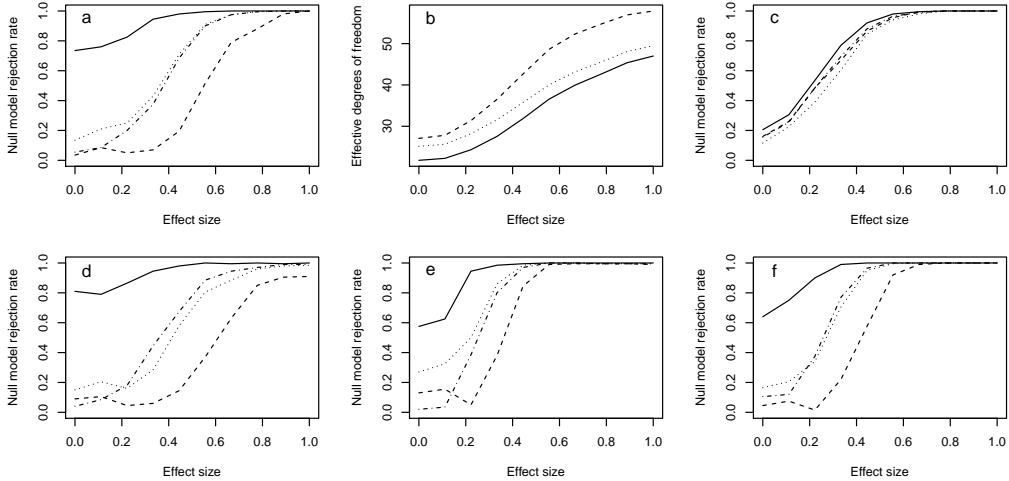}

\vspace*{-.5cm}

\caption{\small Simulation based illustration of the problems with previous AIC type model selection criteria and the relatively good performance of the section \ref{new.aic} version. In all panels: (i) the solid curves are for conventional conditional AIC, (ii) the dotted curves are for the section \ref{new.aic} version, (iii) the middle length dashed curves are for AIC based on the heuristic upper bound degrees of freedom, (iv) the dashed dot curves are for the marginal likelihood based AIC and (v) the long dashed curves are for the \cite{greven.kneib2010} corrected AIC (top row only). (a) Observed probability of selecting the larger model as the effect strength of the differing term is increased from zero, for a 40 level random effect and Gaussian likelihood. (b) whole model effective degrees of freedom used in the alternative conditional AIC scores for the left hand panel as effect size increases. (c) Same as (a), but where the term differing between the two  models was a smooth curve. (d) As (a) but for a Bernoulli likelihood. (e) As (a) for a beta likelihood. (f) As (a) for a Cox proportional hazards partial likelihood.}
 \label{aic-power}
\end{figure}
\renewcommand{\baselinestretch}{\dsp}

\section{Simulation results \label{sim.sec}}

\renewcommand{\baselinestretch}{1}
\begin{figure}
\vspace*{-1cm}

\eps{0}{.5}{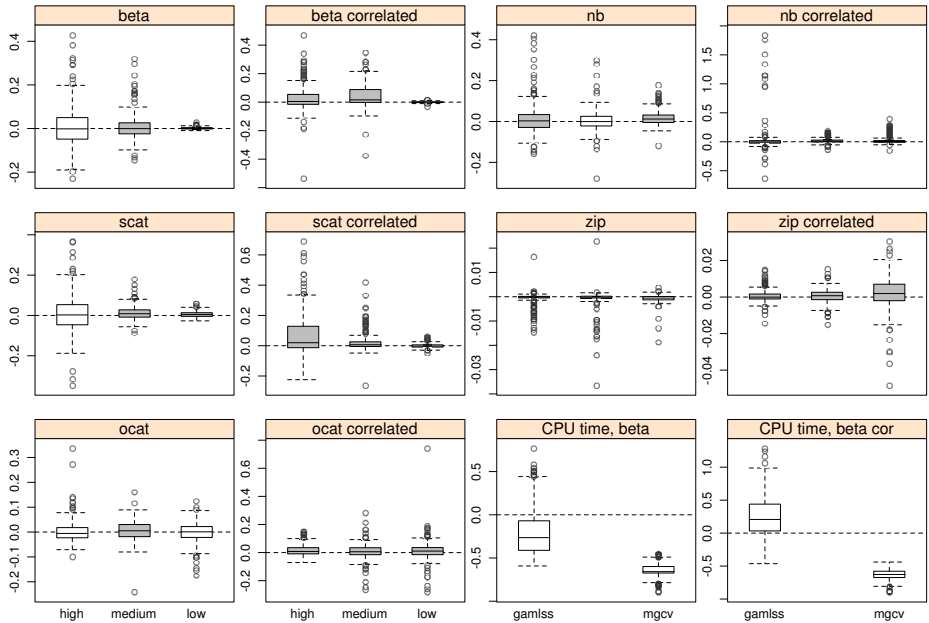} 

\vspace*{-.5cm}
\caption{\small Results of simulation comparison with {\tt gamlss} (beta, nb, scat, zip) and {\tt BayesX} (ocat) packages for one dimensional P-spline models. The two plots at lower right show comparisons of $\log_{10}$ computing times for the case with the smallest time advantage for the new method --- Beta regression. The remaining panels show boxplots of  replicate by replicate difference in MSE/Brier's score each standardized by the average MSE or Brier's score for the particular simulation comparison. Each panel shows three box plots, one for each noise to signal level. Positive values indicate that the new method is doing better than the alternative. Boxplots are shaded grey when the difference is significant at the 5\% level (all three for nb correlated should be grey). In all cases where the difference is significant at 5\% the new method is better than the alternative, except for the zero inflated Poisson with uncorrelated data, where the alternative method is better at all noise levels.  
 \label{sim-res.fig}}
\end{figure}
\renewcommand{\baselinestretch}{\dsp}

The improvement resulting from using the corrected AIC of section \ref{new.aic} can be illustrated by simulation. Simulations were conducted for additive models with true expected values given by 
$\eta = f_0(x_0) +f_1(x_1) + f_2(x_2) + f_3(x_3)$, where the $f_j$ are shown in SA E, and the $x$ covariates are all independent $U(0,1)$ deviates. Two model comparisons were considered. In the first a 40 level Gaussian random effect was added to $\eta$, with the random effect standard deviation being varied from 0 (no effect) to 1. AIC was then used to select between models with or without the random effect included, but where all smooth terms were modelled using penalized regression splines. In the second case models with and without $f_0$ were compared, with the true model being based on $ c f_0$ in place of $f_0$, where the effect strength $c$ was varied from 0 (no effect) to 1. Model selection was based on i) conventional conditional generalized AIC using $\tau_0$ from section \ref{new.aic}, ii) the corrected AIC of section \ref{new.aic}, iii) a version of AIC in which the degrees of freedom penalty is based on $\tau_1$ from section  \ref{new.aic}, iv) AIC based on the marginal likelihood with the number of parameters given by the number of smoothing parameters and variance components plus the number of unpenalized coefficients in the model and v) The \cite{greven.kneib2010} corrected AIC for the Gaussian response case. The marginal likelihood in case (iv) is a version in which un-penalized coefficients are not integrated out (to avoid the usual problems with fixed effect differences and REML, or improper priors and marginal likelihood).

Results are shown in the top row of figure \ref{aic-power} for a sample size of 500 with Gaussian sampling error and standard deviation of 2. For the random effect comparison, conventional conditional AIC is heavily biased towards the more complex model, selecting it on over 70\% of occasions. The ML based AIC is too conservative for an AIC criterion with 3.5\% selection of the larger model when it is not correct, as against the roughly 16\% one might expect from AIC comparison of models differing in 1 parameter. The known underestimation of variance components estimated by this sort of marginal likelihood is partly to blame. The AIC based on $\tau_1$ from section \ref{new.aic} also lacks power, performing even  less well than the ML based version. By contrast, the new corrected AIC performs well, and in this example is a slight improvement on \cite{greven.kneib2010}. For the smooth comparison the different calculations differ much less, although the alternatives are slightly less biased towards the more complex model than the conventional conditional generalized AIC, with the corrected section \ref{new.aic} version showing the smallest bias. The lower row of figure \ref{aic-power} shows equivalent power plots for the same Gaussian random effect and linear predictor $\eta$, but with Bernoulli, beta and Cox proportional hazard (partial) likelihoods (the first two using logit links). 

The purpose of this paper is to develop methods to allow the rich variety of smoothers illustrated in figure \ref{smooth.fig} to be used in models beyond the exponential family, a task for which general methods were not previously available. However for the special case of univariate P-splines \citep{Eilers&Marx96,ME98} some comparison with existing methods is possible, in particular using R package {\tt gamlss} \citep{rigby2005,rigby2013automatic} and the BayesX package \citep[{\tt www.bayesx.org}]{fahrmeir.lang, fahrmeir04,brezger2006,rbayesx,bayesx}. For this special case both packages implement models using essentially the same penalized likelihoods used by the new method, but they optimize localized marginal likelihood scores within the penalized likelihood optimization algorithm to estimate the smoothing parameters.

The comparison was performed using data simulated from models with the linear predictor given above (but without any random effect terms). Comparison of the new method with GAMLSS was only possible for negative binomial, beta, scaled t and simple zero inflated Poisson families, and with BayesX was only possible for the ordered categorical model (BayesX has a negative binomial family, but it is currently insufficiently stable for a sensible comparison to be made). Simulations with both uncorrelated and correlated covariates were considered. Three hundred replicates of the sample size 400 were produced for each considered family at three levels of noise (see SA E for further details). Models were estimated using the correct link and additive structure, and using P-splines with basis dimensions of 10, 10, 15 and 8 which were chosen to avoid any possibility of forced oversmoothing, while keeping down computational time.  

Model performance for the negative binomial (nb), beta, scaled t (scat) and zero inflated Poisson (zip) families was compared via MSE, 
$
n^{-1}\sum_{i=1}^n\left\{\hat{\eta}({\bf x}_i) - \eta_t({\bf x}_i)\right\}^2,
$ on the additive predictor scale.
The Brier score, 
$
\frac{1}{n}\sum_{i=1}^n\sum_{j=1}^R (p_{ij}-\hat{p}_{ij})^2, 
$ was used to measure the performance for the ordered categorical (ocat) family,
where $R$ is a number of categories, $p_{ij}$ are true category probabilities and $\hat{p}_{ij}$ their estimated values. In addition the computational performance (CPU time) of the alternative methods was recorded. Figure \ref{sim-res.fig} summarizes the results. In general the new method provides a small improvement in statistical performance, which is slightly larger when covariates are correlated. The correlated covariate setting is the one in which local approximate smoothness selection methods would be expected to perform less well, relative to `whole model' criteria. In terms of speed and reliability the new method is an improvement, especially for correlated covariates, which tend to lead to reduced numerical stability, leading the alternative methods to fail in up to 4\% of cases.

\section{Example: predicting prostate cancer \label{prostate}}

\begin{figure}
\eps{-90}{.55}{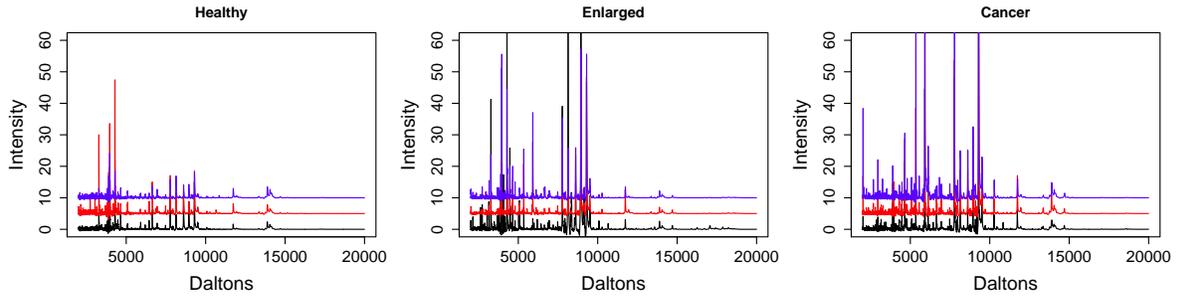}\vspace*{-.5cm}
\caption{\small 3 representative protein mass spectra (centred and normalized) from serum taken from patients with apparently healthy prostate, enlarged prostate and prostate cancer. It would be useful to be able to predict disease status from the spectra. The red and blue spectra have been shifted upward by 5 and 10 units respectively. 
 \label{prostate-raw.fig}}
\end{figure}
\begin{figure}
\eps{-90}{.55}{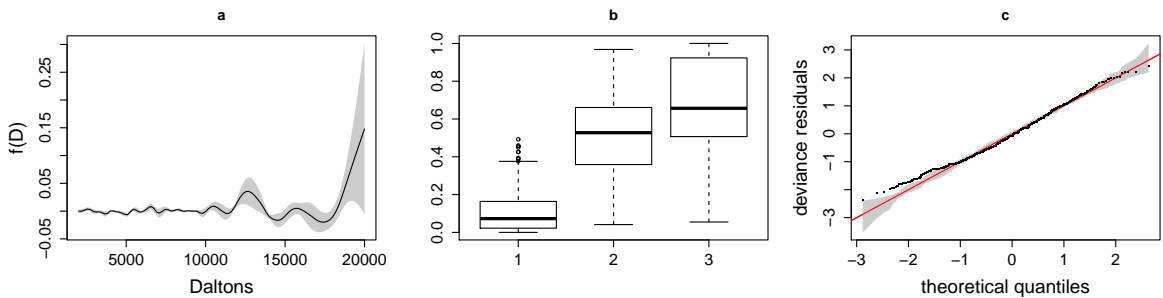}\vspace*{-.5cm}
\caption{\small Results from the ordered categorical prostate model fit. {\bf a}. The estimated coefficient function $f(D)$ with 95\% confidence interval. {\bf b}. Boxplots of the model probability of cancer, for the 3 observed states (1, healthy, 2, enlarged and 3, cancer). {\bf c}. QQ-plot of ordered deviance residuals against simulated theoretical quantiles, indicating some mismatch in the lower tail.  
 \label{prostate-fit.fig}}
\end{figure}

This section and the next provide example applications of the new methods, while SA F provides further examples in survival analysis and animal distribution modelling. Figure \ref{prostate-raw.fig} shows representative protein mass spectra from serum taken from patients with a healthy prostate, relatively benign prostate enlargement and prostate cancer \citep[see][]{adam2002}. To avoid the need for intrusive biopsy there is substantial interest in developing non-invasive screening tests to distinguish cancer, healthy and more benign conditions. One possible model is an ordered categorical signal regression in which the mean of a logistically distributed latent variable $z$ is given by 
$$
\mu_i = \alpha + \int f(D) \nu_i(D) dD
$$
where $f(D)$ is an unknown smooth function of mass $D$ (in Daltons) and $\nu_i(D)$ is the $i^\text{th}$ spectrum . The probability of the patient lying in category 1, 2 or 3 corresponding to `healthy', `benign enlargement' and `cancer' is then given by the probability of $z_i$ lying in the range $(-\infty,-1]$,
$(-1,\theta]$ or $(\theta,\infty)$, respectively (see SA K). 

Given the methods developed in this paper, estimation of this model is routine, as is the exploration of whether an adaptive smooth should be used for $f$, given the irregularity of the spectra. Figure \ref{prostate-fit.fig} shows some results of model fitting. The estimated $f(D)$ is based on a rank 100 thin plate regression spline. Its effective degrees of freedom is 29. An adaptive smooth gives almost identical results. The right panel shows a QQ-plot of ordered deviance residuals against simulated theoretical quantiles \citep{augustin2012qq}. There is modest deviation in the lower tail. The middle panel shows boxplots of the probability of cancer according to the model for the 3 observed categories. Cancer and healthy are quite well separated, but cancer and benign enlargement less so. 
For cases with cancer the model gave cancer a higher probability than normal prostate in 92\% of cases, and a higher probability that either other category in 83\% of cases.  For healthy patients the model gave the normal category higher probability than cancer in 85\% of cases and the highest probability in 77\% of cases. These results are somewhat worse than those reported in \cite{adam2002} for a relatively complex machine learning method which involved first pre-processing the spectra to identify peaks believed to be discriminating. On the other hand the signal regression model here would allow the straightforward inclusion of further covariates, and does automatically supply uncertainty estimates.

\section{Multivariate additive modelling of fuel efficiency \label{mvn.mpg}}
\begin{figure}
\eps{-90}{.45}{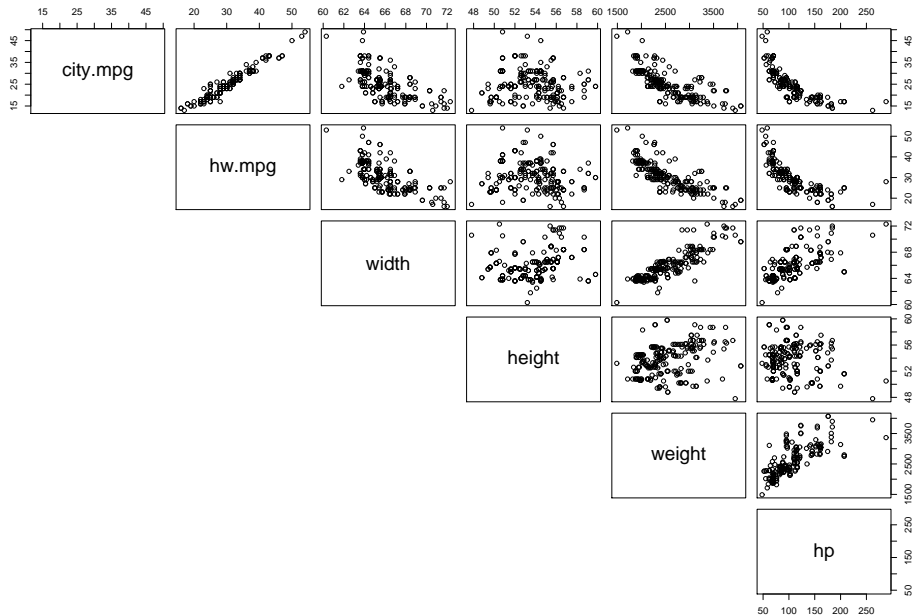}\vspace*{-.5cm}
\caption{\small Part of a data set from the USA on fuel efficiency of cars.    
 \label{mpg-data.fig}}
\end{figure}

Figure \ref{mpg-data.fig} shows part of a dataset on the fuel efficiency of 207 US car models, along with their characteristics \citep{Bache2013}. Two efficiency measures were taken: miles per gallon (MPG) in city driving, and the same for highway driving. One possible model might be a bivariate additive model, as detailed in SA H, where the two {\tt mpg} measurements are modelled as bivariate Gaussian, with means given by separate linear predictors for the two components. A priori it might be expected that city efficiency would be highly influenced by weight and highway efficiency by air resistance and hence by frontal area, or some other combination of height and width of the car. 

The linear predictors for the two components were based on the additive fixed effects of factors `fuel type' (petrol or diesel), `style' of car (hatchback, sedan, etc.) and `drive' (all-, front- or rear-wheel). In addition i.i.d. Gaussian random effects of the 22 car manufacturers were included, as well as smooth additive effects of car weight and horsepower. Additive and tensor product smooths of height and width were tried as well as a smooth of the product of height and width, but there was no evidence to justify their inclusion - term selection penalties \citep{marra.wood2011} remove them, p-values indicate they are not significant and AIC suggests that they are better dropped.

The possibility of smooth interactions between weight and horsepower were also considered, using smooth main effects plus smooth interaction formulations of the form $f_1(h) + f_2(w) + f_3(h,w)$. The smooth  interaction term $f_3$ can readily be constructed in a way that excludes the main effects of $w$ and $h$, by constructing its basis using the usual tensor product construction \citep[e.g.][]{wood2006tensor}, but based on marginal bases into which the constraints $\sum_{i} f_1(h_i) = 0$ and $\sum_{i} f_2(w_i) = 0$ have already been absorbed by linear reparameterization. The marginal smoothing penalties and hence the induced tensor product smoothing penalties are  unaffected by the marginal constraint absorption. This construction is the obvious generalization of the construction of parametric interactions in linear models, and is simpler than the various schemes proposed in the literature. 

\begin{figure}
\eps{-90}{.6}{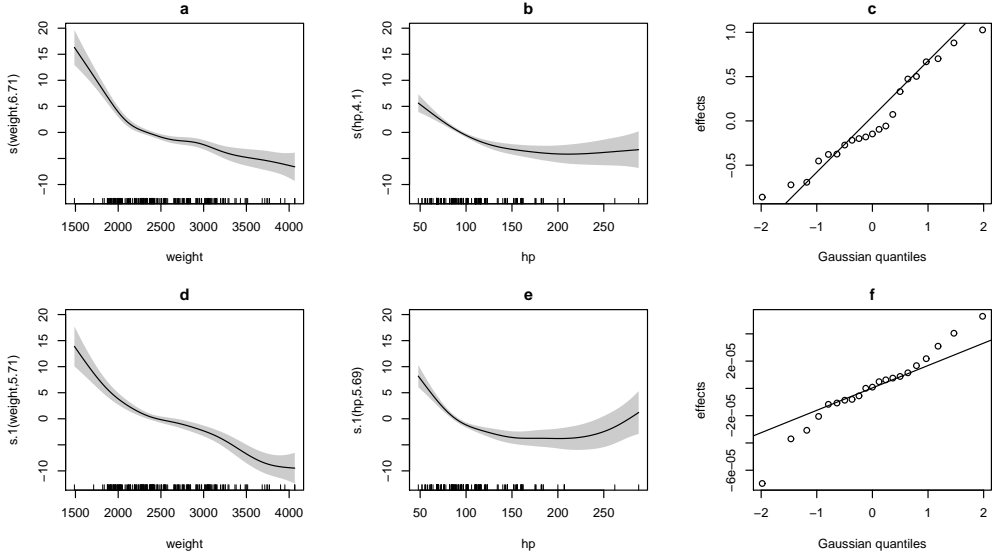}\vspace*{-.5cm}
\caption{\small Fitted smooth and random effects for final car fuel efficiency model. Panels a-c relate to the city fuel consumption, while d-f are for the highway. c and f are normal QQ-plots of the predicted random effects for manufacturer, which in the  case of highway MPG are effectively zero.    
 \label{mpg-fit.fig}}
\end{figure}

The interactions again appear to add nothing useful to the model fit, and we end up with a model in which the important smooth effects are horse power ({\tt hp}) and weight, while the important fixed effects are fuel type and drive, with diesel giving lower fuel consumption than petrol and all wheel drive giving higher consumption than the 2-wheel drives. These effects were important for both city and highway, whereas the random effect of manufacturer was only important for the city.  Figure \ref{mpg-fit.fig} shows the smooth and random effects for the city and highway linear predictors. Notice the surprising similarity between the effects although the city smooth effects are generally slightly less pronounced than those for the highway. The overall $r^2$ for the model is 85\% but with the city and highway error MPG standard deviation estimated as 1.9 and 2.3 MPG respectively. The estimated correlation coefficient is 0.88. 

\section{Discussion}

This paper has outlined a practical framework for smooth regression modelling with reduced rank smoothers, for likelihoods beyond the exponential family. The methods build seamlessly on the existing framework for generalized additive modelling, so that practical application of any of the models implemented as part of this work is immediately accessible to anyone familiar with GAMs via penalized regression splines. The key novel components contributed here are i) general, reliable and efficient smoothing parameter estimation methods based on maximized Laplace Approximate Marginal Likelihood, ii) a corrected AIC, and distributional results incorporating smoothing parameter uncertainty, to aid model selection and further inference and iii) demonstration of the framework's practical utility by provision of the details for some practically  important models.  The proposed methods should be widely applicable in situations in which effects are really smooth, and the methods scale well with the number of smooth model terms. In situations in which some component functions are high rank random fields, then the INLA approach of \citep{rue2009inla} will be much more efficient, however there are trade-offs between efficiency and stability in this case, since pivoting, used by our method to preserve stability, has instead to be employed to preserve sparsity in the INLA method (see SA K).   

The methods are implemented in R package {\tt mgcv} from version 1.8 (see SA M).

\singlespacing

\subsubsection*{Acknowledgements}

We thank the anonymous referees for a large number of very helpful comments that substantially improved the paper and Phil Reiss for spotting an embarrassing error in SA A. SNW and NP were funded by EPSRC grant EP/K005251/1 and NP was also funded by  EPSRC grant EP/I000917/1. BS was funded by the German Research Association (DFG) Research Training Group `Scaling Problems in Statistics' (RTG 1644). SNW is grateful to Carsten Dorman and his research group at the University of Freiburg, where the extended GAM part of this work was carried out.  

%\doublespacing
%\begin{appendices}

\appendix

\section{Implicit differentiation in the extended GAM case \label{egam.imd}}

Let ${\cal D}\indices*{*_i^{\hat \beta}_j^{\hat \beta}}$ denote elements of the inverse of the Hessian matrix (${\bf X}\ts {\bf WX} + {\bf S}^\lambda $) with elements  ${\cal D}\indices*{*^i_{\hat \beta}^j_{\hat \beta}}$ , and note that $\hat {\bm \beta} $ is the solution of 
$
{\cal D}^i_{\hat \beta} = 0.
$   
Finding the total derivative with respect to $\bm \theta$ of both sides of this we have
$$
{\cal D}\indices*{*^i_{\hat \beta}^{\sm k}_{\hat \beta}} \dif{\hat \beta_{\sm k}}{\theta_j} + {\cal D}\indices*{*^i_{\hat \beta}^j_\theta} = 0 ,\text{~~implying~that~~} \dif{\hat \beta_k}{\theta_j} = 
- {\cal D}\indices*{*_k^{\hat \beta}_{\sm i}^{\hat \beta}}{\cal D}\indices*{*^{\sm i}_{\hat \beta}^j_\theta}.
$$
Differentiating once more yields 
$$
\ddif{\hat \beta_i}{\theta_j}{\theta_k} = 
-{\cal D}\indices*{*_i^{\hat \beta}_{\sm l}^{\hat \beta}} \left (
{\cal D}\indices*{*^{\sm l}_{\hat \beta}^{\sm p}_{\hat \beta}^{\sm q}_{\hat \beta}} \dif{\hat \beta_{\sm q}}{\theta_j} \dif{\hat \beta_{\sm p}}{\theta_k} +
{\cal D}\indices*{*^{\sm l}_{\hat \beta}^{\sm p}_{\hat \beta}^j_\theta}\dif{\hat \beta_{\sm p}}{\theta_k} +
{\cal D}\indices*{*^{\sm l}_{\hat \beta}^{\sm p}_{\hat \beta}^k_\theta}\dif{\hat \beta_{\sm p}}{\theta_j} +
 {\cal D}\indices*{*^{\sm l}_{\hat \beta}^j_\theta^k_\theta} \right ).
$$
The required partials are obtained from those generically available for the distribution and link used and by differentiation of the penalty. Generically we can obtain derivatives of $D_i$ w.r.t $\mu_i$ and $\bm \theta$. 

The preceding expressions hold whether $\theta_j$ is a parameter of the likelihood or a log smoothing parameter. Suppose $\Lambda$ denotes the set of log smoothing parameters, then
$$
{\cal D}\indices*{*^i_\beta^j_\theta} = \left \{ \begin{array}{ll} 
2 \exp(\theta_j) S^j_{i{\sm k}} \beta_{\sm k} & \theta_j \in \Lambda \\
D\indices*{*^i_\beta^j_\theta} & \text{otherwise}
\end{array} \right .
$$
where ${\bf S}^j$ here denotes the penalty matrix associated with $\theta_j$. Similarly
$$
{\cal D}\indices*{*^l_\beta^p_\beta^j_\theta} = \left \{ \begin{array}{ll} 
2\exp(\theta_j) S^j_{lp} & \theta_j \in \Lambda \\
D\indices*{*^l_\beta^p_\beta^j_\theta}& \text{otherwise}
\end{array} \right .
\text{ while  }
{\cal D}\indices*{*^l_\beta^j_\theta^k_\theta} = \left \{ \begin{array}{ll} 
2 \exp(\theta_j) S^j_{l{\sm m}} \beta_{\sm m} & j=k; \theta_j, \theta_k \in \Lambda\\
D\indices*{*^l_\beta^j_\theta^k_\theta} & \theta_j, \theta_k \not \in \Lambda\\
0 & \text{otherwise.}
\end{array} \right . 
$$
Derivatives with respect to  $\bm \eta$ are obtained by standard transformations
\beq
D_{\eta}^i = D_\mu^i / h^\prime_i \label{eta.mu1}
\eeq
where $h^\prime_i = h^\prime(\mu_i)$ and more primes indicate higher derivatives. Furthermore
\beq
D\indices*{*^i_\eta^i_\eta} = D\indices*{*^i_\mu^i_\mu} /h^{\prime 2}_i - D_\mu^i  h^{\prime\prime}_i / h^{\prime 3}_i
\eeq
where the expectation of the second term on the right hand side is zero at the true parameter values. 

\beq \text{Also } 
D\indices*{*^i_\eta^i_\eta^i_\eta} = D\indices*{*^i_\mu^i_\mu^i_\mu} / h^{\prime 3}_i - 3 D\indices*{*^i_\mu^i_\mu}  h^{\prime\prime}_i / h^{\prime 4}_i + D_\mu^i  \left ( 
3 h^{\prime\prime 2}_i/h_i^{\prime 5} - h^{\prime\prime\prime}_i/h^{\prime 4}_i
\right ), \text{ and}
\eeq
\beq
D\indices*{*^i_\eta^i_\eta^i_\eta^i_\eta} =
D\indices*{*^i_\mu^i_\mu^i_\mu^i_\mu}/h_i^{\prime 4} -
6 D\indices*{*^i_\mu^i_\mu^i_\mu}  h^{\prime\prime}_i/h^{\prime 5}_i +
D\indices*{*^i_\mu^i_\mu}  (15 h_i^{\prime\prime 2}/h_i^{\prime 6} - 4 h^{\prime\prime\prime}/h_i^{\prime 5}) -
D^i_\mu  (15 h_i^{\prime\prime 3}/h_i^{\prime 7} - 10 h_i^{\prime\prime} h_i^{\prime\prime\prime}/h_i^{\prime 6} + h_i^{\prime\prime\prime\prime}/h_i^{\prime 5} ) \label{eta.mu4}
\eeq
Mixed partial derivatives with respect to $\bm \eta$/$\bm \mu$ and $\bm \theta$ transform in the same way, the formula to use depending on the number of $\eta$ subscripts. The rules relating the derivatives w.r.t $\bm \eta$ to those with respect to $\bm \beta$ are much easier:
$
D_\beta^i = D_\eta^{\sm k} X_{{\sm k}i}, ~~~ D\indices*{*^i_\beta^j_\beta} = D_{\eta\eta}^{{\sm k}{\sm k}} X_{{\sm k}i}X_{{\sm k}j},~~~
D\indices*{*^i_\beta^j_\beta^k_\beta} = D\indices*{*^{\sm l}_\eta^{\sm l}_\eta^{\sm l}_\eta} X_{{\sm l}i}X_{{\sm l}j}X_{{\sm l}k}.
$
Again mixed partials follow the rule appropriate for the number of $\beta$ subscripts present. It is usually more efficient to compute using the definitions, rather than forming the arrays explicitly.

The ingredients so far are sufficient to compute $\hat {\bm \beta}$ and its derivatives with respect to $\bm\theta$. We now need to consider the derivatives of $\cal V$ with respect to $\bm \theta$. Considering $\cal D$ first, the components relating to the penalties are straightforward. The deviance components are then
$$
\dif{D}{\theta_i} = D_{\hat \eta}^{\sm j} \dif{\hat \eta_{\sm j}}{\theta_i} + D_{\hat \theta}^i
\text{ and }
\ddif{D}{{\theta}_i}{{ \theta}_j} = D_{\hat \eta\hat\eta}^{{\sm k}{\sm k}}  \dif{\hat \eta_{\sm k}}{\theta_i}\dif{\hat \eta_{\sm k}}{\theta_j} + D_{\hat \eta}^{\sm k} \ddif{\hat \eta_{\sm k}}{\theta_i}{\theta_j} + 
D_{\hat \eta\theta}^{{\sm k}j} \dif{\hat \eta_{\sm k}}{\theta_i} +
D_{\hat \eta\theta}^{{\sm k}i} \dif{\hat \eta_{\sm k}}{\theta_j} +
D\indices*{*^i_\theta^j_\theta},
$$
where the derivatives of $\hat {\bm \eta}$ are simply $\bf X$ multiplied by the derivatives of $\hat {\bm \beta}$. The partials of $\tilde l$ are distribution specific. The derivatives of the determinant terms are obtainable using \cite{wood2011} once derivatives of $w_i$ with respect to $\bm \theta$ have been obtained. These are
$$
\dif{w_i}{\theta_j} = \frac{1}{2} D\indices*{*^i_{\hat \eta}^i_{\hat \eta}^i_{\hat \eta}}  \dif{\hat \eta_i}{\theta_j} + \frac{1}{2} D\indices*{*^i_{ \eta}^i_{\hat \eta}^j_\theta},
$$
$$
\ddif{w_i}{\theta_j}{\theta_k} = \frac{1}{2} D\indices*{*^i_{\hat \eta}^i_{\hat \eta}^i_{\hat \eta}^i_{\hat \eta}}  \dif{\hat \eta_i}{\theta_j} \dif{\hat \eta_i}{\theta_k} + \frac{1}{2} D\indices*{*^i_{\hat \eta}^i_{\hat \eta}^i_{\hat \eta}} \ddif{\hat \eta_i}{\theta_j}{\theta_k} + \frac{1}{2}D\indices*{*^i_{\hat \eta}^i_{\hat \eta}^i_{\hat \eta}^k_\theta}  \dif{\hat \eta_i}{\theta_j}+ \frac{1}{2}D\indices*{*^i_{\hat \eta}^i_{\hat \eta}^i_{\hat \eta}^j_\theta}  \dif{\hat \eta_i}{\theta_k}
+ \frac{1}{2} D\indices*{*^i_{\hat \eta}^i_{\hat \eta}^j_\theta^k_\theta}.
$$

\singlespacing

\bibliography{/home/sw283/bibliography/journal,/home/sw283/bibliography/simon}
\bibliographystyle{chicago}

\pagebreak
%\widetext
\begin{center}

\textbf{\Large Supplementary material: Smoothing parameter and model selection for general smooth models}
\end{center}
%%%%%%%%%% Merge with supplemental materials %%%%%%%%%%
%%%%%%%%%% Prefix a "S" to all equations, figures, tables and reset the counter %%%%%%%%%%
\setcounter{equation}{0}
\setcounter{figure}{0}
\setcounter{table}{0}
\setcounter{page}{1}
\makeatletter
\renewcommand{\theequation}{S\arabic{equation}}
\renewcommand{\thefigure}{S\arabic{figure}}
\renewcommand{\bibnumfmt}[1]{[S#1]}
\renewcommand{\citenumfont}[1]{S#1}

\appendix

\section{Consistency of regression splines \label{basis.dim}}

There is already a detailed literature on the asymptotic properties of penalized regression splines \citep[e.g.][]{gu.kim, hall2005, kauermann2009, claeskens2009, wang2011asymptotics, yoshida2014asymptotics}. Rather than reproduce that literature, the purpose of this section and the next is to demonstrate the simple way in which the properties of penalized regression splines are related to the properties of regression splines, which in turn follow from the properties of interpolating splines. We will mostly focus on cubic splines and `infill asymptotics' in which the domain of the function of interest remains fixed as the sample size increases. We use the expression `at most $O(n^a)$' as shorthand for `$O(n^b)$ where $b\le a$', and use $O(\cdot)$ to denote stochastic boundedness when referring to random quantities.

\subsection{Cubic interpolating splines}

\begin{figure}
\eps{-90}{.45}{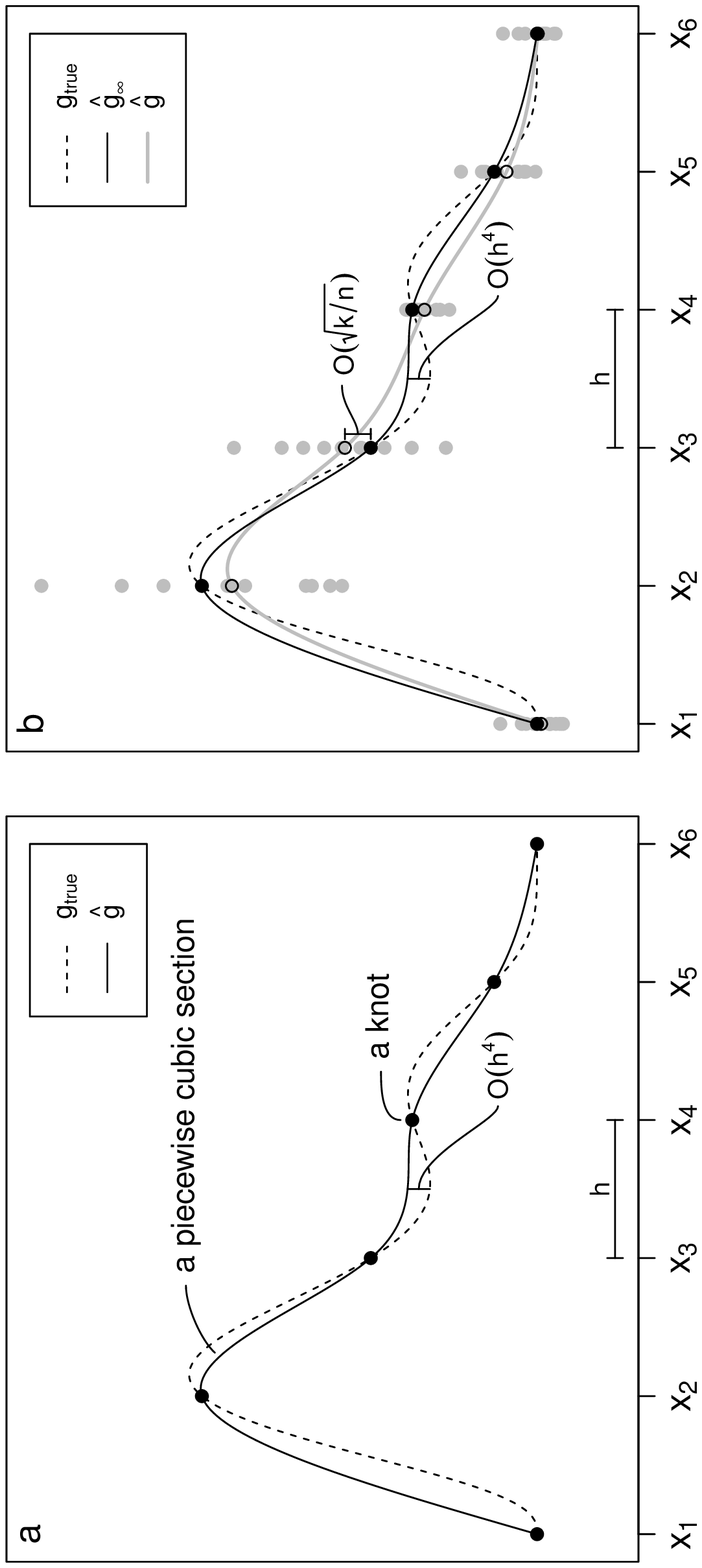}
\vspace*{-.3cm}

\caption{{\bf a.} A cubic interpolating spline (continuous curve), interpolating 6 `knot' points (black dots) with evenly spaced $x$ co-ordinates, from a true function (dashed curve). The spline is made up of piecewise cubic sections between each consecutive pair of knots. The approximation error is $O(h^4)$, where $h$ is the knot spacing on the $x$ axis. {\bf b.} A simple regression spline (grey curve) fitted to $n$ noisy observations (grey dots) of the true function (dashed curve), with $n/k$ data at each of the $k$ knot locations $x_j$. As $n/k \to \infty$ the regression spline tends to the limiting interpolating spline (black curve), which has $O(h^4)=O(k^{-4})$ approximation error.
\label{prs.fig}}
\end{figure}

Let $g(x)$ denote a 4 times differentiable function, observed at $k$ points $x_j, g(x_j)$, where the $x_j$ are strictly increasing with $j$. The cubic spline interpolant, $\hat g(x)$, is constructed from piecewise cubic polynomials on each interval $[x_j,x_{j+1}]$ constructed so that $\hat g(x_j) = g(x_j)$, the first and second derivatives of $\hat g(x)$ are continuous, and two additional end conditions are met. Example end conditions are the `natural' end conditions $\hat g^{\prime\prime}(x_1)=\hat g^{\prime\prime}(x_k)=0$ or the `complete' end conditions $\hat g^{\prime}(x_1)=g^{\prime}(x_1)$, $\hat g^{\prime}(x_k)=g^{\prime}(x_k)$. $\hat g(x)$ is unique given the end conditions. See figure \ref{prs.fig}a. A cubic spline interpolant with natural boundary conditions has the interesting property of being the interpolant minimizing $\int g^{\prime\prime}(x)^2 dx$ \citep[see e.g.][theorem 2.3]{green.silverman}. 

Let $h = \max_j(x_{j+1}-x_j)$, the `knot spacing'. By Taylor's theorem, a piecewise cubic interpolant must have an upper bound on interpolation error  $ O(h^\alpha)$ where $\alpha\ge 4$. In fact if $g^{(i)}(x)$ denotes the $i^\text{th}$ derivative of $g$ with respect to $x$ 
\beq
|\hat g^{(i)}(x) - g^{(i)}(x) | = O(h^{4-i}), ~~~~ i=0,\ldots,3 
\label{ispline.err}
\eeq
where $x$ is anywhere in $[x_1,x_k]$ for complete (or deBoor's `not-a-knot') end conditions, or is sufficiently interior to $[x_1,x_k]$ for natural end conditions. \citet[][chapter 5]{deBoor2001} provides especially clear derivation of these results, while \cite{hall76} provides sharp versions. 

\subsection{Regression splines \label{reg.spline}}

The space of interpolating splines with $k$ knots can be spanned by a set of $k$ basis functions. Various convenient bases can readily be computed: for example the B-spline basis functions have compact support, while the $j^\text{th}$ cardinal basis function takes the value 1 at $x_j$ and 0 at any other knot $x_i$ \citep[see e.g.][]{lancaster.salkauskas, deBoor2001}. For the cardinal basis, the spline coefficients are $g(x_j)$, the values of the spline at the knots.  Given a set of basis functions and $n>k$ noisy observations of $g(x)$, it is possible to perform spline regression. \cite{agarwal1980} and \cite{zhou2000} study this in detail, but a very simple example serves to explain the main results.

Consider the case in which we have $n/k$ noisy observations for each $g(x_j)$, and a model that provides a regular likelihood for $g(x_j)$ such that $|\hat g(x_j) - g(x_j)| = O(\sqrt{k/n})$, where $\hat g(x_j)$ is the MLE for $g(x_j)$ (which depends only on the $n/k$ observations of $g(x_j)$, as is clear from considering the cardinal basis representation). Suppose also that the $x_j$ are equally spaced. In this setting the cubic regression spline estimate of $g(x)$ is just the cubic spline interpolant of $x_j, \hat g(x_j)$, and the large sample limiting $\hat g(x)$ is simply the cubic spline interpolant of $x_j, g(x_j)$. By (\ref{ispline.err}) the limiting approximation error is $O(h^4)=O(k^{-4})$. Since the interpolant is linear in the $\hat g(x_j)$ the standard deviation of $\hat g(x)$ is $O(\sqrt{k/n})$. So if the limiting approximating error is not to eventually dominate the sampling error, we require $O(k^{-4}) \le O(\sqrt{k/n})$, and for minimum sampling error we would therefore choose $k = O(n^{1/9})$, corresponding to a mean square error rate of $O(n^{-8/9})$ for $g(x)$ and $O(n^{-4/9})$ for $g^{\prime\prime}$. See figure \ref{prs.fig}b.

\cite{agarwal1980} shows that the result for $g(x)$ holds when the observations are spread out instead of being concentrated at the knots, while \cite{zhou2000} confirms the equivalent for derivatives. In summary, cubic regression splines are consistent for $g(x)$ and its first 3 derivatives, provided that the maximum knot spacing decreases with sample size, to control the approximation error. Optimal convergence rates are obtained by allowing $h$ to depend on $n$ so that the order of the approximation error and the sampling variability are equal. 

\section{Penalized regression spline consistency under LAML}

Here we show how penalized regression spline estimates retain consistency under LAML estimation of smoothing parameters. To this end it helps to have available a spline basis for which individual coefficients form a meaningful sequence as the basis dimension increases, so we introduce this basis first, before demonstrating consistency and then considering convergence rates.

\subsection{An alternative regression basis}

An alternative spline basis is helpful in understanding how penalization affects consistency of spline estimation. Without loss of generality, restrict the domain of $g(x)$ to $[0,1]$ and consider the spline penalty $\int g^{(m)}(x)^2 dx = \int (\nabla^m g)^2 dx$ where $\nabla^m$ is the $m^\text{th}$ order differential operator. Let $\nabla^{m*}$ be the adjoint of $\nabla^m$ with respect to the inner product $\langle g,h \rangle = \int g(x)h(x)dx$. Then from the definition of an adjoint operator, $\int g^{(m)}(x)^2 dx = \int g {\cal K}^m g dx$, where ${\cal K}^m = \nabla^{m*}\nabla^m$. Now consider the eigenfunctions of ${\cal K}^m$, such that ${\cal K}^m \phi_j(x) = \varLambda_j \phi_j(x)$, $\varLambda_{j+1}>\varLambda_j \ge 0$. Since ${\cal K}^m$ is clearly self adjoint, $\langle \phi_j,\phi_i \rangle = 1$ if $i=j$ and 0 otherwise.  Notice that if $\beta^*_i = \langle g,\phi_i \rangle$, then we can write $g(x) = \sum_{i}\beta_i^* \phi_i(x) $. Finite $\int g^{(m)}(x)^2 dx$ implies that $\beta_i^*\to 0$ as $i \to \infty$. In fact generally we are interested in functions with low $\int g^{(m)}(x)^2 dx$, so it is the low order eigenvalues and their eigenfunctions that are of interest.

To compute discrete approximations to the $\phi_j$, first define $\Delta = (n-1)^{-1}$ for some discrete grid size $n$, and let $\phi_{ji}=\phi_j(i\Delta - \Delta)$ and $g_i = g(i\Delta - \Delta)$. A discrete representation of ${\cal K}^2$ is then ${\bf K} = {\bf D}\ts{\bf D}$ where $D_{ij}=0$ except for $D_{i,i} = D_{i,i+2}=1/\Delta^2$ and $D_{i,i+1}=-2/\Delta^2$ for $i = 1 , \ldots , n-2$ (the approximation for other values of $m$  substitutes $m^\text{th}$ order differences in the obvious way). The (suitably normalized) eigenvectors of $\bf K$ then approximate ${\bm \phi}_1, {\bm \phi}_2, \ldots$. Alternatively we can represent ${\bm \phi}_1 \ldots {\bm \phi}_k$ and any other $\bf g$ using a rank $k$ cubic spline basis. Hence we can write ${\bf g} = {\bf X}\bp$, where $\bf X$ has QR decomposition, ${\bf X} = {\bf QR}$ and  
$\int g^{(m)}(x)^2 dx = \bp\ts {\bf S}\bp = \bp\ts {\bf R}\ts{\bf Q}\ts {\bf QR}\its{\bf SR}^{-1} {\bf Q}\ts {\bf QR}\bp$. So the approximation of ${\cal K}^2$ is 
${\bf QR}\its{\bf SR}^{-1} {\bf Q}\ts$, which has eigenvectors $\bf QU$ where $\bf U$ is from the eigen-decomposition ${\bf U}\tilde {\bm \Lambda} {\bf U}\ts = {\bf R}\its{\bf SR}^{-1}$. 

Now if we reparameterize the regression spline basis so that $\bp^* = \Delta^{1/2}{\bf U}\ts {\bf R}\bp$, we obtain a normalized version of the Demmler-Reinsch basis \citep[][\S 4.10.4]{DemmlerReinsch75,nychka96DR,wood2006igam}, where $\bf S$ becomes ${\bm \Lambda} = \tilde {\bm \Lambda}\Delta^{-1}$ (the numerical approximation to the first $k$, $\varLambda_i$) and $\bf X$ becomes ${\bf QU}\Delta^{-1/2}$: but the latter is simply the numerical approximation to ${\bm \phi}_1 \ldots {\bm \phi}_k$. 

\begin{figure}
\eps{-90}{.7}{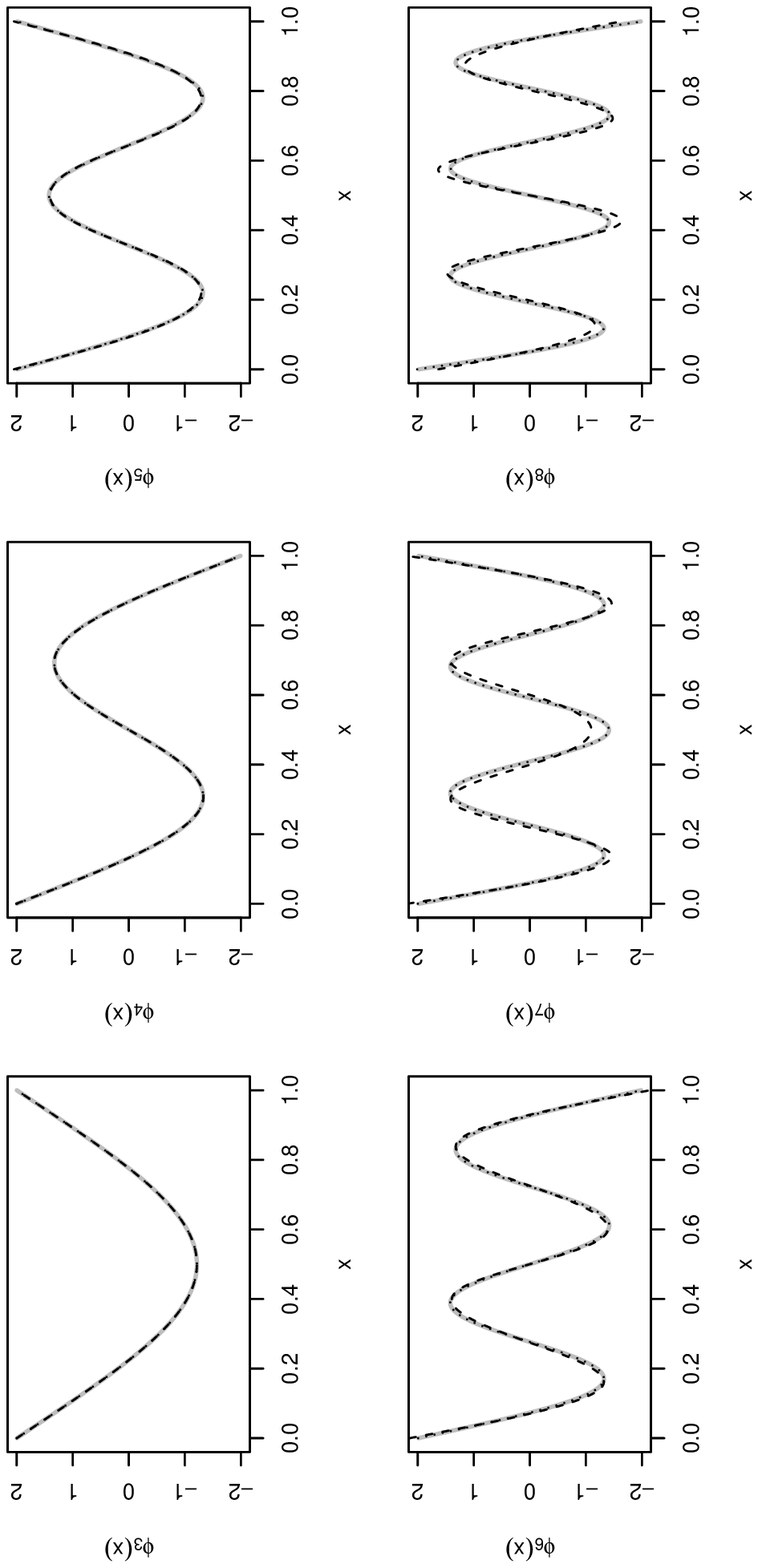}\vspace*{-.5cm}

\caption{Eigenfunctions of ${\cal K}^2$ shown in grey, with Demmler-Reinch spline basis functions overlaid in black. The first two linear functions are not shown. The dashed curves are for a rank 8 cubic spline basis, while the dotted curve, exactly overlaying the grey curves, are for a rank 16 cubic spline basis. 
\label{eigenfunc.fig}}
\end{figure}

Figure \ref{eigenfunc.fig} shows  the first 6 non-linear eigenfunctions of ${\cal K}^2$ computed by `brute-force' discretization in grey, with the normalized cubic Demmler-Reinsch spline basis approximations shown in black for a rank 8 basis (dashed) and a rank 16 basis (dotted). Notice that the rank 8 basis approximation gives visible approximation errors for $\phi_6 \ldots \phi_8$, which have vanished for the rank 16 approximation. (Actually if we use the rank 8 thin plate regression spline basis of \cite{tprs} then the approximation is accurate to graphical accuracy.) 

In summary each increase in regression spline basis dimension can be viewed as refining the existing normalized Demmler-Reinsch basis functions, while adding a new one. Hence in this parameterization the notion of a sequence of estimates of a coefficient $\beta_j$ is meaningful even when the basis dimension is increasing.  

\subsection{Consistency of penalized regression splines \label{laml.cons}}

This section explains why the consistency of unpenalized regression splines carries over to penalized regression splines with smoothing parameters estimated by Laplace approximate marginal likelihood. Use of Laplace approximation introduces the extra restriction $k=O(n^\alpha)$, $\alpha\le 1/3$.

Since consistency and convergence rates of regression splines tell us nothing about what basis size to use at any finite sample size, it is usual to use a basis dimension expected to be too large, and to impose smoothing penalties to avoid overfit. In the cubic spline basis case the coefficient estimates become
$$
\hat \bp = \underset{\beta}{\text{argmax}} ~~ l(\bp) - \frac{\lambda}{2} \int g^{\prime \prime}(x)^2 dx
$$
where $\lambda$ is a smoothing parameter and the penalty can be written as $\lambda \int g^{\prime \prime}(x)^2 dx = \lambda \bp\ts {\bf S}\bp$, for known coefficient matrix $\bf S$. From a Bayesian viewpoint the penalty arises from an improper Gaussian prior $\bp \sim N\{{\bf 0},(\lambda {\bf S})^-\}$. 

Consistency of the unpenalized regression spline estimate for $g$ and $g^{\prime\prime}$ implies consistency of penalized estimates when the smoothing parameter is estimated by Laplace approximate marginal likelihood (again assuming  a regular likelihood and that the true $g$ is 4 time differentiable). To see this, first set the smoothing parameter to 
$$
\lambda^* = \frac{k-2}{\int g^{\prime\prime}(x)^2 dx}, 
$$
where the basis size $k = O(n^\alpha)$ for $\alpha \in (0,1/3)$. Routine calculation shows that this is the value of $\lambda$ that maximizes the prior density at the true $P(g)=\int g^{\prime \prime}(x)^2 dx$, although we do not need this fact. Because the regression spline is consistent for $g^{\prime\prime}$ it is also consistent for $P(g)$. So in the unpenalized case the evaluated $P(\hat g)$ would be $O(\int g^{\prime \prime}(x)^2 dx)$, while in the penalized case it must be at most $O(\int g^{\prime \prime}(x)^2 dx)$. Hence with the given $\lambda^*$ the penalty is at most $O(k)$, while the log likelihood is $O(n)$. Intuitively this suggests that the penalty is unlikely to alter the consistency of the unpenalized maximum likelihood estimates.

To see that this intuition is correct, we first reparameterize using the normalized Demmler-Reinsch basis of the previous section. Then the penalized estimate of $ \bm \beta $ must satisfy 
\beq
\pdif{l}{\bm \beta}  - \lambda^* {\bm \Lambda} \bp = 0. \label{ndr.ne}
\eeq
It turns out that if we linearize this equation about the unpenalized $\hat \bp$, then in the large sample limit the solution of the linearized version is at the unpenalized $\hat \bp$, implying that (\ref{ndr.ne}) must have a root at $\hat \bp$ in the large sample limit. Specifically, defining $\Delta \bp = \bp - \hat \bp$, and then solving the linearized version of (\ref{ndr.ne}) for $\Delta \bp$ yields
$$
\Delta \bp = -\left ({\bf H} + \lambda^* {\bm \Lambda}  \right )^{-1} \lambda^* {\bm \Lambda}\hat \bp. \text{~~where~~} {\bf H} = -\pddif{l}{\bp}{\bp\ts}.
$$
Given the reparameterization the elements of $({\bf H} + \lambda^* {\bm \Lambda} )^{-1}$ are at most $O(n^{\delta-1}) $ where $0\le\delta\le \alpha$, while $\lambda^* \hat \bp \ts {\bm \Lambda}\hat \bp = O(n^\alpha)$. Hence if all the $|\hat \beta_i|$ are bounded below then the $\lambda^* \Lambda_{ii} \hat \beta_i$ are at most $O(n^\alpha)$ and the elements of $\Delta \bp$ are at most $O(n^{2\alpha+\delta-1})$ (since each $\Delta \beta_i$ is the sum of $O(n^\alpha) $ terms each of which is the product of an $O(n^{\delta-1})$ and an $O(n^\alpha)$ term).  Alternatively, $\hat \beta_i = O(n^{(\delta-1)/2})$, in which case $\lambda^* \Lambda_{ii} = O(n^{\gamma})$ where $\alpha < \gamma \le \alpha + 1 - \delta$. If $\gamma \le 1 - \delta$ then the elements of $\Delta \bp$ will be at most $O(n^{\alpha +(\delta-1)/2})$. Otherwise the $i^\text{th}$ row and column of $({\bf H} + \lambda^* {\bm \Lambda} )^{-1}$ are $O(n^{-\gamma})$, but then the elements of $\Delta \bp$ are also $O(n^{\alpha +(\delta-1)/2})$.
 So $\Delta \bp \to 0$ given the assumption that $\alpha < 1/3$ (of course this is only sufficient here).

Since the true $g$ is unknown we can not use $\lambda^*$ in practice. Instead $\lambda$ is chosen to maximize the Laplace approximate marginal likelihood (LAML),
$$
{\cal V} = \log f({\bf y}|\hat {\bm \beta}_\lambda) + \log f_\lambda (\hat {\bm \beta}_\lambda) - \frac{1}{2} \log |{\cal H}_\lambda | + \frac{k}{2} \log(2 \pi) \simeq \log \int f({\bf y}| {\bm \beta}) f_\lambda ( {\bm \beta})d {\bm \beta}
$$ 
where $\hat {\bm \beta}_\lambda$ denotes the posterior mode/ penalized MLE of ${\bm \beta}$ for a given $\lambda$, and ${\cal H}_\lambda$ is the Hessian of the negative log of $f({\bf y}|\hat {\bm \beta}) f_\lambda (\hat {\bm \beta})$. \cite{shun1995laplace} show that in general we require $k = O(n^\alpha)$ for $\alpha \le 1/3$ for the Laplace approximation to be well founded. %Hence penalized regression splines with first derivative penalties appear to be at the limit of what can be handled by Laplace approximate marginal likelihood, in the penalized smoothing asypmtotic regime, and inference about full smoothing splines via LAML would require more specific theory. 
If $g= \alpha_0 + \alpha_1 x$ for finite real constants $\alpha_0 $ and $\alpha_1$, then the smoothing penalty is 0 for the true $g$ and consistency follows from the consistency in the un-penalized case, irrespective of $\lambda$. 

Now suppose that $g$ is not linear. A maximum of ${\cal V} $ must satisfy
\beq
\dif{{\cal V}}{\lambda} =
\left ( \left . \pdif{\log f({\bf y}| {\bm \beta})}{ {\bm \beta}} \right |_{\hat {\bm \beta}_\lambda} + 
\left .\pdif{\log f_\lambda ( {\bm \beta})}{{\bm \beta}} \right |_{\hat {\bm \beta}_\lambda}  \right ) 
\dif{\hat {\bm \beta}_\lambda}{\lambda} + 
\pdif{\log f_\lambda (\hat {\bm \beta}_\lambda)}{\lambda} - 
\frac{1}{2} \tr{{\cal H}_\lambda^{-1} {\bf S}} - \frac{1}{2} \tr{{\cal H}_\lambda^{-1} \dif{{\bf H}}{\lambda}} =  0 \label{laml.deriv}
\eeq

The first term in brackets is zero by definition, so the maximizer of $\cal V$ must satisfy $2\ilpdif{\log f_\lambda (\hat {\bm \beta}_\lambda)}{\lambda} - \tr{{\cal H}_\lambda^{-1} {\bf S}} - \tr{{\cal H}_\lambda^{-1} \ildif{{\bf H}}{\lambda}}=  0$ implying (after some routine manipulation) that the maximiser, $\hat \lambda$, must satisfy $\lambda^\prime(\hat \lambda) = \hat \lambda$, where 
\beq
\lambda^\prime(\lambda) = \frac{k-2}{\hat {\bm \beta}_\lambda\ts {\bf S}\hat {\bm \beta}_\lambda + \tr{{\cal H}_\lambda^{-1} {\bf S}} + \tr{{\cal H}_\lambda^{-1} \ildif{{\bf H}}{\lambda}}}. \label{laml.sp}
\eeq
$\left . \ilpdif{{\cal V}}{\lambda}\right |_{\epsilon} \le 0 $ for arbitrarily small $\epsilon>0$ would imply a LAML optimal smoothing parameter $\lambda = 0$, otherwise $\left . \ilpdif{{\cal V}}{\lambda}\right |_\epsilon > 0 $ implying that the right hand side of (\ref{laml.sp}) is positive at $\lambda=\epsilon$. Hence if  $\lambda^\prime \le \lambda^*$ when $\lambda=\lambda^*$, then LAML must have a turning point in $(0,\lambda^*)$\footnote{Consider plotting $\lambda^\prime(\lambda)$ against $\lambda$ for $0<\lambda<\lambda^*$. The $\lambda^\prime(\lambda)$ curve will start above the line $\lambda^\prime = \lambda$ and finish below it.}. In fact $\tr{{\cal H}_{\lambda^*}^{-1} \ildif{{\bf H}}{\lambda^*}} \to 0$ as $n \to \infty$ (see \ref{dH.vanish}), while consistency of $\hat {\bm \beta}_{\lambda^*}$ implies that the limiting value of $\hat {\bm \beta}_{\lambda^*}\ts {\bf S}\hat {\bm \beta}_{\lambda^*}$ is $ \int g^{\prime \prime}(x)^2 dx$. Hence in the large sample limit, since $\tr{{\cal H}_\lambda^{-1} {\bf S}}>0$, we have that $\lambda^\prime < \lambda^*$ as required (the latter is equivalent to $\left . \ilpdif{\cal V}{\lambda} \right |_{\lambda^*}<0$ confirming that there is a {\em maximum} in $(0,\lambda^*)$).  Notice how straightforward this is relative to what is needed for full spline smoothing where $k = O(n)$ and much more work is required. 

The result is unsurprising of course. Restricted marginal likelihood is known to smooth less that Generalized Cross Validation \citep{wahba85}, but the latter is a prediction error criterion and smoothing parameters resulting in consistent estimates are likely to have lower prediction error than smoothing parameters that result in inconsistent estimation, at least asymptotically. 

\subsubsection{$\tr{{\cal H}_\lambda^{-1} \ildif{{\bf H}}{\lambda}}$ \label{dH.vanish}}

In section \ref{laml.cons} we require that $\tr{{\cal H}_{\lambda^*}^{-1} \ildif{{\bf H}}{\lambda^*}} \to 0$ in the large sample limit. Unfortunately there are too many summations over $k$ elements involved in this computation for the simple order bounding calculations used in section \ref{laml.cons} to yield satisfactory bounds for all $\alpha \in (0,1/3)$. This can be rectified by another change of basis, to a slightly modified  normalized Demmler-Reinsch type basis in which $\bf H + \lambda {\bf S}$ is diagonal. Specifically let ${\bf H} = {\bf R}\ts {\bf R}$, ${\bf U}{\bm \Lambda}{\bf U}\ts = {\bf R}\its {\bf S}{\bf R}^{-1}$, and let the reparameterization be $\bp^* = n^{-1/2}{\bf U}\ts{\bf R}\bp$. In the remainder of this section we work with this basis.

We have
$$
\dif{H_{ij}}{\lambda} = \sum_k \ptdif{l}{\beta_{i}}{\beta_j}{\beta_k} \dif{\hat \beta_k}{\lambda}
$$
where the third derivative terms are $O(n)$ (at most). By implicit differentiation we also have
$$
\dif{\hat \bp}{\lambda} = -({\bf I}n + \lambda^* {\bm \Lambda})^{-1} {\bm \Lambda} \hat \bp
$$
in the new parameterization. As in section \ref{laml.cons}, for $\hat \beta_i$ bounded away from zero, the fact that $\hat \bp\ts {\bm \Lambda} \hat \bp = O(1)$ leads easily to the required result, but again the $\hat \beta_i = O(n^{-1/2})$ case makes the bounds slightly less easy to find. In that case $\Lambda_{ii} = O(n^{\gamma})$, $0<\gamma \le 1$, while   $\lambda^* \Lambda_{ii} = O(n^{\gamma+\alpha})$. Then if $\gamma + \alpha \ge 1$ the $i^\text{th}$ leading diagonal element of  $({\bf I}n + \lambda^* {\bm \Lambda})^{-1}$ is $O(n^{-\gamma-\alpha})$, and $\ilpdif{\hat \beta_i}{\lambda} = O(n^{-\alpha - 1/2})$. Otherwise $\ilpdif{\hat \beta_i}{\lambda} = O(n^{\gamma - 3/2})$, which is less than or equal to  $O(n^{-\alpha - 1/2})$ if $\gamma + \alpha < 1$. In consequence $\ilpdif{H_{ij}}{\lambda} = O(n^{1/2})$, at most. It then follows that $\tr{{\cal H}_\lambda^{-1} \ildif{{\bf H}}{\lambda}} = -\tr{ ({\bf I}n + \lambda^* {\bm \Lambda})^{-1}\ildif{{\bf H}}{\lambda}} = O(n^{\alpha-1/2})$.

\subsection{Convergence rates}

The preceding consistency results reveal nothing about convergence rates. For a cubic spline with evenly spaced knots parameterised using a cardinal spline basis, ${\bf S} = O(k^3)$ \cite[see e.g. ][\S 4.1.2]{wood2006igam}, so $\lambda {\bf S}$ has elements of at most $ O(k^4)$, while the Hessian of the log likelihood has elements $O(n/k)$. In consequence if $k=O(n^{\alpha})$, $\alpha<1/5$, $\lambda {\bf S}$ is completely dominated by the Hessian of the log likelihood in the large sample limit (the elements of the score vector also dominate the elements of the penalty gradient vector), so that the penalty has no effect on any model component. Hence in the limit we have an un-penalized regression spline, and the asymptotic mean square error convergence rate is $O(n^{-8\alpha})$ (bias/approximation error dominated) for $\alpha \le 1/9$ and $O(n^{\alpha - 1})$ (variance dominated) otherwise. Notice that at the $\alpha \to 1/5$ edge of this `asymptotic regression' regime the convergence rate tends to $O(n^{-4/5})$.%, the optimal non-parametric rate for a cubic spline.

\begin{figure}
\eps{-90}{.55}{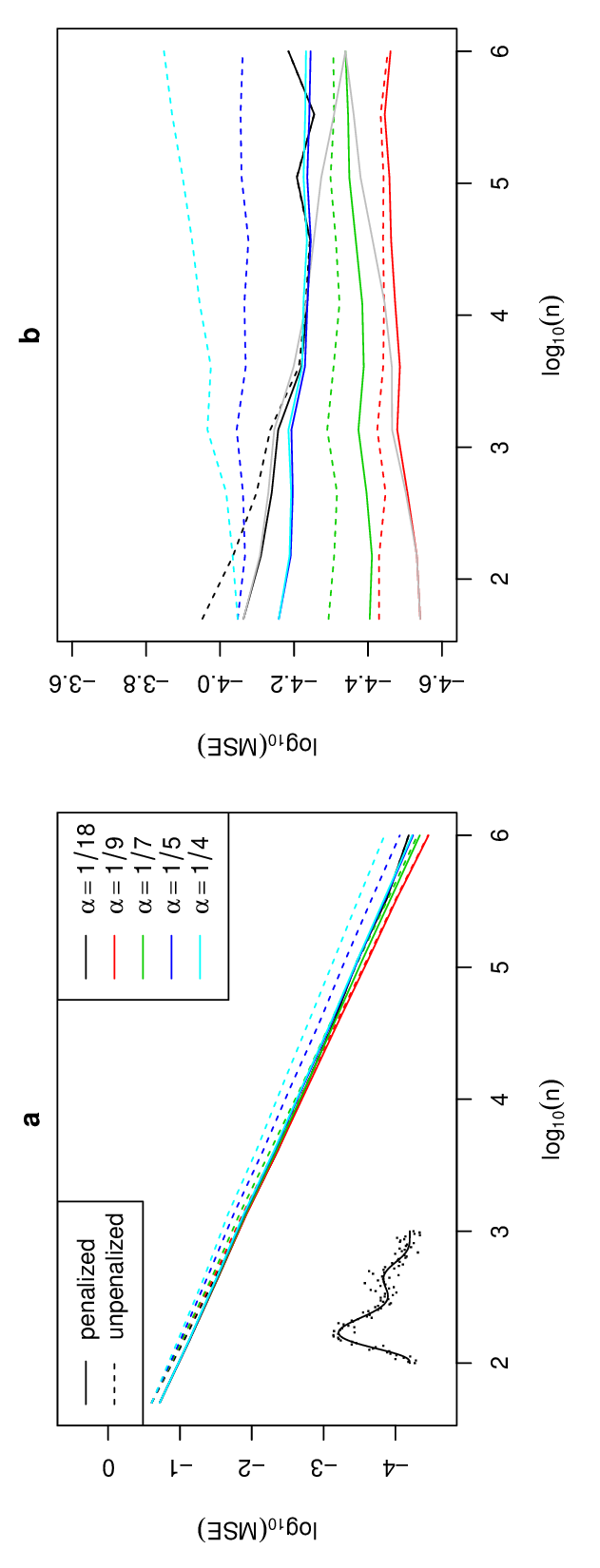}\vspace*{-.5cm}

\caption{{\bf a.} Example of MSE convergence for simple Gaussian smoothing. The true function is shown at lower left, with 100 noisy samples also shown. The coloured lines show log MSE averaged over 100 replicates against log sample size when the basis size $k \propto n^\alpha$ for various $\alpha$ values (all starting from $k=12$ at $n=50$). Dashed lines are for unpenalized regression and solid for penalized. For $\alpha=1/18 $ we eventually see an approximation error dominated rate. For $\alpha < 1/5$ the penalized and unpenalized curves converge, while for $\alpha \ge 1/5$ the penalty always improves the convergence rate. {\bf b.} The same data, but de-trended by subtraction of the log MSE that would have occurred under the theoretical asymptotic convergence rate, if the observed MSE at $n=10^6$ is correct. The theoretical rate used for $\alpha\ge1/5$ was $n^{-4/5}$. For reference, the grey curves show curves obtained for $\alpha=1/7$ if we incorrectly use the theoretical rates for $\alpha=1/18, 1/9$.     
\label{convergence.fig}}
\end{figure}

For $\alpha \ge 1/5$ the total dominance of $\lambda {\bf S}$ by the Hessian ceases: i.e. as $n \to \infty$ the penalty can suppress overfit, in principle suppressing spurious components of the fit more rapidly than the likelihood alone would do. We do not know how to obtain actual convergence rates in this regime under LAML, although we expect them to lie between $O(n^{-4/5}) $ and $O(n^{\alpha-1})$, with simulation evidence suggesting rates close to $O(n^{-4/5})$. Figure \ref{convergence.fig} shows observed convergence rates for a simple Gaussian smoothing example (a binary example gives a similar plot, but with slower convergence of the penalized case to the unpenalized case for $\alpha < 1/5$). 

The best mean square error rate possible for a non-parametric estimator of a $C^4$ function is $O(n^{-8/9})$ \citep{cox1983}, which a cubic smoothing spline can achieve under certain assumptions on the the rate of change of $\lambda$ with $n$ \citep{stone1982, speckman1985}.  \citet{hall2005} obtain the same rate for penalized cubic regression splines as considered here. However obtaining rates under smoothing parameter selection (by REML,GCV or whatever) is more difficult. \citet{kauermann2009} consider inference under LAML selection of smoothing parameters, but assume $k=O(n^{1/9})$ (in the cubic case). As we have seen, under LAML smoothness selection, this leads to penalized regression simply tending to  unpenalized regression in the limit. \cite{claeskens2009} recognise the existence of 2 asymptotic regimes, corresponding to penalizing in the limit and not, but do not treat the estimation of smoothing parameters. 

% by a penalized smoother with two continuous derivatives is $O(n^{-4/5})$ \citep{stone1982, speckman1985}, which \citet{hall2005} show also applies to penalized cubic regression splines as considered here. This rate suggests using $k=O(n^\alpha)$, $\alpha \ge 1/5$. However, in quite general settings, \citet{kauermann2009} demonstrate the theoretical well-foundedness of penalized regression splines with LAML selection of smoothing parameters when $k = O(n^{1/9})$ for the cubic case. The corresponding mean square error rate is $O(n^{-8/9})$. The apparent inconsistency arises from the consideration of different asymptotic regimes in the \citet{kauermann2009} and \citet{hall2005} papers. As discussed above, $\alpha < 1/5$ yields a regime in which penalized regression splines `relax' back to unpenalized regression splines in the large sample limit, while for $\alpha \ge 1/5$ the penalty may continue to suppress overfit even in the large sample limit \citep[see also][]{claeskens2009}. 

It could be argued that in practice a statistician would tend to view a model fit with very low penalization as an indication of possible underfit, and to increase the basis dimension in response, which implies that under LAML the $\alpha \ge 1/5$ regime (penalizing in the large sample limit) is more informative in practice. The counter argument is that it is odd to choose the regime that gives the lower asymptotic convergence rates. A third point of view simply makes the modelling assumption that the truth is in the space spanned by a finite set of spline basis functions, in which case unpenalized consistency follows from standard maximum likelihood theory, and the effect of penalization with LAML smoothing parameter selection is readily demonstrated to vanish in the large sample limit. In any case the use of penalized regression splines seems to be reasonably well justified. 

\subsection{Large sample posterior under penalization}

Consider a regular log likelihood with second and third derivatives $O(n/k)$, so that we are interested in values of the model parameters such that $|\beta_i - \hat \beta_i| = O(\sqrt{k/n})$. By Taylor's theorem \citep[e.g.][\S 2.3.4]{gill.murray.wright} we have
\begin{eqnarray}
\log f(\bp|{\bf y} ) &\propto& \log f({\bf y}|\bp) - \frac{1}{2}\bp \ts{\bf S}^\lambda \bp \nonumber \\ 
&=& \log f({\bf y}|\hat \bp ) - \frac{1}{2}\hat\bp\ts{\bf S}^\lambda\hat \bp - 
\frac{1}{2} (\bp - \hat\bp)\ts ( \hat {\bm{{\cal I}}}+{\bf S}^\lambda) (\bp - \hat \bp) + R \label{post.taylor}
\end{eqnarray}
where
$$
R = \frac{1}{6}\sum_{ijk} \left . \ptdif{\log f({\bf y}|\bp)}{\beta_i}{\beta_j}{\beta_k} \right |_{\bp^*} (\beta_i - \hat \beta_i) (\beta_j - \hat \beta_j) (\beta_k - \hat \beta_k) 
$$
and $\bp^* = t\bp + (1-t)\hat \bp$ for some $t \in (0,1)$. (\ref{post.taylor}) can be re-written as 
$$
\log f(\bp|{\bf y} ) \propto \log f({\bf y}|\hat \bp ) - \frac{1}{2}\hat\bp\ts{\bf S}^\lambda\hat \bp - 
\frac{1}{2} (\bp - \hat\bp)\ts ( \hat {\bm{{\cal I}}}+{\bf S}^\lambda + {\bf R}) (\bp - \hat \bp)
$$
where 
$$
R_{ij} = \frac{1}{3}\sum_{k} \left . \ptdif{\log f({\bf y}|\bp)}{\beta_i}{\beta_j}{\beta_k} \right |_{\bp^*} (\hat \beta_k -  \beta_k).
$$
In the region of interest for $\bp$, $R_{ij}$ are at most $O(\sqrt{kn})$, whereas the elements of $\hat {\bm{{\cal I}}}+{\bf S}^\lambda$ are at least $O(n/k)$. Hence if $k = O(n^\alpha)$, $\alpha<1/3$ then $\hat {\bm{{\cal I}}}+{\bf S}^\lambda$ dominates $\bf R$ in the $n\to \infty $ limit, and $\log f(\bp|{\bf y} )$ tends to the p.d.f. of $N(\hat \bp, (\hat {\bm{{\cal I}}}+{\bf S}^\lambda)^{-1})$. Again this is much simpler than would be required for full spline smoothing where $k=O(n)$.

\section{LAML derivation and log determinants}

Consider a model with log likelihood $l = \log f({\bf y}|\bp)$ and improper prior $f(\bp) = |{\bf S}^\lambda|_+^{1/2} \exp \{-\bp \ts {\bf S}^\lambda \bp/2\}/\sqrt{2 \pi}^{p-M_p}$ where $p = \text{dim}(\bp)$. By Taylor expansion of $\log\{f({\bf y}|\bp)f(\bp) \}$ about $\hat \bp$,
\begin{eqnarray*}
\int f({\bf y}|\bp) f(\bp) d \bp &\simeq&
\int \exp \left \{  
l(\hat \bp) -(\bp-\hat \bp)^T\bm{{\cal H}} (\bp-\hat \bp)/2 - 
\hat \bp \ts {\bf S}^\lambda\hat \bp/2 + \log |{\bf S}^\lambda|_+^{1/2} - 
\log (2 \pi)(p - M_p)/2
\right \} d \bp \\
&=& \exp\{ {\cal L}(\hat\bp)\}|{\bf S}^\lambda|_+^{1/2} \sqrt{2 \pi}^{M_p-p} \int \exp \{-(\bp-\hat \bp)^T\bm{{\cal H}} (\bp-\hat \bp)/2\}   d \bp  \\
&=& \exp\{ {\cal L}(\hat\bp)\} \sqrt{2 \pi}^{M_p} |{\bf S}^\lambda|_+^{1/2} / |\bm{{\cal H}}|^{1/2}
\end{eqnarray*}
where $\bm{{\cal H}}$ is the negative Hessian of the penalized log likelihood, $\cal L$.

\subsection{The problem with log determinants \label{ldet.prob}}

Unstable determinant computation is the central constraint on the development of practical fitting methods, and it is necessary to understand the issues in order to understand the structure of the numerical fitting methods. A very simple example provides adequate illustration of the key problem. Consider the real $5 \times 5$ matrix ${\bf C}$ with QR decomposition ${\bf C} = {\bf QR}$ so that $|{\bf C}| = |{\bf R}| = \prod_i R_{ii}$. Suppose that ${\bf C} = {\bf A} + {\bf B}$ where $\bf A$ is rank 2 with non-zero elements of size $O(a)$, $\bf B  $ is rank 3 with non-zero elements of size $O(b)$ and $a\gg b$. Let the schematic non-zero structure of ${{\bf C} = {\bf A} + {\bf B}}$ be 
{\matsize
$$
\bmat{ccccc}
\bullet & \bullet & \bullet & & \\ 
\bullet & \bullet & \bullet & & \\ 
\bullet & \bullet & \bullet & \cdot & \cdot 
\\&&\cdot&\cdot & \cdot 
\\&&\cdot&\cdot&\cdot
\emat = 
\bmat{ccccc} 
\bullet & \bullet & \bullet & & \\ \bullet & \bullet & \bullet & & \\ \bullet & \bullet & \bullet & & \\&&&& \\&&&&
\emat + 
\bmat{ccccc} 
&&&&\\
&&&&\\
&&\cdot&\cdot&\cdot\\
&&\cdot&\cdot&\cdot\\
&&\cdot&\cdot&\cdot\\
\emat
$$
}\normalsize
where $\bullet$ shows the $O(a)$ elements and $\cdot$ those of $O(b)$. 
Now QR decomposition \citep[see][]{GvL4} operates by applying successive householder reflections to $\bf C$, each in turn zeroing the subdiagonal elements of successive columns of $\bf C$. Let the product of the first 2 reflections be ${\bf Q}_2\ts$ and consider the state of the QR decomposition after 2 steps. Schematically ${\bf Q}_2\ts {\bf C} = {\bf Q}_2\ts{\bf A} +{\bf Q}_2\ts{\bf B}$ is 
{\matsize
$$
\bmat{ccccc}
\bullet & \bullet & \bullet &\cdot & \cdot\\ 
 & \bullet & \bullet & \cdot & \cdot \\ 
 &  & d_1 & \cdot & \cdot 
\\&& d_2 &\cdot & \cdot 
\\&& d_3 &\cdot&\cdot
\emat = 
\bmat{ccccc} 
\bullet & \bullet & \bullet & & \\ 
 & \bullet & \bullet & & \\ 
 &  & d^\prime_1 & & \\&& d^\prime_2 && \\&& d^\prime_3 &&
\emat + 
\bmat{ccccc} 
&&\cdot&\cdot&\cdot\\
&&\cdot&\cdot&\cdot\\
&&d^{\prime\prime}_1&\cdot&\cdot\\
&&d^{\prime\prime}_2&\cdot&\cdot\\
&&d^{\prime\prime}_3&\cdot&\cdot\\
\emat
$$
}\normalsize
Because $\bf A$ is rank 2, $\bf d^\prime_j$ should be 0, and $d_j$ should be $d_j^{\prime\prime}$ but computationally $d^\prime_j = O(\epsilon a)$ where $\epsilon $ is the machine precision. Hence if $b$ approaches $O(\epsilon a)$, we suffer catastrophic loss of precision in $\bf d$, which will be inherited by $R_{33}$ and the computed value of $|{\bf C}|$. Matrices such as $\sum_j \lambda_j  \ts {\bf S}^j $ can suffer from exactly this problem, since some $\lambda_j$ can legitimately tend to infinity while others remain finite, and the ${\bf S}^j$ are usually of lower rank than the dimension of their non-zero sub-block: hence both log determinant terms in the LAML score are potentially unstable.

One solution is based on similarity transform. In the case of our simple example, consider the similarity transform ${\bf U} {\bf CU}\ts = {\bf U}{\bf A}{\bf U}\ts + {\bf U}{\bf B} {\bf U}\ts$ constructed to produce the following schematic
{\matsize
$$
\bmat{ccccc}
\bullet & \bullet & \cdot & \cdot& \cdot \\ 
\bullet & \bullet & \cdot & \cdot&\cdot \\ 
\cdot & \cdot & \cdot & \cdot & \cdot 
\\ \cdot & \cdot &\cdot&\cdot & \cdot 
\\ \cdot & \cdot &\cdot&\cdot&\cdot
\emat = 
\bmat{ccccc} 
\bullet & \bullet &  & & \\ 
\bullet & \bullet &  & & 
 \\&&&& \\&&&&\\&&&&
\emat + 
\bmat{ccccc} 
\cdot &\cdot&\cdot&\cdot&\cdot\\
\cdot &\cdot&\cdot&\cdot&\cdot\\
\cdot &\cdot&\cdot&\cdot&\cdot\\
\cdot &\cdot&\cdot&\cdot&\cdot\\
\cdot&\cdot &\cdot&\cdot&\cdot\\
\emat.
$$
}\normalsize
${\bf U} {\bf CU}\ts$ can then be computed by adding ${\bf U}{\bf B} {\bf U}\ts$ to ${\bf U}{\bf A} {\bf U}\ts$ with the theoretically zero elements set to exact zeroes. $|{\bf U} {\bf CU}\ts| = |{\bf C}|$, but computation based on the similarity transformed version no longer suffers from the precision loss problem, no-matter how disparate $a $ and $b$ are in magnitude. \cite{wood2011} discusses the issues in more detail and provides a practical generalized version of the similarity transform  approach, allowing for multiple rank deficient components where the dominant blocks may be anywhere on the diagonal. 

\section{Smoothing parameter uncertainty \label{spu.append}}

\noindent $\ilpdif{{\bf R}}{\rho_k}$: Computation of the ${\bf V}^{\prime\prime}$ term requires $\ilpdif{{\bf R}^\prime}{\rho}$ where ${\bf R}^{\prime {\sf T}} {\bf R}^\prime = {\bf V}_\beta$. Generally we have access to $\ilpdif{\bf A}{\rho}$ where ${\bf A} = {\bf V}_\beta^{-1}$. Given Cholesky factorization ${\bf R}\ts{\bf R} = {\bf A}$ then ${\bf R}^\prime = {\bf R}\its$, and $\ilpdif{{\bf R}^{\prime \sf T}}{\rho} = -{\bf R}^{-1} \ilpdif{{\bf R}}{\rho}{\bf R}^{-1}$. Applying the chain rule to the Cholesky factorization yields
$$
\pdif{R_{ii}}{\rho} = \frac{1}{2} R_{ii}^{-1} B_{ii},~~
\pdif{R_{ij}}{\rho} = R_{ii}^{-1} \left ( B_{ij} - R_{ij} \pdif{R_{ii}}{\rho} \right ),~~ B_{ij} = \pdif{A_{ij}}{\rho} - \sum_{k=1}^{i-1} \pdif{R_{ki}}{\rho} R_{kj} + R_{ki} \pdif{R_{kj}}{\rho},
$$
and $\sum_{k=1}^{0} x_i$ is taken to be 0. The equations are used starting from the top left of the matrices and working across the columns of each row before moving on to the next row, at approximately double the floating point cost of the original Cholesky factorization, but with no square roots. 

\subsubsection*{Ratio of the first order correction terms}

In the notation of section 4, %\ref{dist.sec}, 
we now show that for any smooth $g_j$, $\ilpdif{\hat{\bm \beta}}{{\rho}_j}$ tends to dominate $\ilpdif{{\bf R}_{\hat \rho}\ts {\bm z}}{{\rho}_j}$ for those components of the smooth that are detectably non zero. First rewrite ${\bf S}^\rho = {\bf S}_{-j} + \lambda_j {\bf S}_j$, by definition of ${\bf S}_{-j}$, and then form the spectral decomposition ${\bm{{\cal I}}} + {\bf S}_{-j} = {\bf VDV}\ts$. Form a second spectral decomposition ${\bf D}^{-1/2}{\bf V}\ts {\bf S}_j {\bf V}{\bf D}^{-1/2}={\bf U}{\bm \Lambda}{\bf U}\ts$, so that 
${\bm{{\cal I}}}+{\bf S}^\rho = {\bf VD}^{1/2}{\bf U}({\bf I} + \lambda_j {\bm \Lambda}) 
{\bf U}\ts {\bf D}^{1/2}{\bf V}\ts
$. Now linearly re-parameterize so that ${\bf S}_j$ becomes $\bm \Lambda$ and ${\bm{{\cal I}}}+{\bf S}^\rho$ becomes ${\bf I} + \lambda_j{\bm \Lambda}$, while ${\bf R}\ts = ({\bf I} + \lambda_j{\bm \Lambda})^{-1/2}$. By the implicit function theorem, in the new parameterization 
$
\ildif{\hat {\bm \beta}}{\rho_j} = \lambda_j({\bf I} + \lambda_j {\bm \Lambda})^{-1}{\bm \Lambda} \hat {\bm \beta}. 
$ 
Notice that $\bm \Lambda$ has only $\text{rank}({\bf S}_j)$ non-zero entries, corresponding to the parameters in the new parameterization representing the penalized component of $g_j$. Furthermore
$
\ildif{{\bf R}\ts {\bf z}}{\rho_j} \simeq \lambda_j ({\bf I} + \lambda_j {\bm \Lambda})^{-3/2}{\bm \Lambda} {\bf z}, 
$
where we have neglected the indirect dependence on smoothing parameters via the curvature of $\bm{{\cal I}}$ changing as $\bm \bp$ changes with $\bm \rho$. Hence for any penalized parameter $\beta_i$ of $g_j$
$$
\frac{\ildif{\hat {\beta_i}}{\rho_j}}{\ildif{({\bf R}\ts {\bf z})_i}{\rho_j}} \simeq 
\frac{\hat \beta_i}{(1+\lambda_j \Lambda_{ii})^{-1/2}z_i},
$$
but $(1+\lambda_j \Lambda_{ii})^{-1/2}$ is the (posterior) standard deviation of $\beta_i$. So the more clearly non-zero is $\beta_i$, the more $\ildif{\hat {\beta_i}}{\rho_j}$  dominates $\ildif{({\bf R}\ts {\bf z})_i}{\rho_j}$. The dominance increases with sample size (provided that the data are informative), for all components except those heavily penalized towards zero.

\noindent {\bf Proof of lemma 1} %Note that $\E_\pi ({\bm \beta} \bp\ts) = {\bf S}$.
Form eigen-decompositions  $\hat{\bm{{\cal I}}} = {\bf VDV}\ts$ and ${\bf D}^{-1/2}{\bf V}\ts {\bf SVD}^{-1/2} =  {\bf U}{\bm \Lambda}{\bf U}\ts$, and linearly re-parameterize $\bp^\prime = {\bf U}\ts{\bf D}^{-1/2}{\bf V}\ts \bp$, so that in the new parameterization $\hat{\bm{{\cal I}}} $ becomes an identity matrix, while the prior becomes $\bp^\prime \sim N({\bf 0}, {\bm \Lambda}^-)$, ${\bf V}_{\hat \beta^\prime} = ({\bf I} + {\bm \Lambda})^{-2}$ and ${\bf V}_{\beta^\prime} = ({\bf I} + {\bm \Lambda})^{-1}$.
\begin{eqnarray*}
{\bf V}_{\hat \beta^\prime} + \E_\pi (\tilde {\bm \Delta}_{\beta^\prime} \tilde {\bm \Delta}_{\beta^\prime}\ts) &=& ({\bf I} + {\bm \Lambda})^{-2} + \E_\pi [\{({\bf I} + {\bm \Lambda})^{-1} - {\bf I} \}
\bp^\prime \bp^{\prime {\sf T}}\{({\bf I} + {\bm \Lambda})^{-1} - {\bf I} \}]\\
&=& ({\bf I} + {\bm \Lambda})^{-2} + \{({\bf I} + {\bm \Lambda})^{-1} - {\bf I} \}
{\bm \Lambda}^-\{({\bf I} + {\bm \Lambda})^{-1} - {\bf I} \}\\
&=& ({\bf I} + {\bm \Lambda})^{-1} [({\bf I} + {\bm \Lambda})^{-1} + 
{\bm \Lambda}^- \{({\bf I} + {\bm \Lambda})^{-1} - {\bf I} \} - ({\bf I} + {\bm \Lambda}){\bm \Lambda}^- \{({\bf I} + {\bm \Lambda})^{-1} - {\bf I} \}] \\
&=& ({\bf I} + {\bm \Lambda})^{-1} [({\bf I} + {\bm \Lambda})^{-1} - \bm{ \Lambda \Lambda}^-\{({\bf I} + {\bm \Lambda})^{-1} - {\bf I} \}] = ({\bf I} + {\bm \Lambda})^{-1}{\bf I} = {\bf V}_{\beta^\prime}.
\end{eqnarray*}

\section{Further simulation details}

\begin{figure}
\eps{-90}{.7}{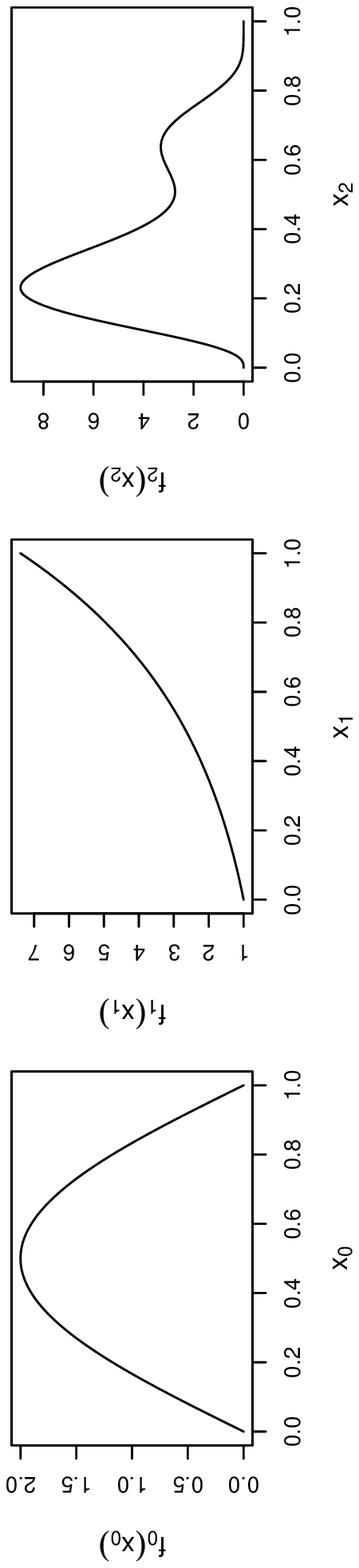}
\vspace*{-.6cm}
 
     \caption{Shapes of the functions used for the simulation study \citep[from][]{gu.wahba.msp}. $f_3(x_3)=0$.}
     \label{GW-shapes}
\end{figure}

Figure \ref{GW-shapes} shows the functions used in the simulation study in the main paper. In the uncorrelated covariate case $x_{0i}$, $x_{1i}$, $x_{2i}$ and $x_{3i}$ were all i.i.d. $U(0,1)$. Correlated covariates were marginally uniform, but were generated as $x_{ji} = \Phi^{-1}(z_{ji})$ where $\Phi$ is the standard normal c.d.f. and $(z_{0i},z_{1i}, z_{2i}, z_{3i}) \sim N({\bf 0},{\bm \Sigma})$ with $\bm \Sigma$ having unit diagonal and 0.9 for all other elements. The noise level was set by either using the appropriate values of the distribution parameters or by multiplying the linear predictor by the appropriate scale factor as indicated in the second column of table \ref{table-para} (the scale factor is denoted by $d$). The simulation settings and failure rates are given in table \ref{table-para} 

\renewcommand{\baselinestretch}{1}
\begin{table}[h] 
  \begin{center}
     \begin{tabular}{|c|c|c|c|c|}
        \hline
        \multicolumn{3}{|c|}{Simulation setting} &  Alternative  & \multicolumn{1}{|c|}{MSE/Biers diff. }\\
        \hline                     
        \rm{Family} & parameters   & \rm{approx.} $r^2$ & \% failure & \rm{p-value} \\
         \hline
         \rm{nb} & $\theta=3,d=.12$  &  0.25  &  -(.3) & 0.0015($0.0013$) \\   
                 &  $\theta=3,d=.2$   &  0.45   & -(.7) & 0.087($<10^{-5}$) \\    
                 & $\theta=3,d=.4$  & 0.79   &  -(.3) & $<10^{-5}$($<10^{-5}$) \\   
         \hline        
         \rm{beta} & $\theta=0.02$  & 0.3   & -(1.3)  &  0.40($<10^{-5}$) \\   
                  &  $\theta=0.01$   &  0.45   & -(1.3)  & 0.16($<10^{-5}$) \\ 
                  & $\theta=0.001$  &  0.9  & -(.7) & $<10^{-5}$($0.044$)\\
          \hline      
           \rm{scat} & $\nu = 5,\sigma=2.5$  & 0.5  &  -(2)  &  $.021$($<10^{-5}$)\\
                         &  $\nu = 3,\sigma=1.3$   &  0.7   & .3(-)  &  $<10^{-5}$($<10^{-5}$)\\
                         & $\nu = 4,\sigma=0.9$  &  0.85  &  -(.3)  & $<10^{-5}$($0.41$)\\
           \hline  
           \rm{zip} & ${\bm\theta} =(-2,0 ), d=2$  & 0.5    & 2(3.3)   & $<10^{-5}$($0.004$) \\
                         &  ${\bm\theta} =(-2,0 ),d=2.5$                   & 0.67       & 4.3(3.7) & $<10^{-5}$($0.001$) \\
                        & ${\bm\theta} =(-2,0 ), d=3$                    & 0.8        & 8.3(4.3)  & $<10^{-5}$($<10^{-5}$) \\
           \hline 
           \rm{ocat}     & ${\bm\theta} =(-1,0,3),d=.3$ & 0.4  &  - & 0.388($<10^{-5}$) \\
                         &  ${\bm\theta} =(-1,0,3),d=1$  & 0.7  & - &  0.0025(0.0023) \\
                         & ${\bm\theta} =(-1,0,3),d=2$ & 0.85  & - &   0.191($7.3\times 10^{-4}$)\\
           \hline
\end{tabular}
\end{center}\caption{Simulation settings, failure rates and p-values for performance differences when comparing the new methods to existing software. The approximate $r^2$ column gives the approximate proportion of the variance explained by the linear predictor, for each scenario. The fit failure rates for the alternative procedure are also given (for the correlated covariate case in brackets): the new method produced no failures. The p-values for the difference between MSE or Briers scores between the methods are also reported. The new method had the better average scores in all cases that were significant at the 5\% level, except for the zip model on uncorrelated data, where the GAMLSS methods achieved slightly lower MSE.\label{table-para}}
\end{table} 
\renewcommand{\baselinestretch}{\dsp}

\section{Some examples}

This section presents some example applications which are all routine given the framework developed here, but would not have been without it. See appendix \ref{software.append} for a brief description of the software used for this.

\subsection{Kidney renal clear cell carcinoma: Cox survival modelling with smooth interactions}
\begin{figure}
\eps{-90}{.55}{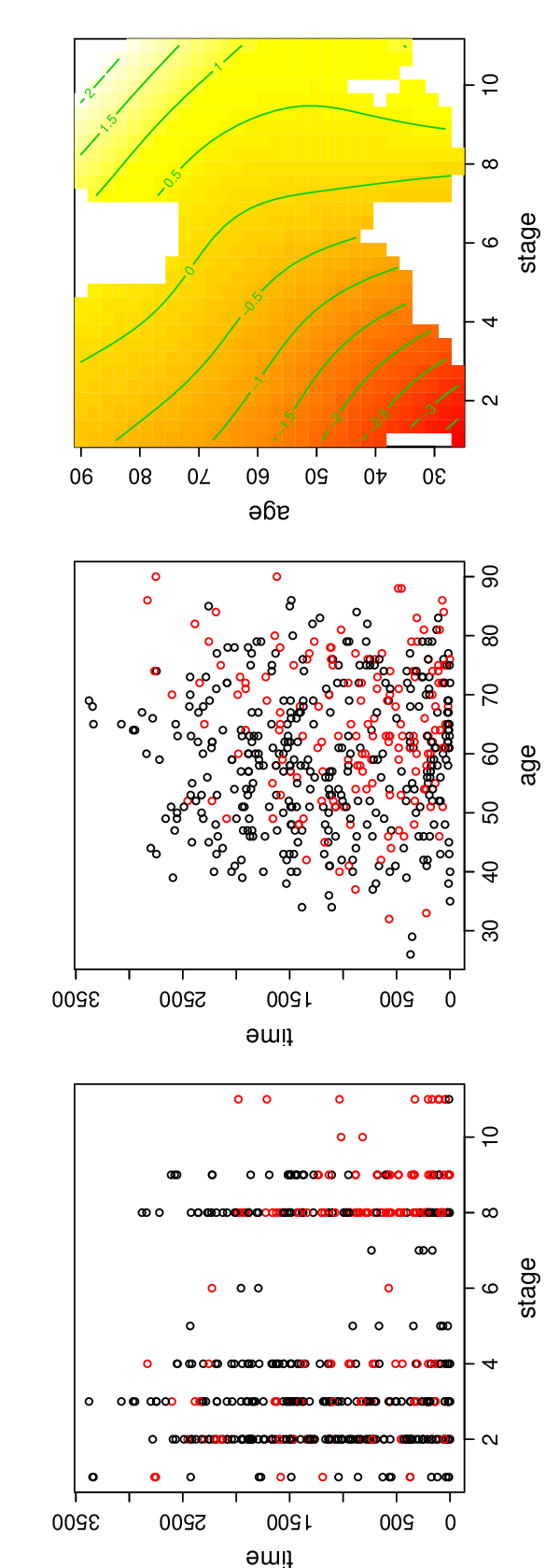}\vspace*{-.5cm}
\caption{Left: Survival times (red) and censoring times (black) against disease stage for patients with kidney renal clear cell carcinoma. Middle: times against patient age. Right: the combined smooth effect of age and stage on the linear predictor scale from a Cox Proportional hazards survival model estimated by maximum penalized partial likelihood. Higher values indicate higher hazard resulting in shorter survival times.    
 \label{kirc.fig}}
\end{figure}

The left 2 panels of figure \ref{kirc.fig} show survival times of patients with kidney renal clear cell carcinoma, plotted against disease stage at diagnosis and age, with survival time data in red and censoring time data in black (available from \verb+https://tcga-data.nci.nih.gov/tcga/+). Other recorded variables include race, previous history of malignancy and laterality (whether the left or right kidney is affected). A possible model for the survival times would be a Cox proportional hazards model with linear predictor dependent on parametric effects of the factor predictors and smooth effects of age and stage. Given the new methods this model can readily be estimated, as detailed in appendix \ref{cox.append}. A model with smooth main effects plus an interaction has a marginally lower AIC than the main effects only and the combined effect of age and stage is shown in the right panel of figure \ref{kirc.fig}. Broadly it appears that both age and stage increase the hazard, except at relatively high stage where age matters little below ages in the mid sixties. Disease in the right kidney leads to  significantly reduced hazard (p=.005) relative to disease in the left kidney: the reduction on the linear predictor scale being 0.45. This effect is likely to relate to the asymmetry in arrangement of other internal organs. There was no evidence of an effect of race or previous history of malignancy. 

\subsection{Overdispersed Horse Mackerel eggs \label{horse}}

\begin{figure}
\eps{-90}{.45}{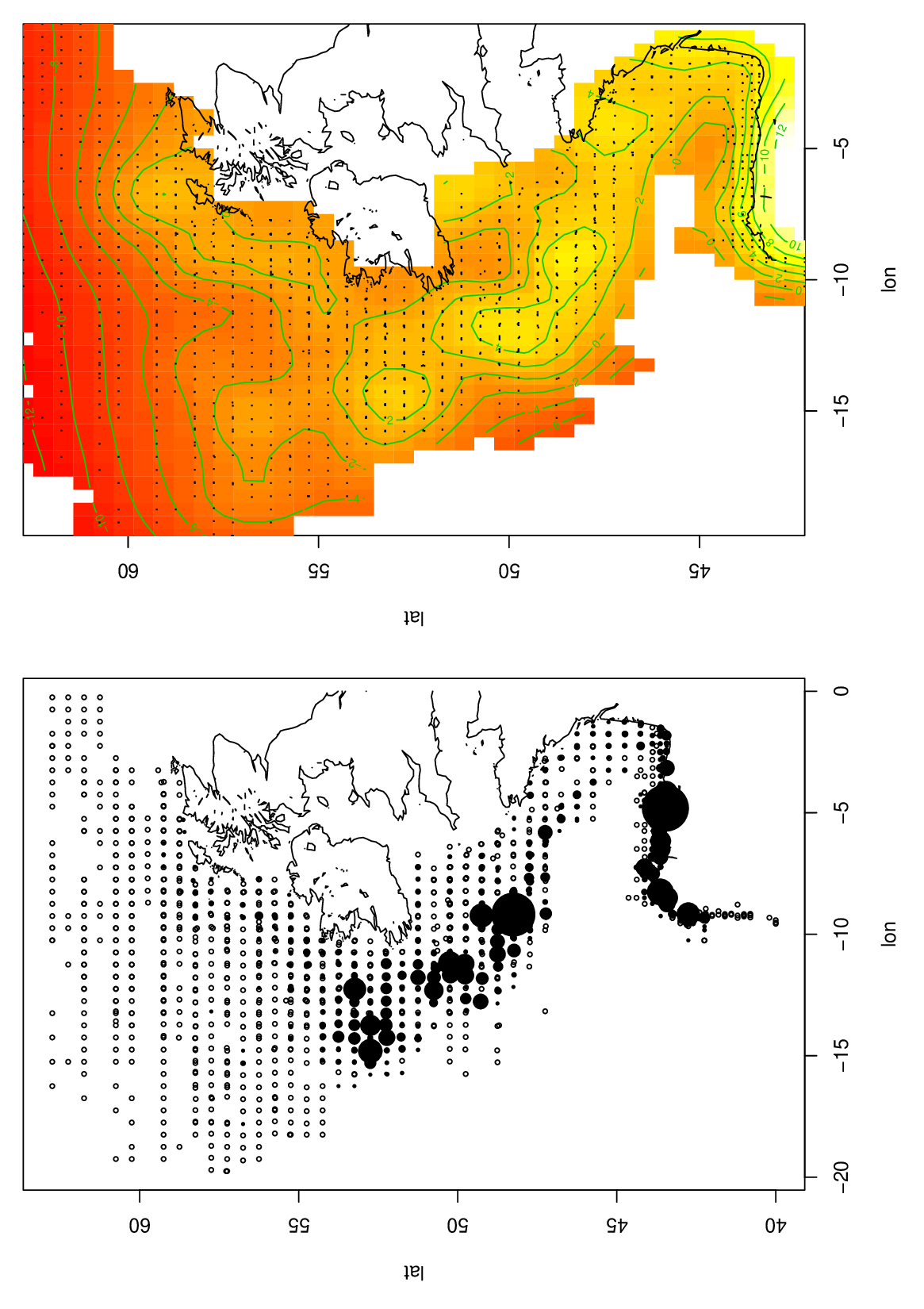}\vspace*{-.5cm}
\caption{Left: 2010 Horse Mackerel egg survey data. Open circles are survey stations with no eggs, while solid symbols have area proportional to number of eggs sampled. Right: Fitted spatial effect from the best fit negative binomial based model.    
 \label{egg-data.fig}}
\end{figure}

Figure \ref{egg-data.fig} shows data from a 2010 survey of Horse Mackerel eggs. The data are from the WGMEGS working group (\verb+http://www.ices.dk/marine-data/data-portals/Pages/Eggs-and-larvae.aspx+). Egg surveys are commonly undertaken to help in fish stock assessment and are attractive because unbiased sampling of eggs is much easier than unbiased sampling of adult fish. The eggs are collected by ship based sampling and typically show over-dispersion relative to Poisson and a high proportion of zeroes. The high proportion of zeroes is often used to justify the use of zero inflated models, although reasoning based on the marginal distribution of eggs is clearly incorrect, and the zeroes are often highly clustered in space, suggesting a process with a spatial varying mean, rather than zero inflation. 

The new methods make it straightforward to rapidly compare several possible models for the data, in particular Poisson, zero-inflated Poisson, Tweedie and negative binomial distributions. A common structure for the expected number of eggs, $\mu_i$, (or Poisson parameter in the zero inflated case) was :
$$
\log(\mu_i) = \log({\tt vol}_i) + b_{s(i)} + f_1({\tt lo}_i,{\tt la}_i) + f_2({\tt T.20}_i) + 
f_3({\tt T.surf}_i) + f_4({\tt sal.20}_i)
$$
where ${\tt vol}_i$ is the volume of water sampled, $b_{s(i)}$ is an independent Gaussian random effect for the ship that obtained sample $i$, ${\tt lo}_i$ and ${\tt la}_i$ are longitude and latitude  (actually converted onto a square grid for modelling), ${\tt T.20}_i$ and ${\tt T.surf}_i$ are water temperature at 20m depth and the surface, respectively and ${\tt sal.20}_i$ is salinity at 20m depth. Univariate smooth effects were modelled using rank 10 thin plate regression splines, while the spatial effect was modelled using a rank 50 Duchon spline, with a first order derivative penalty and $s = 1/2$ \citep{duchon77, miller2014}.

An initial Poisson fit of this model structure was very poor with clear over-dispersion. We therefore tried negative binomial, Tweedie and two varieties of zero inflated Poisson models. The details of the zero inflated Poisson models are given in Appendix \ref{zip.append}. The extended GAM version has the zero inflation rate depending on a logistic function of the linear predictor controlling the Poisson mean, with the restriction that zero inflation must be non-increasing with the Poisson mean. The more general GAMLSS formulation (section 3.2 and appendix \ref{zip.append}) has a linear predictor for the probability, $p_i$, of potential presence of eggs 
$$
\text{logit}(p_i) = f_1({\tt lo}_i,{\tt la}_i) + f_2({\tt T.20}_i)
$$
with the same model as above for the Poisson mean given potential presence. 

\begin{figure}
\eps{-90}{.45}{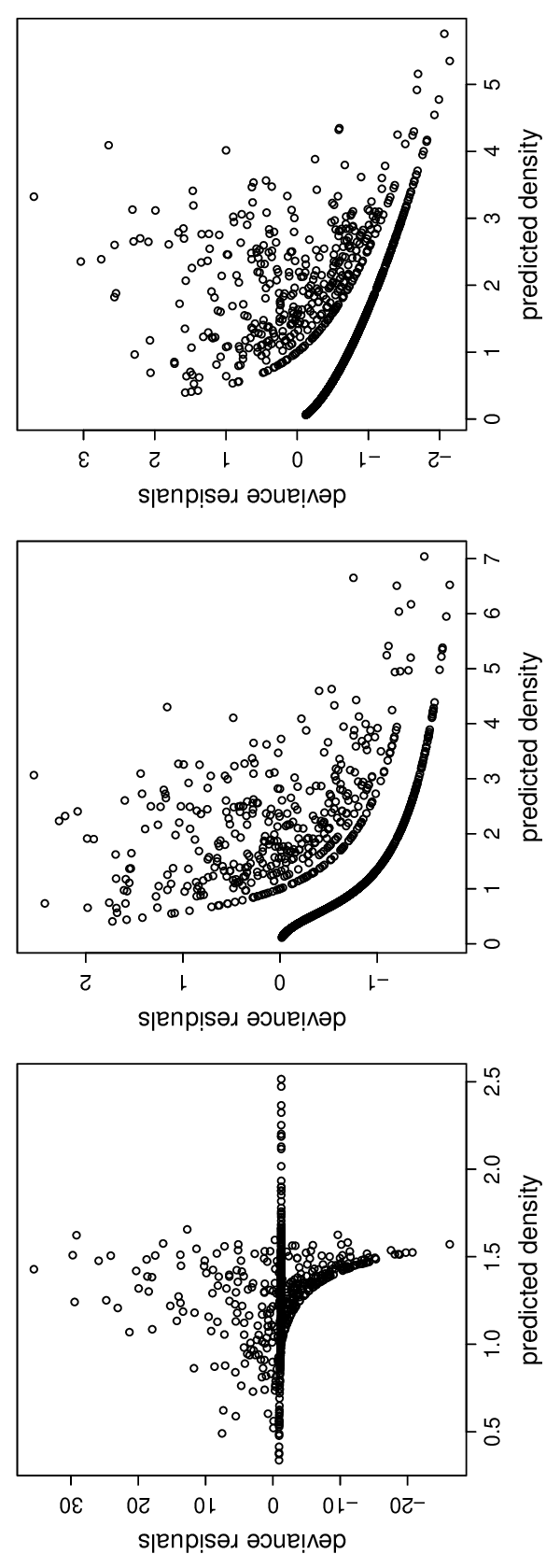}\vspace*{-.5cm}
\caption{Residual plots for three Horse Mackerel egg models. Deviance residuals have been scaled by the scale parameter so that they should have unit variance for a good model. The fourth root of fitted values is used to best show the structure of the residuals. Left is for a zero inflated Poisson model: the zero inflation has served to reduce the variability in the fitted values, allowed substantial over prediction of a number of zeroes, and has not dealt with over-dispersion. Middle and right are for the equivalent negative binomial and Tweedie models. The right two are broadly acceptable, although there is some over-prediction of very low counts evident at the right of both plots.    
 \label{egg-resid.fig}}
\end{figure}

Figure \ref{egg-resid.fig} shows simple plots of scaled deviance residuals against 4th root of fitted values. The plot for the extended GAM version of the zero inflated model is shown in the left panel and makes it clear that zero inflation is not the answer to the over-dispersion problem in the Poisson model; the GAMLSS zero inflated plot is no better. The negative binomial and Tweedie plots are substantially better, so that formal model selection makes sense in this case. The AIC of section 5 selects the negative binomial model with an AIC of 4482 against 4979 for the Tweedie (the Poisson based models have much higher AIC, of course). 

\begin{figure}
\eps{-90}{.45}{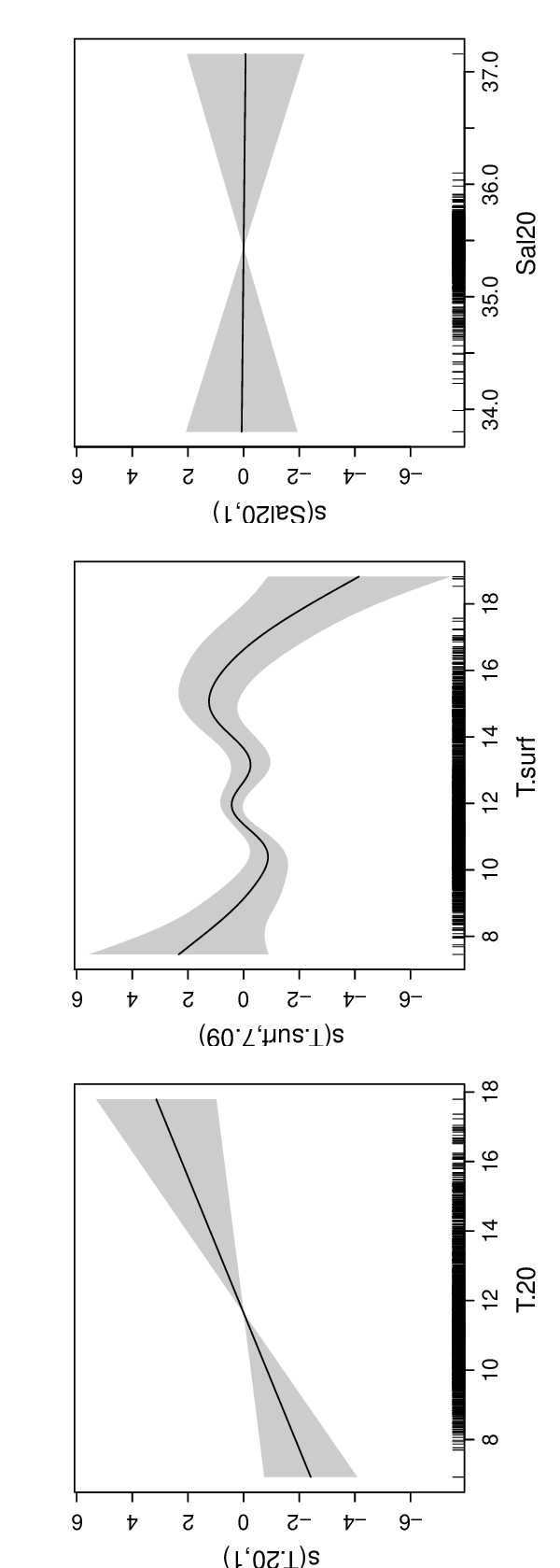}\vspace*{-.5cm}
\caption{Horse Mackerel model univariate effect estimates.  
 \label{egg-effects.fig}}
\end{figure}

Further model checking then suggested increasing the basis dimension of the spatial smooth and changing from a Duchon spline to a thin plate spline, so that the final model spatial effect estimate, plotted on the right hand side of figure \ref{egg-data.fig}, uses a thin plate regression spline with basis dimension 150 (although visually the broad structure of the effect estimates is very similar to the original fit). The remaining effect estimates are plotted in figure \ref{egg-effects.fig}. Clearly   
there is no evidence for an effect of salinity, while the association of density with water temperature is real, but given the complexity of the surface affect, possibly acting as a surrogate for the causal variables here. The final fitted model explains around 70\% of the deviance in egg count.

\section{Cox proportional hazards model \label{cox.append}}

The Cox proportional Hazards model \citep{cox1972} is an important example of a general smooth model requiring the methods of section 3.1, at least if the computational cost is to be kept linear in the sample size, rather than quadratic. With some care in the structuring of the computations, the computational cost can be kept to $O(Mnp^2)$. Let the $n$ data be of the form $(\tilde t_i, \X_i, \delta_i)$, i.e. an event time, model matrix row (there is no intercept in the Cox model) and an indicator of death (1) or censoring (0). Assume w.l.o.g. that the data are ordered so that the $t_i$ are non-increasing with $i$. The time data can conveniently be replaced by a vector $\bf t$ of $n_t$ unique decreasing event times, and an $n$ vector of indices, $r$, such that $t_{r_i} = \tilde t_i$.  

The log likelihood, as in \cite{h&t90}, is
$$
l(\bp) = \sum_{j=1}^{n_t} \left [
\sum_{\{i:r_i=j\}} \delta_i \X_i \bp - d_j \log \left \{ \sum_{\{i:r_i \le j\}} \exp(\X_i \bp)   \right \} \right ].
$$
Now let $\eta_i \equiv \X_i \bp$, $\gamma_i \equiv \exp(\eta_i)$ and $d_j = \sum_{\{i:r_i=j\}} \delta_i$ (i.e the count of deaths at this event time). Then
$$
l(\bp) = \sum_{j=1}^{n_t} \left [
\sum_{\{i:r_i=j\}} \delta_i \eta_i - d_j \log \left \{ \sum_{\{i:r_i \le j\}} \gamma_i   \right \} \right ].
$$
Further define $\gamma^+_j = \sum_{\{i:r_i \le j\}} \gamma_i$, so that we have the recursion 
$$
\gamma_j^+ = \gamma^+_{j-1} + \sum_{\{i:r_i=j\}} \gamma_i
$$
where $\gamma_0^+=0$. Then 
$$
l(\bp) = \sum_{i=1}^n \delta_i \eta_i - \sum_{j=1}^{n_t} d_j \log(\gamma^+_j).
$$
Turning to the gradient $g_k = \ilpdif{l}{\beta_k}$, we have
$$
{\bf g} = \sum_{i=1}^n \delta_i \X_i - \sum_{j=1}^{n_t} d_j {\bf b}^+_j / \gamma^+_j
$$
where ${\bf b}^+_j = {\bf b}^+_{j-1} + \sum_{\{i:r_i=j\}} {\bf b}_i$, ${\bf b}_i = \gamma_i\X_i$. and ${\bf b}^+_0 = {\bf 0}$.
Finally the Hessian $H_{km} = \ilpddif{l}{\beta_k}{\beta_m} $ is given by
$$
{\bf H} = \sum_{j=1}^{n_t} d_j {\bf b}^+_j {\bf b}^{+{\sf T}}_{j} / \gamma^{+2}_j - d_j {\bf A}^+_j / \gamma^+_j
$$
where ${\bf A}^+_j = {\bf A}^+_{j-1} + \sum_{\{i:r_i=j\}} {\bf A}_i $, ${\bf A}_i = \gamma_i \X_i \X_i\ts $ and ${\bf A}^+_0 = {\bf 0}$.

\subsubsection*{Derivatives with respect to smoothing parameters}

To obtain derivatives it will be necessary to obtain expressions for the derivatives of $l$ and $\bf H$ with respect to $\rho_k = \log(\lambda_k)$. Firstly we have
$$
\pdif{\eta_i}{\rho_k} = \X_i \pdif{\hat \bp}{\rho_k},~~  \pdif{\gamma_i}{\rho_k} = \gamma_i \pdif{\eta_i}{\rho_k}, ~~ 
\pdif{{\bf b}_i}{\rho_k} = \pdif{\gamma_i}{\rho_k} \X_i
{\rm ~~~ and ~~~ } 
\pdif{{\bf A}_i}{\rho_k} = \pdif{\gamma_i}{\rho_k} \X_i \X_i\ts.
$$ 
Similarly
$$
\pddif{\eta_i}{\rho_k}{\rho_m} = \X_i \pddif{\hat \bp}{\rho_k}{\rho_m},~~
\pddif{\gamma_i}{\rho_k}{\rho_m} = \gamma_i \pdif{\eta_i}{\rho_k}\pdif{\eta_i}{\rho_m}  + \gamma_i \pddif{\eta_i}{\rho_k}{\rho_m}, ~~ 
\pddif{{\bf b}_i}{\rho_k}{\rho_m} = \pddif{\gamma_i}{\rho_k}{\rho_m} \X_i.
$$
Derivatives sum in the same way as the terms they relate to. 
$$
\pdif{l}{\rho_k} = \sum_{i=1}^n \delta_i \pdif{\eta_i}{\rho_k} - \sum_{j=1}^{n_t} \frac{d_j}{\gamma^+_j} \pdif{\gamma^+_j}{\rho_k},
$$
and 
$$
\pddif{l}{\rho_k}{\rho_m} = \sum_{i=1}^n \delta_i \pddif{\eta_i}{\rho_k}{\rho_m} + 
\sum_{j=1}^{n_t} \left (
\frac{d_j}{\gamma^{+2}_j} \pdif{\gamma^+_j}{\rho_m} \pdif{\gamma^+_j}{\rho_k} - 
\frac{d_j}{\gamma_j^+}\pddif{\gamma^+_j}{\rho_k}{\rho_m}
 \right ),
$$
while 

$$
\pdif{\bf H}{\rho_k} = \sum_{j=1}^{n_t} \frac{d_j}{\gamma^{+2}_j} \left \{ 
 {\bf A}^+_j \pdif{\gamma^+_j}{\rho_k} + \pdif{{\bf b}^+}{\rho_k} {\bf b}^{+{\sf T}} +
{\bf b}^+ \pdif{{\bf b}^{+{\sf T}}}{\rho_k}  \right \} -
\frac{d_j}{\gamma^+_j} \pdif{{\bf A}^+_j}{\rho_k} -\frac{2 d_j}{\gamma^{+3}_j} {\bf b}^+ {\bf b}^{+ {\sf T}} \pdif{\gamma^+_j}{\rho_k}
$$
and
\begin{multline*}
\pddif{\bf H}{\rho_k}{\rho_m} = \sum_{j=1}^{n_t}  \frac{-2d_j}{\gamma^{+3}_j} \pdif{\gamma^+_j}{\rho_m} \left \{ 
 {\bf A}^+_j \pdif{\gamma^+_j}{\rho_k} + \pdif{{\bf b}^+}{\rho_k} {\bf b}^{+{\sf T}} +
{\bf b}^+ \pdif{{\bf b}^{+{\sf T}}}{\rho_k}  \right \} 
+ \frac{d_j}{\gamma^{+2}_j} \left \{ \pdif{{\bf A}^+_j}{\rho_m} \pdif{\gamma^+_j}{\rho_k} \right .\\ \left . 
+  {\bf A}^+_j \pddif{{\bf \gamma}^+_j}{\rho_k}{\rho_m} + \pddif{{\bf b}^+}{\rho_k}{\rho_m} {\bf b}^{+ {\sf T}} + \pdif{{\bf b}^+}{\rho_k} \pdif{{\bf b}^{+ {\sf T}}}{\rho_m} +
\pdif{{\bf b}^+}{\rho_m} \pdif{{\bf b}^{+ {\sf T}}}{\rho_k} + {\bf b}^+ \pddif{{\bf b}^{+ {\sf T}}}{\rho_k}{\rho_m} \right\} \\ +
\frac{d_j}{\gamma^{+2}_j} \pdif{\gamma^+_j}{\rho_m} \pdif{{\bf A}^+_j}{\rho_k} - \frac{d_j}{\gamma^+_j} \pddif{{\bf A}_j^+}{\rho_k}{\rho_m} + \frac{6 d_j}{\gamma^{+4}_j} \pdif{\gamma^+_j}{\rho_m} {\bf b}^+ {\bf b}^{+\sf T} \pdif{\gamma^+_j}{\rho_k} \\
- \frac{2 d_j}{\gamma^{+3}_j} \left \{ 
\pdif{{\bf b}^+}{\rho_m} {\bf b}^{+\sf T} \pdif{\gamma^+_j}{\rho_k} + {\bf b}^+ \pdif{{\bf b}^{+\sf T}}{\rho_m} \pdif{\gamma^+_j}{\rho_k} + {\bf b}^+ {\bf b}^{+\sf T} \pddif{\gamma^+_j}{\rho_k}{\rho_m}
\right \}.
\end{multline*}

In fact with suitable reparameterization it will only be necessary to obtain the second derivatives of the leading diagonal of $\bf H$, although the full first derivative of $\bf H$ matrices will be needed. All that is actually needed is ${\rm tr} \left ( \bm{{\cal H}}^{-1} \ilpddif{{\bf H}}{\rho_k}{\rho_m}  \right ) $. Consider the eigen-decomposition $\bm{{\cal H}}^{-1} = {\bf V}{\bm \Lambda}{\bf V}\ts$. We have
$$
\text{tr} \left (\bm{{\cal H}}^{-1}\pdif{{\bf H}}{\theta} \right ) = \text{tr} \left ({\bm \Lambda}\pdif{{\bf V}\ts{\bf HV}}{\theta} \right ), ~~~
\text{tr} \left (\bm{{\cal H}}^{-1} \pddif{\bf H}{\theta_k}{\theta_m} \right ) = \text{tr} \left (  {\bm \Lambda}\pddif{{\bf V}\ts{\bf HV}}{\theta_k}{\theta_m} \right ).
$$
Since $ {\bm \Lambda}$ is diagonal only the leading diagonal of the derivative of the reparameterized Hessian ${\bf V}\ts{\bf HV}$ is required, and this can be efficiently computed by simply using the reparameterized model matrix $\bf XV$. So the total cost of all derivatives is kept to $O(Mnp^2)$.

\subsubsection*{Prediction and the baseline hazard}

\citet[pages 283, 359, 381]{klein2003} gives the details. Here we simply restate the required expressions in forms suitable for efficient computation, using the notation and assumptions of the previous sections. 
\begin{enumerate}
\item The estimated baseline hazard is
$$
h_0(t) = \left \{ \begin{array}{ll}
h_j & t_j \le t < t_{j-1} \\
0 & t < t_{n_t}\\
h_1 & t\ge t_1
\end{array} \right .
$$
where the following back recursion defines $h_j$
$$
h_j = h_{j+1} + \frac{d_j}{\gamma^+_j}, ~~~~~ h_{n_t} = \frac{d_{n_t}}{\gamma^+_{n_t}}.
$$
\item The variance of the estimated hazard is given by the back recursion
$$
q_j = q_{j+1} + \frac{d_j}{\gamma^{+2}_j}, ~~~~~ q_{n_t} = \frac{d_{n_t}}{\gamma^{+2}_{n_t}}.
$$
\item The estimated survival function for time $t$, covariate vector $\bf x$, is
$$
\hat S(t,{\bf x}) = \exp\{-h_0(t)\}^{\exp({\bf x}\ts{\bm \beta})}
$$ 
and consequently $\log \hat S(t,{\bf x}) = -h_0(t)\exp({\bf x}\ts{\bm \beta})$. Let $\hat S_i$ denote the estimated version for the $i^{\rm th}$ subject, at their event time.
\item The estimated variance of $\hat S(t,{\bf x}) $ is 
$$
\hat S(t,{\bf x})^2 (q_i + {\bf v}_i\ts {\bf V}_\beta {\bf v}_i),~~~\text{if}~~~ t_i \le t < t_{i-1}
$$
where ${\bf v}_i = {\bf a}_i - {\bf x} h_i$, and the vector ${\bf a}_i$ is defined by the back recursion 
$$
{\bf a}_i = {\bf a}_{i+1} + {\bf b}_i^+ \frac{d_i}{\gamma^{+2}_i}, ~~~~
{\bf a}_{n_t} =  {\bf b}_{n_t}^+ \frac{d_{n_t}}{\gamma^{+2}_{n_t}}.
$$
For efficient prediction with standard errors, there seems to be no choice but to compute the $n_t$, ${\bm a}_i$ vectors at the end of fitting and store them. 
\item Martingale residuals are defined as 
$$
\hat M_j = \delta_j + \log \hat S_j,
$$
and deviance residuals as 
$$
\hat D_j = \text{sign}(\hat M_j) [-2\{\hat M_j + \delta_j \log(-\log \hat S_j)  \}  ]^{1/2}.
$$
The latter also being useful for computing a deviance.
\end{enumerate}

\section{Multivariate additive model example \label{mvn.append}}

Consider a model in which independent observations $\bf y$ are $m$ dimensional multivariate Gaussian with precision matrix ${\bm \Sigma}^{-1} = {\bf R}\ts {\bf R}$, $\bf R $ being a Cholesky factor of the form 
$$
{\bf R} = \bmat{cccc} 
e^{\theta_1} & \theta_2 & \cdot & \cdot \\
0 & e^{\theta_{m+1}} & \theta_{m+2} & \cdot \\
\cdot & \cdot & \cdot & \cdot \\
\cdot & \cdot & \cdot & \cdot 
\emat.
$$ 
Let $\mathbb{D}$ denote the set of $\theta_i$'s giving the diagonal elements of $\bf R$, with corresponding indicator function $\mathbb{I}_{\mathbb{D}}(i)$ taking value 1 if $\theta_i$ is in $\mathbb{D}$ and 0 otherwise. The mean vector $\bm \mu$ has elements $\mu_i = {\bf x}^i {\bm \beta}^i$, where ${\bf x}^i$ is a model matrix row for the $i^{\rm th}$ component with corresponding coefficient vector ${\bm \beta}^i$. In what follows it will help to define $\bar {\bf x}_i^l$ as an $m $ vector of zeroes except for element $l$ which is $x^l_i$ .

Consider the log likelihood for a single $\bf y$
$$
l = - \frac{1}{2} ({\bf y} - {\bm \mu})\ts {\bf R}\ts {\bf R} ({\bf y} - {\bm \mu}) + \sum_{\theta_i \in \mathbb{D}} \theta_i,
$$
where $\sum_{\theta_i \in \mathbb{D}} \theta_i =\log |{\bf R}|$.
For Newton estimation of the model coefficients we need gradients 
$$
l\indices*{*^i_\theta} = - ({\bf y} - {\bm \mu})\ts {\bf R}\ts {\bf R}\indices*{*^i_\theta} ({\bf y} - {\bm \mu}) + \mathbb{I}_\mathbb{ D}(i)
$$
and
$$
l\indices*{*^i_{\beta^l}} = \bar {\bf x}\indices*{*^l_i}{}\ts {\bf R}\ts {\bf R} ({\bf y} - {\bm \mu}).
$$
Then we need Hessian blocks
$$
l\indices*{*^i_{\beta^l}^j_{\beta^k}} = - \bar {\bf x}^l_i{}\ts{\bf R}\ts {\bf R}\bar {\bf x}^k_j,
$$
$$
l\indices*{*^i_{\beta^l}^j_{\theta}} =\bar {\bf x}^l_i{}\ts ({\bf R}\indices*{*^j_\theta}{}\ts {\bf R} + {\bf R}\ts {\bf R}\indices*{*^j_\theta})({\bf y} - {\bm \mu}),
$$
$$
l\indices*{*^i_\theta^j_\theta} = -({\bf y} - {\bm \mu})\ts {\bf R}\indices*{*^j_\theta}{}\ts {\bf R}\indices*{*^i_\theta} ({\bf y} - {\bm \mu}) 
-({\bf y} - {\bm \mu})\ts {\bf R}\ts {\bf R}\indices*{*^i_\theta^j_\theta} ({\bf y} - {\bm \mu}). 
$$
For optimization with respect to log smoothing parameters $\rho$ we need further derivatives, but note that the third derivatives with respect to ${\bm \beta}^l$ are zero. The non-zero 3rd derivatives are
$$
l\indices*{*^i_{\beta^l}^j_{\beta^m}_\theta^k} = - \bar {\bf x}^l_i{}\ts 
({\bf R}\indices*{*^k_\theta}{}\ts {\bf R} + {\bf R}\ts {\bf R}\indices*{*^k_\theta}) 
\bar {\bf x}^m_j,
$$
$$
l\indices*{*^i_{\beta^l}^j_\theta^k_\theta} =  \bar {\bf x}^l_i{}\ts 
({\bf R}\indices*{^j_\theta^k_\theta}{}\ts {\bf R} +
{\bf R}\indices*{^j_\theta}{}\ts{\bf R}\indices*{^k_\theta} + 
{\bf R}\indices*{^k_\theta}{}\ts{\bf R}\indices*{^j_\theta} +
 {\bf R}\ts {\bf R}\indices*{^j_\theta^k_\theta})({\bf y} - {\bm \mu}),
$$
$$
l\indices*{*^i_\theta^j_\theta^k_\theta} = - ({\bf y} - {\bm \mu})\ts 
({\bf R}\indices*{^j_\theta^k_\theta}{}\ts {\bf R}\indices*{^i_\theta} + 
{\bf R}\indices*{^j_\theta}{}\ts {\bf R}\indices*{^i_\theta^k_\theta} +
{\bf R}\indices*{^k_\theta}{}\ts {\bf R}\indices*{^i_\theta^j_\theta} +
{\bf R}\ts {\bf R}\indices*{^i_\theta^j_\theta^k_\theta} ) 
({\bf y} - {\bm \mu}).
$$
These are useful for computing the following\ldots
$$
l\indices*{*^i_{\beta^l}^j_{\beta^m}_\rho^k} = l\indices*{*^i_{\beta^l}^j_{\beta^m}_\theta^q} \pdif{\hat \theta_q}{\rho_k},
$$
$$
l\indices*{*^i_{\theta}^j_{\theta}_\rho^k} = 
l\indices*{*^i_{\theta}^j_{\theta}_\theta^q} \pdif{\hat \theta_q}{\rho_k} +
l\indices*{*^q_{\beta^l}^i_{\theta}_\theta^j} \pdif{\hat \beta_q^l}{\rho_k},
$$
$$
l\indices*{*^i_{\beta^l}^j_{\theta}_\rho^k} =
l\indices*{*^i_{\beta^l}^j_{\theta}_\theta^q} \pdif{\hat \theta_q}{\rho_k}+
l\indices*{*^i_{\beta^l}^q_{\beta^m}_\theta^j} \pdif{\hat \beta_q^m}{\rho_k}.
$$
This is sufficient for Quasi-Newton estimation of smoothing parameters. 

Sometimes models with multiple linear predictors should share some terms across predictors. In this case the general fitting and smoothing parameter methods should work with the vector of unique coefficients, $\bar \bp$, say, to which corresponds a model matrix $\bar {\bf X}$. The likelihood derivative computations on the other hand can operate as if each linear predictor had unique coefficients, with the derivatives then being summed over the copies of each unique parameter. Specifically, let $i_{kj}$ indicate which column of $\bar {\bf X}$ gives column $j$ of ${\bf X}^k$, and let ${\cal J}_i = \{k,j: i_{kj} = i\}$, i.e. the set of $k,j$ pairs identifying the replicates of column $i$ of $\bar X$ among the ${\bf X}^k$, and the replicates of $\bar \beta_i$ among the $\bp^k$. Define a `${\cal J}$ contraction over $x^k$' to be an operation of the form 
$$
\bar x_i = \sum_{k,j \in {\cal J}_i } x^k_j ~~~ \forall i.
$$
Then the derivative vectors with respect to $\bar \bp$ are obtained by a ${\cal J}$ contraction over the derivative vectors with respect to $\bp^k$. Similarly the Hessian with respect to $\bar \bp$ is obtained by consecutive  ${\cal J}$ contractions over the rows and columns of the Hessian with respect to the $\bp^k$. For the computation (4) in section 3.2 we would apply ${\cal J}$ contractions to the columns of the two matrices in round brackets on the right hand side of (4) ({\bf B} would already be of the correct dimension, of course). The notion of ${\cal J}$ contraction simplifies derivation and coding in the case when different predictors reuse terms, but note that computationally it is simpler and more efficient to implement ${\cal J}$ contraction based only on the index vector $i_{kj}$, rather than by explicitly forming the ${\cal J}_i$.

\section{Zero inflated Poisson models \label{zip.append}}

Zero inflated Poisson models are popular in ecological abundance studies when one process determines whether a species is (or could be) present, and the number observed, given presence (suitability), is a Poisson random variable. Several alternatives are possible, but the following `hurdle model' tends to minimise identifiability problems. 
$$
f(y) = \left \{ 
\begin{array}{ll}
1-p & y=0\\
{p \lambda^y}/\{(e^\lambda - 1 )y!\} & \text{otherwise.}
\end{array}
\right .
$$
So observations greater than zero follow a zero truncated Poisson. Now adopt the unconstrained parameterization, $\gamma=\log \lambda$ and $\eta = \log\{-\log(1-p)\}$ (i.e. using log and complementary log-log links). If $\gamma = \eta$ this recovers an un-inflated Poisson model. The log likelihood is now
$$
l = \left \{ \begin{array}{ll} 
- e^\eta & y=0 \\
\log (1-e^{-e^\eta}) + y \gamma - \log(e^{e^\gamma}-1) - \log y! & y>0. 
\end{array} \right .
$$
Some care is required to evaluate this without unnecessary overflow, since it is easy for the $1-e^{-e^\eta}$ and $e^{e^\gamma}-1$ to evaluate as zero in finite precision arithmetic. Hence the limiting results $\log (1-e^{-e^\eta}) \to \log (e^\eta - e^{2\eta}/2 + e^{3\eta})/6) \to \eta$ as $\eta \to - \infty$ and $\log (e^{e^\gamma}-1) \to \log(e^\gamma + e^{2\gamma}/2 + e^{3\gamma})/6) \to \gamma$ as $\gamma \to -\infty$ can be used. The first pair of limits is useful as the arguments of the $\log$s becomes too close to 1 and the second pair as the exponential of $\eta$ or $\gamma$ approaches underflow to zero. (The log gamma function of $y+1$ computes $\log y!$)

The derivatives for this model are straightforward as all the mixed derivatives are zero. For the $y>0$ part,
$$
l_\eta = \frac{e^{\eta}}{e^{e^\eta}-1},~~
l_\gamma = y - \alpha,~~ \text{where}~~ \alpha = \frac{e^\gamma}{1-e^{-e^\gamma}},~~
l_{\eta\eta} = (1-e^\eta) l_\eta - l_\eta^2, ~~
l_{\gamma\gamma} = \alpha^2 - (e^\gamma + 1)\alpha,
$$
$$
l_{\eta\eta\eta} = -e^\eta l_\eta + (1-e^\eta)^2 l_\eta - 3(1-e^\eta)l^2_\eta + 2 l^3_\eta,
~~
l_{\gamma\gamma\gamma} = -2 \alpha^3 + 3(e^\gamma+1)\alpha^2 - e^\gamma \alpha - (e^\gamma + 1)^2 \alpha,
$$
$$
l_{\eta\eta\eta\eta} = (3e^\eta -4)e^\eta l_\eta + 4e^\eta l_\eta^2 + (1-e^\eta)^3 l_\eta -
7 (1-e^\eta)^2 l_\eta^2 + 12(1-e^\eta)l_\eta^3 - 6l_\eta^4
$$
$$
\text{and}~~~l_{\gamma\gamma\gamma\gamma} = 6 \alpha^4 - 12 (e^\gamma+1)\alpha^3 + 4 e^\gamma \alpha^2 +
7(e^\gamma+1)^2 \alpha^2 - (4+3e^\gamma)e^\gamma \alpha - (e^\gamma+1)^3\alpha.
$$
As with $l$ itself, some care is required to ensure that the derivatives evaluate accurately and without overflow over as wide a range of $\gamma$ and $\eta$ as possible. To this end note that as $\eta \to \infty$ all derivatives with respect to $\eta$ tend to zero, while as $\gamma \to \infty $, $l_{\gamma\gamma\gamma}\to l_{\gamma\gamma\gamma\gamma} \to -e^\gamma$. As $\eta \to - \infty$ accurate evaluation of the derivatives with respect to $\eta$ rests on $l_\eta \to 1-e^\eta/2 - e^{2 \eta}/12$. Substituting this into the derivative expressions, the terms of $O(1)$ can be cancelled analytically: the remaining terms then evaluate the derivatives without cancellation error problems. For $\gamma \to - \infty$ the equivalent approach uses $\alpha \to 1 + e^\gamma/2 + e^\gamma/12$.

An extended GAM version of this model is also possible, in which $\eta$ is a function of $\gamma$ and extra parameters, $\theta$, for example via $\eta = \theta_1 + e^{\theta_2} \gamma$. The idea is that the degree of zero inflation is a non-increasing function of $\gamma$, with $\theta_1=\theta_2=0 $ recovering the Poisson model. The likelihood expressions are obtained by transformation. Let $\bar l_\gamma$ denote the total derivative with respect to $\gamma$ in such a model.
$$
\bar l_\gamma = l_\gamma + l_\eta \eta_\gamma, ~~~
\bar l_{\gamma\gamma}= l_{\gamma\gamma} + l_{\eta\eta} \eta_\gamma \eta_\gamma + l_\eta \eta_{\gamma\gamma},~~~
\bar l_{\theta_i} = l_\eta \eta_{\theta_i},~~~ 
\bar l_{\gamma \theta_i} = l_{\eta\eta} \eta_{\theta_i} \eta_\gamma + l_\eta \eta_{\theta\gamma}
$$
$$
\bar l_{\gamma \gamma \theta_i} = 
l_{\eta\eta\eta} \eta_{\theta_i} \eta_\gamma^2 + l_{\eta\eta}  (2\eta_{\gamma\theta_i}\eta_\gamma+\eta_{\gamma\gamma}\eta_{\theta_i}) + l_\eta \eta_{\gamma\gamma \theta_i},~~~
\bar l_{\gamma\gamma\gamma} = l_{\gamma\gamma\gamma} +   l_{\eta\eta\eta} \eta_\gamma^3 + 3 l_{\eta\eta} \eta_\gamma \eta_{\gamma\gamma} + l_\eta \eta_{\gamma\gamma\gamma}
$$
$$
\bar l_{\theta_i\theta_j} = l_{\eta\eta} \eta_{\theta_i} \eta_{\theta_j}+ l_\eta \eta_{\theta_i\theta_j},~~~~
\bar l_{\gamma \theta_i\theta_j} = 
l_{\eta\eta\eta}\eta_{\theta_i} \eta_{\theta_j} \eta_\gamma + l_{\eta\eta} (\eta_{\theta_i\theta_j} \eta_\gamma +  \eta_{\theta_i\gamma} \eta_{\theta_j}+ \eta_{\theta_j\gamma} \eta_{\theta_i}) + 
l_\eta \eta_{\theta_i\theta_j\gamma}
$$
\begin{multline*}
\bar l_{\gamma\gamma\theta_i\theta_j} =  l_{\eta\eta\eta\eta} \eta_{\theta_i} \eta_{\theta_j} \eta_\gamma^2 +
l_{\eta\eta\eta}(\eta_{\theta_i\theta_j} \eta_\gamma^2 + 
2 \eta_{\theta_i} \eta_\gamma \eta_{\theta_j\gamma} + 2 \eta_{\theta_j} \eta_\gamma \eta_{\theta_i\gamma}
+ \eta_{\theta_i} \eta_{\theta_j} \eta_{\gamma\gamma})\\ +
l_{\eta\eta}(2 \eta_{\gamma \theta_i}\eta_{\gamma \theta_j} + 2 \eta_\gamma \eta_{\gamma \theta_i\theta_j} + 
 \eta_{\theta_i} \eta_{\gamma\gamma \theta_j} + \eta_{\theta_j} \eta_{\gamma\gamma \theta_i}+ \eta_{\theta_i\theta_j}\eta_{\gamma\gamma}) + l_\eta \eta_{\gamma\gamma\theta_i\theta_j}   
\end{multline*}
$$
\bar l_{\gamma\gamma\gamma\theta_i} =  
l_{\eta\eta\eta\eta} \eta_{\theta_i} \eta_\gamma^3 +
3 l_{\eta\eta\eta} (\eta_\gamma^2 \eta_{\theta_i \gamma} + \eta_{\theta_i} \eta_\gamma \eta_{\gamma\gamma}) +
l_{\eta\eta} (3 \eta_{\theta_i\gamma}\eta_{\gamma\gamma} + 3 \eta_\gamma \eta_{\gamma\gamma\theta_i} + 
\eta_{\theta_i} \eta_{\gamma\gamma\gamma}) + l_\eta \eta_{\gamma\gamma\gamma\theta_i}
$$
$$
\bar l_{\gamma\gamma\gamma\gamma} = l_{\gamma\gamma\gamma\gamma} + 
l_{\eta\eta\eta\eta}\eta^4_\gamma + 6 l_{\eta\eta\eta} \eta_\gamma^2 \eta _{\gamma\gamma}
+l_{\eta\eta} (3 \eta^2_{\gamma\gamma} + 4 \eta_\gamma \eta_{\gamma\gamma\gamma})
+l_\eta \eta_{\gamma\gamma\gamma\gamma}
$$
If $\eta = \theta_1 + e^{\theta_2} \gamma$ then $\eta_{\theta_1}=1$, $\eta_{\gamma \theta_2\theta_2} = \eta_{\gamma\theta_2}  = \eta_\gamma = e^{\theta_2}$, $\eta_{\gamma} = \eta_{\theta_2\theta_2} = e^{\gamma} e^{\theta_2}$: other required derivatives are 0.

Computationally it makes sense to define the deviance as $-2l$ and the saturated log likelihood as $\tilde l = 0$ during model estimation, and only to compute the true $\tilde l$ and deviance at the end of fitting, since there is no closed form for $\tilde l$ in this case (the same is true for beta regression). 

\section{Tweedie model details \label{tweedie.append}} 

This example illustrates an extended GAM case where the likelihood is not available in closed form. The Tweedie distribution \citep{tweedie1984} has a single $\theta$ parameter, $p$, and a scale parameter, $\phi$. We have $V(\mu) = \mu^p$, and a density. 
$$
f(y) = a(y,\phi,p) \exp \{ \mu^{1-p}(y /(1-p) - \mu/(2-p))/\phi \}.
$$
We only consider $p$ in (1,2). The difficulty is that 
$$
a(y, \phi, p) = \frac{1}{y} \sum_{j=1}^\infty W_j
$$ 
where, defining $\alpha = (2-p)/(1-p)$, 
$$
\log W_j =  j \left \{ 
\alpha \log(p-1) -
\log(\phi)/(p-1) - \log(2-p)
\right \} - \log \Gamma(j+1) - \log \Gamma(-j\alpha)- j \alpha \log y.
$$ 
The sum is interesting in that the early terms are near zero, as are the later terms, so that it has to be summed-from-the-middle, which can be a bit involved: \citet{dunn.smyth2005}, give the details, but basically they show that the series maximum is around $j_{\rm max} = y^{2-p}/\{\phi(2-p)\}$. 

Let $\omega = \sum_{j=1}^\infty W_j$. We need derivatives of $\log \omega$ with respect to $\rho = \log \phi$ and $p$, or possibly $\theta$ where 
$$
p = \{a + b\exp(\theta)\}/\{1+ \exp(\theta)\}
$$
and $1 < a < b < 2$. For optimization this transformation is necessary since the density becomes discontinuous at $p=1$ and the series length becomes infinite at $p=2$.
It is very easy to produce derivative schemes that overflow, underflow or have cancellation error problems, but the following avoids the worst of these issues. We use the identities
$$
\pdif{\log \omega}{x} = \text{sign} \left ( \pdif{\omega}{x}\right ) \exp \left (
\log\left | \pdif{\omega}{x}\right |- \log \omega
 \right )
$$ 
and
$$
 \pddif{\log \omega}{x}{z} = \text{sign}\left (\pddif{\omega}{x}{z} \right ) \exp\left (
\log \left | \pddif{\omega}{x}{z} \right | - \log \omega \right ) -  
 \text{sign}\left (\pdif{\omega}{x} \right )  \text{sign}\left (\pdif{\omega}{z} \right ) 
 \exp \left (
\log \left | \pdif{\omega}{x} \right | + \log \left | \pdif{\omega}{x} \right | - 2 \log \omega
 \right ).
$$
Now 
$$
\pdif{\omega}{x} = \sum_i W_i \pdif{\log W_i}{x}
$$
while 
$$
\pddif{\omega}{x}{z} = \sum_i W_i \pdif{\log W_i}{x} \pdif{\log W_i}{z} + W_i  \pddif{\log W_i}{z}{x},
$$
but note that to avoid over or underflow we can use $W_i^\prime = W_i-\text{max}(W_i)$ in place of $W_i$ in these computations, without changing $\ilpdif{\log \omega}{x}$ or $\ilpddif{\log \omega}{x}{z}$. Note also that 
$$
\log \omega = \log(\sum_i W_i^\prime) + \text{max}(W_i).
$$
All that remains is to find the actual derivatives of the $\log W_j$ terms.
$$
\pdif{\log W_j}{\rho} = \frac{-j}{p-1} ~~~\text{and}~~~\pdif{^2\log W_j}{\rho^2} = 0.
$$
It is simplest to find the derivatives with respect to $p$ and then transform to those with respect to $\theta$: 
$$
\pddif{\log W_j}{\rho}{\theta} = \pdif{p}{\theta} \frac{j}{(p-1)^2}.
$$
The remaining derivatives are a little more complicated
$$
\pdif{\log W_j}{p} = j \left \{ 
\frac{\log(p-1) + \log \phi}{(1-p)^2} + \frac{\alpha }{p-1} + \frac{1}{2-p} 
\right \} + 
\frac{j \psi_0 (-j\alpha)}{(1-p)^2} - \frac{j \log y}{(1-p)^2} 
$$
and
$$
\pdif{^2 \log W_i}{p^2} = j\left \{ \frac{2\log (p-1) + 2\log \phi}{(1-p)^3} - 
 \frac{(3 \alpha - 2)}{(1-p)^2} + \frac{1}{(2-p)^2} \right \} + 
  \frac{2j\psi_0(-j \alpha)}{(1-p)^3} -  \frac{j^2\psi_1(-j\alpha)}{(1-p)^4} - \frac{2j\log y}{(1-p)^3}
$$
where $\psi_0$ and $\psi_1$ are digamma and trigamma functions respectively. These then transform according to 
$$
\pdif{\log W_j}{\theta} = \pdif{p}{\theta} \pdif{\log W_j}{p} ~~~ \text{and}~~~
\pdif{^2\log W_j}{\theta^2} = \pdif{^2 p}{\theta^2} \pdif{\log W_j}{p} + 
\left ( \pdif{p}{\theta} \right )^2\pdif{^2\log W_j}{p^2}.
$$
The required transform derivatives are
$$
\pdif{p}{\theta} = \frac{e^\theta(b-a)}{(e^\theta+1)^2}  ~~~ \text{and}~~~
\pdif{^2 p}{\theta^2} = \frac{e^{2\theta}(a-b)+e^\theta(b-a)}{(e^\theta+1)^3}.
$$

\section{Ordered categorical model details \label{ocat.append}}

This example provides a useful illustration of an extended GAM model where the number of $\theta$ parameters varies from model to model. The basic model is that $y$ takes a value from $r=1, \ldots, R$, these being ordered category labels. Given 
$-\infty = \alpha_0 < \alpha_1,\ldots < \alpha_R = \infty$ we have that $y = r$ if a latent variable $u=\mu + \epsilon$ is such that $\alpha_{r-1} < u \le \alpha_r$, which occurs with probability
$$
{\rm Pr}(Y = r) = F(\alpha_r - \mu) - F(\alpha_{r-1} - \mu)
$$
where $F$ is the c.d.f. of $\epsilon$. See \cite{kneib2006ocat} and \cite[section 6.3.1]{fahrmeir2013} for a particularly clear exposition. 
$$
F(x) = \exp(x)/(1+\exp(x))
$$
is usual. For identifiability reasons $\alpha_1 = -1$, or any other constant, so there are $R-2$ free parameters to choose to control the thresholds. Generically we let 
$$
\alpha_r = \alpha_1 + \sum_{i=1}^{r-1} \exp(\theta_i), ~~~ 1 < r < R.
$$
Note that the cut points in this model {\em can} be treated as linear parameters in a GLM PIRLS iteration, but this is not a good approach if smoothing parameter estimates are required. The problem is that the cut points are then not forced to be correctly ordered, which means that the PIRLS iteration has to check for this as part of step length control. Worse still, if a category is missing from the data then the derivative of the likelihood with respect to the cut points can be non zero at the best fit, causing implicit differentiation to fail. 

Direct differentiation of the $ \log {\rm Pr}(Y = r) = F(\alpha_r - \mu) - F(\alpha_{r-1} - \mu) $ in terms of $\theta_i$ is ugly, and it is better to work with derivatives with respect to the $\alpha_r$ and use the chain rule. The saturated log likelihood can then be expressed as
$$
\tilde l = \log [ F\{(\alpha_r-\alpha_{r-1})/2\} - F\{(\alpha_{r-1}-\alpha_{r})/2\}]
$$
while the deviance is 
$$
D = 2 [l_s - \log \{ F(\alpha_r-\mu) - F(\alpha_{r-1}-\mu)\}].
$$
Define $f_1 = F(\alpha_r-\mu)$ and $f_0 = F(\alpha_{r-1}-\mu)$, $f=f_1 - f_0$. Similarly
$$
a_1 = f_1^2 - f_1, ~~~ a_0 = f_0^2 - f_0,~~~ a = a_1 - a_0,
$$ 
$$
b_j = f_j - 3 f_j^2 + 2 f_j^3, ~~~ b = b_1 - b_0 
$$
$$
c_j = -f_j + 7 f_j^2 - 12 f_j^3 + 6 f_j^4, ~~~ c = c_1 - c_0
$$
and 
$$
d_j = f_j - 15 f_j^2 + 50 f_j^3 - 60 f_j^4 + 24 f_j^5, ~~~ d = d_1 - d_0. 
$$

The sharp eyed reader will have noticed that all these expressions are prone to severe cancellation error problems as $f_j \to 1$. Stable expressions are required. For $f$, note that if $b>a$
$$
\frac{e^b}{1+e^b} - \frac{e^a}{1+e^a} = \frac{e^{-a} - e^{-b}}{(e^{-b}+1)(e^{-a}+1)} = \frac{1 - e^{a-b}}{(e^{-b}+1)(1 + e^a)}
$$
The first is used if $0>b>a$, the second if $b>a>0$ and the last if $b>0>a$. Now writing $x$ as a generic argument of $F$, we have
\begin{multline*}
a_j = \frac{-e^x}{(1+e^x)^2} = \frac{-e^{-x}}{(e^{-x}+1)^2},~~~ b_j = \frac{e^x-e^{2x}}{(1+e^x)^3} = \frac{e^{-2x}-e^{-x}}{(e^{-x}+1)^3},~~~
c_j = \frac{-e^{3x} + 4e^{2x} - e^x}{(1+e^x)^4} \\ = \frac{-e^{-x} + 4e^{-2x} - e^{-3x}}{(e^{-x}+1)^4},~~~
d_j = \frac{-e^{4x} + 11e^{3x}-11e^{2x}+e^x}{(1+e^x)^5} = 
\frac{-e^{-x} + 11e^{-2x}-11e^{-3x}+e^{-4x}}{(e^{-x}+1)^5}
\end{multline*}
These are useful by virtue of involving terms of order 0, rather than 1.

Then
$$
D_\mu = -2a/f, ~~~ D_{\mu\mu} = 2a^2/f^2 - 2 b / f,~~~ D_{\mu\mu\mu} = - 2c/f - 4 a^3/f^3 + 6ab/f^2.
$$
Note that $D_{\mu\mu}\ge 0$.
$$
D_{\mu\mu\mu\mu} = 6 b^2/f^2 + 8ac/f^2 + 12 a^4/f^4 - 24 a^2 b / f^3 -2d/f,
$$
$$
D_{\mu\alpha_{r-1}} = 2a_0 a/f^2 - 2 b_0/f,~~~ D_{\mu\alpha_{r}} = -2 a_1 a /f^2 + 2 b_1/f,
$$
$$
D_{\mu\mu\alpha_{r-1}} = -2 c_0/f + 4 b_0 a/f^2 - 4 a_0 a^2/f^3 + 2 a_0 b / f^2,~~~ 
D_{\mu\mu\alpha_{r}} = 2  c_1 / f - 4 b_1 a / f^2 + 4 a_1 a^2 / f^3 - 2 a_1 b / f^2,
$$
$$
D_{\mu\mu\mu\alpha_{r-1}} = -2d_0/f + 2 a_0c/f^2 +6 c_0a/f^2 - 12 b_0 a^2/f^3 + 12 a_0 a^3/f^4 + 6 b_0 b /f^2 - 12 a_0 ab/f^3,
$$
$$
D_{\mu\mu\mu\alpha_{r}} = 2d_1/f - 2 a_1c/f^2 -6 c_1a/f^2 + 12 b_1 a^2/f^3 - 12 a_1 a^3/f^4 - 6 b_1 b /f^2 + 12 a_1 ab/f^3.
$$
Furthermore,
$$
D_{\mu\alpha_{r-1}\alpha_{r-1}} = 2 c_0/f - 2 b_0a/f^2 + 4 a_0b_0/f^2 - 4 a_0^2a/f^3,~~~
D_{\mu\alpha_{r}\alpha_{r}} = -2 c_1/f + 2 b_1a/f^2 + 4 a_1b_1/f^2 - 4 a_1^2a/f^3,
$$
$$
D_{\mu\alpha_{r-1}\alpha_{r}} = - 2 a_0 b_1/f^2 - 2 b_0 a_1/f^2 + 4 a_0a_1a/f^3,
$$
while 
$$
D_{\mu\mu\alpha_{r-1}\alpha_{r-1}} = 2 d_0/f - 4 c_0 a/f^2 + 4 b_0^2/f^2 + 4 a_0 c_0/f^2 + 4 b_0 a^2/f^3 
- 16 a_0 b_0 a / f^3 + 12 a_0^2 a^2/f^4 - 2 b_0 b/f^2 - 4 a_0^2 b / f^3,
$$
$$
D_{\mu\mu\alpha_{r}\alpha_{r}} = -2 d_1/f + 4 c_1 a/f^2 + 4 b_1^2/f^2 + 4 a_1 c_1/f^2 - 4 b_1 a^2/f^3 
- 16 a_1 b_1 a / f^3 + 12 a_1^2 a^2/f^4 + 2 b_1 b/f^2 - 4 a_1^2 b / f^3,
$$
$$
D_{\mu\mu\alpha_{r-1}\alpha_{r}} = 0. %% 8a_0b_1a/f^3 + 8 b_0a_1a/f^3 - 12a_0a_1a/f^4 - 4 b_0b_1/f^2 + 4 a_0a_1b/f^3 - 2 c_0 a_1/f^2 - 2 c_1 a_0/f^2 = 0??
$$
%%Problem: the last term appears to be zero in fact, but I can't see why. 

Finally there are some derivatives not involving $\mu$ and hence involving the terms in $\tilde l$. First define
$$
\bar \alpha = (\alpha_r - \alpha_{r-1})/2, ~~~ \gamma_1 = F(\bar \alpha), ~~~ \gamma_0 = F(-\bar \alpha),
$$
$$
A = \gamma_1 - \gamma_0, ~~~ B = \gamma_1^2 - \gamma_1 + \gamma_0^2 - \gamma_0, ~~~
C = 2 \gamma_1^3 - 3 \gamma_1^2 + \gamma_1 - 2 \gamma_0^3 + 3 \gamma_0^2 - \gamma_0.
$$
Then
$$
D_{\alpha_{r-1}} = B/A - 2 a_0/f, ~~~~ D_{\alpha_r} = -B/A + 2 a_1/f
$$
and
$$
D_{\alpha_{r-1}\alpha_{r-1}} = 2 b_0/f + 2 a_0^2/f^2 + C/(2A) - B^2/(2A^2),~~~
D_{\alpha_{r}\alpha_{r}} = -2 b_1/f + 2 a_1^2/f^2 + C/(2A) - B^2/(2A^2),~~~
$$
$$
D_{\alpha_{r}\alpha_{r-1}} = -2 a_0a_1/f^2 - C/(2A) + B^2/(2A^2).
$$
The derivatives of $\tilde l$ can be read from these expressions.

Having expressed things this way, it is necessary to transform to derivatives with respect to $\bm \theta$. 
$$
\pdif{D}{\theta_k} = \left \{ \begin{array}{ll}
0 & r \le k \\
\exp(\theta_k) \ilpdif{D}{\alpha_r} & r = k+1\\
\exp(\theta_k) (\ilpdif{D}{\alpha_r} +\ilpdif{D}{\alpha_{r-1}}) & r > k+1 \\
\exp(\theta_k) \ilpdif{D}{\alpha_{r-1}} & r=R.
\end{array} \right .
$$

\section{Example comparison with INLA and JAGS \label{inla.append}}

\begin{figure}
\eps{-90}{.55}{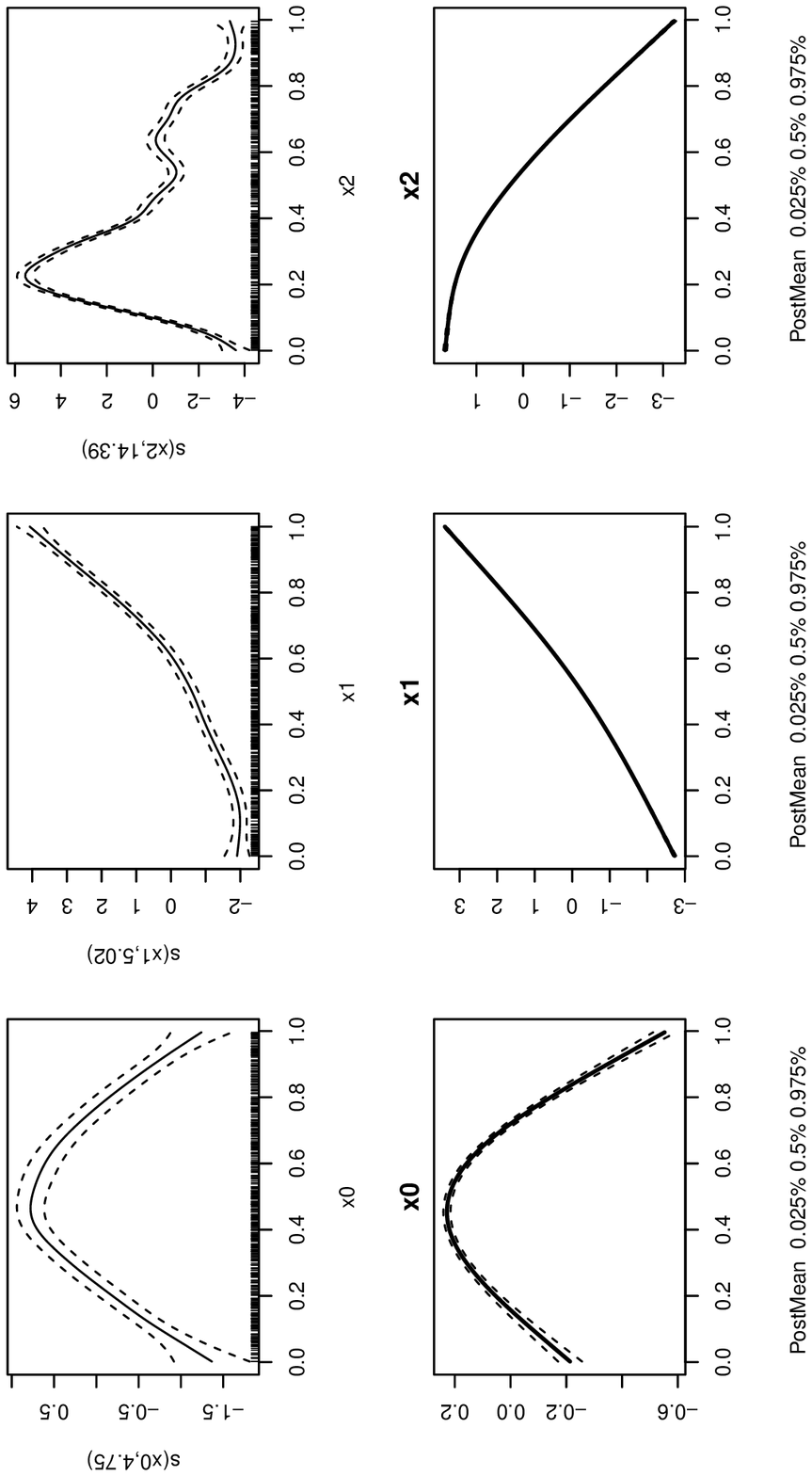}\vspace*{-.5cm}
\caption{{\tt mgcv} (top row) and INLA (lower row) fits (with 95\% credible intervals) to the simple 3 term additive model simulated in Appendix \ref{inla.append}.  Each row is supposed to be reconstructing the same true function, which in reality looks like the estimate in the upper row.     On this occasion INLA encounters numerical stability problems and the estimates are poor.
 \label{inla-eg.fig}}
\end{figure}

As mentioned in the main text, for models that require high rank random fields INLA offers a clearly superior approach to the methods proposed here, but at the cost of requiring sparse matrix methods, which preclude stabilizing reparameterization or pivoting for stability. On occasion this has noticeable effects on inference. For example the following code is adapted from the first example in the {\tt gam} helpfile in R package {\tt mgcv}. 
\begin{verbatim}
library(mgcv);library(INLA)
n <- 500; set.seed(0) ## simulate some data... 
dat <- gamSim(1,n=n,dist="normal",scale=1)
k=20;m=2
b <- gam(y~s(x0,k=k,m=m)+s(x1,k=k,m=m)+s(x2,k=k,m=m),
                 data=dat,method="REML")
md <- "rw2"
b2 <- inla(y~f(x0,model=md)+ f(x1,model=md)+ 
             f(x2,model=md),data=dat,verbose=TRUE)})
\end{verbatim}
On the same dual core laptop computer the {\tt gam} fit took 0.2 seconds and the INLA fit 40.9 seconds. Figure \ref{inla-eg.fig} compares the function estimates: INLA has encountered numerical stability problems and the reconstructions, which should look like those on the top row of the figure, are poor. Replicate simulations often give INLA results close to the truth, indistinguishable from the {\tt mgcv} results and computed in less than 1 second, but the shown example is not unusual. For this example we can fix the problem by binning the covariates, in which case the estimates and intervals are almost indistinguishable from the {\tt gam} estimates. However the necessity of doing this does emphasise that the use of sparse matrix methods precludes the use of pivoting to alleviate the effects of poor model conditioning. 

The same example can be coded in JAGS, for example using the function {\tt jagam} from {\tt mgcv} to auto-generate the JAGS model specification and starting values. To obtain samples giving comparable results to the top row of figure \ref{inla-eg.fig} took about 16 seconds, emphasising that simulation is relatively expensive for these models.  

\section{Software implementation \label{software.append}}

We have implemented the proposed framework in R \citep{rcore}, by extending package {\tt mgcv} (from version 1.8-0), so that the {\tt gam} function can estimate all the models mentioned in this paper, in a manner that is intuitively straightforward for anyone familiar with GAMs for exponential family distributions. Implementation was greatly facilitated by use of \citet{bravington.debug}. For the beta, Tweedie, negative binomial, scaled t, ordered categorical, simple zero inflated Poisson and Cox proportional hazards models, the user simply supplies one of the families {\tt betar}, {\tt tw}, {\tt nb}, {\tt scat}, {\tt ocat}, {\tt ziP} or {\tt cox.ph} to {\tt gam} as the {\tt family} argument, in place of the usual exponential family {\tt family}. For example the call to fit a Cox proportional hazards model is something like.
\begin{verbatim}
gam(time ~ s(x) + s(z),family=cox.ph,weights=censor)
\end{verbatim} 
where {\tt censor} would contain a 0 for a censored observation, and a 1 otherwise. Model {\tt summary} and {\tt plot} functions work exactly as for any GAM, while {\tt predict} allows for prediction on the survival scale.

Linear functionals of smooths are incorporated as model components using a simple summation convention. Suppose that \verb+X+ and \verb+L+ are matrices. Then a model term \verb+S(X,by=L)+ indicates a contribution to the $i^\text{th}$ row of the linear predictor of the form $ \sum_j f(X_{ij})L_{ij}$. This is the way that the section 8 model is estimated.

For models with multiple linear predictors {\tt gam} accepts a list of model formulae. For example a GAMLSS style zero inflated Poisson model would be estimated with something like
\begin{verbatim}
gam(list(y ~ s(x) + s(z),~ s(v)+s(w)),family=ziplss)
\end{verbatim} 
where the first formula specifies the response and the linear predictor the Poisson parameter given presence, while the second, one sided, formula specifies the linear predictor for presence. {\tt gaulss} and {\tt multinom} families provide further examples.

Similarly a multivariate normal model is fit with something like 
\begin{verbatim}
gam(list(y1 ~ s(x) + s(z),y2 ~ s(v)+s(w)),family=mvn(d=2))
\end{verbatim}
where each formula now specifies one component of the multivariate response and the linear predictor for its mean. There are also facilities to allow terms to be shared by different linear predictors, for example
\begin{verbatim}
gam(list(y1 ~ s(x),y2 ~ s(v),y3 ~ 1 ,1 + 3 ~ s(z) - 1),family=mvn(d=3))
\end{verbatim}
specifies a multivariate normal model in which the linear predictors for the first (\verb+y1+) and third (\verb+y3+) components of the response share the same dependence on a smooth of $z$.

The software is general and can accept an arbitrary number of formulae as well as dealing with the identifiability issues that can arise between parametric components when linear predictors share terms. Summary and plotting functions label model terms by component, and prediction produces matrix predictions when appropriate.

\bibliography{/home/sw283/bibliography/journal,/home/sw283/bibliography/simon}

\begin{thebibliography}{}

\bibitem[\protect\citeauthoryear{Adam, Qu, Davis, Ward, Clements, Cazares,
  Semmes, Schellhammer, Yasui, Feng, et~al.}{Adam et~al.}{2002}]{adam2002}
Adam, B.-L., Y.~Qu, J.~W. Davis, M.~D. Ward, M.~A. Clements, L.~H. Cazares,
  O.~J. Semmes, P.~F. Schellhammer, Y.~Yasui, Z.~Feng, et~al. (2002).
\newblock Serum protein fingerprinting coupled with a pattern-matching
  algorithm distinguishes prostate cancer from benign prostate hyperplasia and
  healthy men.
\newblock {\em Cancer Research\/}~{\em 62\/}(13), 3609--3614.

\bibitem[\protect\citeauthoryear{Agarwal and Studden}{Agarwal and
  Studden}{1980}]{agarwal1980}
Agarwal, G.~G. and W.~Studden (1980).
\newblock Asymptotic integrated mean square error using least squares and bias
  minimizing splines.
\newblock {\em The Annals of Statistics\/}, 1307--1325.

\bibitem[\protect\citeauthoryear{Akaike}{Akaike}{1973}]{akaike}
Akaike, H. (1973).
\newblock Information theory and an extension of the maximum likelihood
  principle.
\newblock In B.~Petran and F.~Csaaki (Eds.), {\em International symposium on
  information theory}, Budapest: Akadeemiai Kiadi, pp.\  267--281.

\bibitem[\protect\citeauthoryear{Anderssen and Bloomfield}{Anderssen and
  Bloomfield}{1974}]{anderssen1974}
Anderssen, R. and P.~Bloomfield (1974).
\newblock A time series approach to numerical differentiation.
\newblock {\em Technometrics\/}~{\em 16\/}(1), 69--75.

\bibitem[\protect\citeauthoryear{Augustin, Sauleau, and Wood}{Augustin
  et~al.}{2012}]{augustin2012qq}
Augustin, N.~H., E.-A. Sauleau, and S.~N. Wood (2012).
\newblock On quantile quantile plots for generalized linear models.
\newblock {\em Computational Statistics \& Data Analysis\/}~{\em 56\/}(8),
  2404--2409.

\bibitem[\protect\citeauthoryear{Bache and Lichman}{Bache and
  Lichman}{2013}]{Bache2013}
Bache, K. and M.~Lichman (2013).
\newblock {UCI} machine learning repository.

\bibitem[\protect\citeauthoryear{Belitz, Brezger, Kneib, Lang, and
  Umlauf}{Belitz et~al.}{2015}]{bayesx}
Belitz, C., A.~Brezger, T.~Kneib, S.~Lang, and N.~Umlauf (2015).
\newblock Bayesx: Software for bayesian inference in structured additive
  regression models.

\bibitem[\protect\citeauthoryear{Bravington}{Bravington}{2013}]{bravington.debug}
Bravington, M.~V. (2013).
\newblock {\em debug: MVB's debugger for R}.
\newblock R package version 1.3.1.

\bibitem[\protect\citeauthoryear{Breslow and Clayton}{Breslow and
  Clayton}{1993}]{breslow.clayton}
Breslow, N.~E. and D.~G. Clayton (1993).
\newblock Approximate inference in generalized linear mixed models.
\newblock {\em Journal of the American Statistical Association\/}~{\em 88},
  9--25.

\bibitem[\protect\citeauthoryear{Brezger and Lang}{Brezger and
  Lang}{2006}]{brezger2006}
Brezger, A. and S.~Lang (2006).
\newblock Generalized structured additive regression based on bayesian
  p-splines.
\newblock {\em Computational Statistics \& Data Analysis\/}~{\em 50\/}(4),
  967--991.

\bibitem[\protect\citeauthoryear{Claeskens, Krivobokova, and Opsomer}{Claeskens
  et~al.}{2009}]{claeskens2009}
Claeskens, G., T.~Krivobokova, and J.~D. Opsomer (2009).
\newblock Asymptotic properties of penalized spline estimators.
\newblock {\em Biometrika\/}~{\em 96\/}(3), 529--544.

\bibitem[\protect\citeauthoryear{Cline, Moler, Stewart, and Wilkinson}{Cline
  et~al.}{1979}]{cline1979}
Cline, A.~K., C.~B. Moler, G.~W. Stewart, and J.~H. Wilkinson (1979).
\newblock An estimate for the condition number of a matrix.
\newblock {\em SIAM Journal on Numerical Analysis\/}~{\em 16\/}(2), 368--375.

\bibitem[\protect\citeauthoryear{Cox}{Cox}{1972}]{cox1972}
Cox, D. (1972).
\newblock Regression models and life tables.
\newblock {\em Journal of the Royal Statistical Society. Series B
  (Methodological)\/}~{\em 34\/}(2), 187--220.

\bibitem[\protect\citeauthoryear{Cox}{Cox}{1983}]{cox1983}
Cox, D.~D. (1983).
\newblock Asymptotics for m-type smoothing splines.
\newblock {\em The Annals of Statistics\/}, 530--551.

\bibitem[\protect\citeauthoryear{Davison}{Davison}{2003}]{Davison}
Davison, A.~C. (2003).
\newblock {\em Statistical models}.
\newblock Cambridge: Cambridge University Press.

\bibitem[\protect\citeauthoryear{de~Boor}{de~Boor}{2001}]{deBoor2001}
de~Boor, C. (2001).
\newblock {\em A Practical Guide to Splines\/} (Revised ed.).
\newblock New York: Springer.

\bibitem[\protect\citeauthoryear{Demmler and Reinsch}{Demmler and
  Reinsch}{1975}]{DemmlerReinsch75}
Demmler, A. and C.~Reinsch (1975).
\newblock Oscillation matrices with spline smoothing.
\newblock {\em Numerische Mathematik\/}~{\em 24\/}(5), 375--382.

\bibitem[\protect\citeauthoryear{Duchon}{Duchon}{1977}]{duchon77}
Duchon, J. (1977).
\newblock Splines minimizing rotation-invariant semi-norms in {S}olobev spaces.
\newblock In W.~Schemp and K.~Zeller (Eds.), {\em Construction Theory of
  Functions of Several Variables}, Berlin, pp.\  85--100. Springer.

\bibitem[\protect\citeauthoryear{Dunn and Smyth}{Dunn and
  Smyth}{2005}]{dunn.smyth2005}
Dunn, P.~K. and G.~K. Smyth (2005).
\newblock Series evaluation of {T}weedie exponential dispersion model
  densities.
\newblock {\em Statistics and Computing\/}~{\em 15\/}(4), 267--280.

\bibitem[\protect\citeauthoryear{Eilers and Marx}{Eilers and
  Marx}{1996}]{Eilers&Marx96}
Eilers, P. H.~C. and B.~D. Marx (1996).
\newblock Flexible smoothing with {B}-splines and penalties.
\newblock {\em Statistical Science\/}~{\em 11\/}(2), 89--121.

\bibitem[\protect\citeauthoryear{Fahrmeir, Kneib, and Lang}{Fahrmeir
  et~al.}{2004}]{fahrmeir04}
Fahrmeir, L., T.~Kneib, and S.~Lang (2004).
\newblock Penalized structured additive regression for space-time data: A
  {B}ayesian perspective.
\newblock {\em Statistica Sinica\/}~{\em 14\/}(3), 731--761.

\bibitem[\protect\citeauthoryear{Fahrmeir, Kneib, Lang, and Marx}{Fahrmeir
  et~al.}{2013}]{fahrmeir2013}
Fahrmeir, L., T.~Kneib, S.~Lang, and B.~Marx (2013).
\newblock {\em Regression Models}.
\newblock Springer.

\bibitem[\protect\citeauthoryear{Fahrmeir and Lang}{Fahrmeir and
  Lang}{2001}]{fahrmeir.lang}
Fahrmeir, L. and S.~Lang (2001).
\newblock Bayesian inference for generalized additive mixed models based on
  markov random field priors.
\newblock {\em Applied Statistics\/}~{\em 50}, 201--220.

\bibitem[\protect\citeauthoryear{Gill, Murray, and Wright}{Gill
  et~al.}{1981}]{gill.murray.wright}
Gill, P.~E., W.~Murray, and M.~H. Wright (1981).
\newblock {\em Practical optimization}.
\newblock London: Academic Press.

\bibitem[\protect\citeauthoryear{Golub and Van~Loan}{Golub and
  Van~Loan}{2013}]{GvL4}
Golub, G.~H. and C.~F. Van~Loan (2013).
\newblock {\em Matrix computations\/} (4th ed.).
\newblock Baltimore: Johns Hopkins University Press.

\bibitem[\protect\citeauthoryear{Green and Silverman}{Green and
  Silverman}{1994}]{green.silverman}
Green, P.~J. and B.~W. Silverman (1994).
\newblock {\em Nonparametric Regression and Generalized Linear Models}.
\newblock Chapman \& Hall.

\bibitem[\protect\citeauthoryear{Greven and Kneib}{Greven and
  Kneib}{2010}]{greven.kneib2010}
Greven, S. and T.~Kneib (2010).
\newblock On the behaviour of marginal and conditional {AIC} in linear mixed
  models.
\newblock {\em Biometrika\/}~{\em 97\/}(4), 773--789.

\bibitem[\protect\citeauthoryear{Gu and Kim}{Gu and Kim}{2002}]{gu.kim}
Gu, C. and Y.~J. Kim (2002).
\newblock Penalized likelihood regression: general approximation and efficient
  approximation.
\newblock {\em Canadian Journal of Statistics\/}~{\em 34\/}(4), 619--628.

\bibitem[\protect\citeauthoryear{Gu and Wahba}{Gu and
  Wahba}{1991}]{gu.wahba.msp}
Gu, C. and G.~Wahba (1991).
\newblock Minimizing {GCV/GML} scores with multiple smoothing parameters via
  the {N}ewton method.
\newblock {\em SIAM Journal on Scientific and Statistical Computing\/}~{\em
  12\/}(2), 383--398.

\bibitem[\protect\citeauthoryear{Hall and Meyer}{Hall and Meyer}{1976}]{hall76}
Hall, C.~A. and W.~W. Meyer (1976).
\newblock Optimal error bounds for cubic spline interpolation.
\newblock {\em Journal of Approximation Theory\/}~{\em 16\/}(2), 105--122.

\bibitem[\protect\citeauthoryear{Hall and Opsomer}{Hall and
  Opsomer}{2005}]{hall2005}
Hall, P. and J.~D. Opsomer (2005).
\newblock Theory for penalised spline regression.
\newblock {\em Biometrika\/}~{\em 92\/}(1), 105--118.

\bibitem[\protect\citeauthoryear{Hastie and Tibshirani}{Hastie and
  Tibshirani}{1986}]{h&t86}
Hastie, T. and R.~Tibshirani (1986).
\newblock Generalized additive models (with discussion).
\newblock {\em Statistical Science\/}~{\em 1}, 297--318.

\bibitem[\protect\citeauthoryear{Hastie and Tibshirani}{Hastie and
  Tibshirani}{1990}]{h&t90}
Hastie, T. and R.~Tibshirani (1990).
\newblock {\em Generalized Additive Models}.
\newblock Chapman \& Hall.

\bibitem[\protect\citeauthoryear{Kass and Steffey}{Kass and
  Steffey}{1989}]{kass1989}
Kass, R.~E. and D.~Steffey (1989).
\newblock Approximate bayesian inference in conditionally independent
  hierarchical models (parametric empirical bayes models).
\newblock {\em Journal of the American Statistical Association\/}~{\em
  84\/}(407), 717--726.

\bibitem[\protect\citeauthoryear{Kauermann, Krivobokova, and
  Fahrmeir}{Kauermann et~al.}{2009}]{kauermann2009}
Kauermann, G., T.~Krivobokova, and L.~Fahrmeir (2009).
\newblock Some asymptotic results on generalized penalized spline smoothing.
\newblock {\em Journal of the Royal Statistical Society: Series B (Statistical
  Methodology)\/}~{\em 71\/}(2), 487--503.

\bibitem[\protect\citeauthoryear{Klein and Moeschberger}{Klein and
  Moeschberger}{2003}]{klein2003}
Klein, J. and M.~Moeschberger (2003).
\newblock {\em Survival analysis: techniques for censored and truncated data\/}
  (2nd ed.).
\newblock New York: Springer.

\bibitem[\protect\citeauthoryear{Klein, Kneib, Klasen, and Lang}{Klein
  et~al.}{2014}]{klein2014dr}
Klein, N., T.~Kneib, S.~Klasen, and S.~Lang (2014).
\newblock Bayesian structured additive distributional regression for
  multivariate responses.
\newblock {\em Journal of the Royal Statistical Society: Series C (Applied
  Statistics)\/}~{\em 64}, 569--591.

\bibitem[\protect\citeauthoryear{Klein, Kneib, Lang, and Sohn}{Klein
  et~al.}{2015}]{klein2015dr}
Klein, N., T.~Kneib, S.~Lang, and A.~Sohn (2015).
\newblock Bayesian structured additive distributional regression with an
  application to regional income inequality in germany.
\newblock {\em Annals of Applied Statistics\/}~{\em 9}, 1024--1052.

\bibitem[\protect\citeauthoryear{Kneib and Fahrmeir}{Kneib and
  Fahrmeir}{2006}]{kneib2006ocat}
Kneib, T. and L.~Fahrmeir (2006).
\newblock Structured additive regression for categorical space--time data: A
  mixed model approach.
\newblock {\em Biometrics\/}~{\em 62\/}(1), 109--118.

\bibitem[\protect\citeauthoryear{Laird and Ware}{Laird and
  Ware}{1982}]{laird.ware}
Laird, N.~M. and J.~H. Ware (1982).
\newblock Random-effects models for longitudinal data.
\newblock {\em Biometrics\/}~{\em 38}, 963--974.

\bibitem[\protect\citeauthoryear{Lancaster and \v{S}alkauskas}{Lancaster and
  \v{S}alkauskas}{1986}]{lancaster.salkauskas}
Lancaster, P. and K.~\v{S}alkauskas (1986).
\newblock {\em Curve and Surface Fitting: An Introduction}.
\newblock London: Academic Press.

\bibitem[\protect\citeauthoryear{Liang, Wu, and Zou}{Liang
  et~al.}{2008}]{liang2008AIC}
Liang, H., H.~Wu, and G.~Zou (2008).
\newblock A note on conditional {AIC} for linear mixed-effects models.
\newblock {\em Biometrika\/}~{\em 95\/}(3), 773--778.

\bibitem[\protect\citeauthoryear{Marra and Wood}{Marra and
  Wood}{2011}]{marra.wood2011}
Marra, G. and S.~N. Wood (2011).
\newblock Practical variable selection for generalized additive models.
\newblock {\em Computational Statistics \& Data Analysis\/}~{\em 55\/}(7),
  2372--2387.

\bibitem[\protect\citeauthoryear{Marra and Wood}{Marra and
  Wood}{2012}]{marra.wood2012}
Marra, G. and S.~N. Wood (2012).
\newblock Coverage properties of confidence intervals for generalized additive
  model components.
\newblock {\em Scandinavian Journal of Statistics\/}~{\em 39\/}(1), 53--74.

\bibitem[\protect\citeauthoryear{Marx and Eilers}{Marx and Eilers}{1998}]{ME98}
Marx, B.~D. and P.~H. Eilers (1998).
\newblock Direct generalized additive modeling with penalized likelihood.
\newblock {\em Computational Statistics and Data Analysis\/}~{\em 28},
  193--209.

\bibitem[\protect\citeauthoryear{Miller and Wood}{Miller and
  Wood}{2014}]{miller2014}
Miller, D.~L. and S.~N. Wood (2014).
\newblock Finite area smoothing with generalized distance splines.
\newblock {\em Environmental and Ecological Statistics\/}, 1--17.

\bibitem[\protect\citeauthoryear{Nocedal and Wright}{Nocedal and
  Wright}{2006}]{nocedal.wright}
Nocedal, J. and S.~Wright (2006).
\newblock {\em Numerical optimization\/} (2nd ed.).
\newblock New York: Springer verlag.

\bibitem[\protect\citeauthoryear{Nychka}{Nychka}{1988}]{nychka88}
Nychka, D. (1988).
\newblock Bayesian confidence intervals for smoothing splines.
\newblock {\em Journal of the American Statistical Association\/}~{\em
  83\/}(404), 1134--1143.

\bibitem[\protect\citeauthoryear{Nychka and Cummins}{Nychka and
  Cummins}{1996}]{nychka96DR}
Nychka, D. and D.~Cummins (1996).
\newblock Comment on `{F}lexible smoothing with {B}-splines and penalties' by
  {PHC} {E}ilers and {BD} {M}arx.
\newblock {\em Statist. Sci\/}~{\em 89}, 104--5.
\newblock Demmler Reinsch basis and P-splines.

\bibitem[\protect\citeauthoryear{{R Core Team}}{{R Core Team}}{2014}]{rcore}
{R Core Team} (2014).
\newblock {\em R: A Language and Environment for Statistical Computing}.
\newblock Vienna, Austria: R Foundation for Statistical Computing.

\bibitem[\protect\citeauthoryear{Reiss and Ogden}{Reiss and
  Ogden}{2009}]{reiss.ogden2009}
Reiss, P.~T. and T.~R. Ogden (2009).
\newblock Smoothing parameter selection for a class of semiparametric linear
  models.
\newblock {\em Journal of the Royal Statistical Society: Series B (Statistical
  Methodology)\/}~{\em 71\/}(2), 505--523.

\bibitem[\protect\citeauthoryear{Rigby and Stasinopoulos}{Rigby and
  Stasinopoulos}{2005}]{rigby2005}
Rigby, R. and D.~M. Stasinopoulos (2005).
\newblock Generalized additive models for location, scale and shape.
\newblock {\em Journal of the Royal Statistical Society: Series C (Applied
  Statistics)\/}~{\em 54\/}(3), 507--554.

\bibitem[\protect\citeauthoryear{Rigby and Stasinopoulos}{Rigby and
  Stasinopoulos}{2013}]{rigby2013automatic}
Rigby, R.~A. and D.~M. Stasinopoulos (2013).
\newblock Automatic smoothing parameter selection in {GAMLSS} with an
  application to centile estimation.
\newblock {\em Statistical methods in medical research\/}.

\bibitem[\protect\citeauthoryear{Rue, Martino, and Chopin}{Rue
  et~al.}{2009}]{rue2009inla}
Rue, H., S.~Martino, and N.~Chopin (2009).
\newblock Approximate {B}ayesian inference for latent {G}aussian models by
  using integrated nested {L}aplace approximations.
\newblock {\em Journal of the royal statistical society: Series B\/}~{\em
  71\/}(2), 319--392.

\bibitem[\protect\citeauthoryear{Ruppert, Wand, and Carroll}{Ruppert
  et~al.}{2003}]{ruppert.wand.carroll}
Ruppert, D., M.~P. Wand, and R.~J. Carroll (2003).
\newblock {\em Semiparametric Regression}.
\newblock Cambridge University Press.

\bibitem[\protect\citeauthoryear{S\"afken, Kneib, van Waveren, and
  Greven}{S\"afken et~al.}{2014}]{saefken2014}
S\"afken, B., T.~Kneib, C.-S. van Waveren, and S.~Greven (2014).
\newblock A unifying approach to the estimation of the conditional {A}kaike
  information in generalized linear mixed models.
\newblock {\em Electronic Journal of Statistics\/}~{\em 8}, 201--225.

\bibitem[\protect\citeauthoryear{Shun and McCullagh}{Shun and
  McCullagh}{1995}]{shun1995laplace}
Shun, Z. and P.~McCullagh (1995).
\newblock Laplace approximation of high dimensional integrals.
\newblock {\em Journal of the Royal Statistical Society. Series B
  (Methodological)\/}~{\em 57\/}(4), 749--760.

\bibitem[\protect\citeauthoryear{Silverman}{Silverman}{1985}]{silverman85}
Silverman, B.~W. (1985).
\newblock Some aspects of the spline smoothing approach to non-parametric
  regression curve fitting.
\newblock {\em Journal of the Royal Statistical Society B\/}~{\em 47\/}(1),
  1--53.

\bibitem[\protect\citeauthoryear{Speckman}{Speckman}{1985}]{speckman1985}
Speckman, P. (1985).
\newblock Spline smoothing and optimal rates of convergence in nonparametric
  regression models.
\newblock {\em The Annals of Statistics\/}, 970--983.

\bibitem[\protect\citeauthoryear{Stone}{Stone}{1982}]{stone1982}
Stone, C.~J. (1982).
\newblock Optimal global rates of convergence for nonparametric regression.
\newblock {\em The Annals of Statistics\/}, 1040--1053.

\bibitem[\protect\citeauthoryear{Tweedie}{Tweedie}{1984}]{tweedie1984}
Tweedie, M. (1984).
\newblock An index which distinguishes between some important exponential
  families.
\newblock In {\em Statistics: Applications and New Directions: Proc. Indian
  Statistical Institute Golden Jubilee International Conference}, pp.\
  579--604.

\bibitem[\protect\citeauthoryear{Umlauf, Adler, Kneib, Lang, and
  Zeileis}{Umlauf et~al.}{2015}]{rbayesx}
Umlauf, N., D.~Adler, T.~Kneib, S.~Lang, and A.~Zeileis (2015).
\newblock Structured additive regression models: An r interface to bayesx.
  journal of statistical software.
\newblock {\em Journal of Statistical Software\/}~{\em 63(21)}, 1--46.

\bibitem[\protect\citeauthoryear{Wahba}{Wahba}{1983}]{wahba83}
Wahba, G. (1983).
\newblock Bayesian confidence intervals for the cross validated smoothing
  spline.
\newblock {\em Journal of the Royal Statistical Society B\/}~{\em 45},
  133--150.

\bibitem[\protect\citeauthoryear{Wahba}{Wahba}{1985}]{wahba85}
Wahba, G. (1985).
\newblock A comparison of {GCV} and {GML} for choosing the smoothing parameter
  in the generalized spline smoothing problem.
\newblock {\em The Annals of Statistics\/}, 1378--1402.

\bibitem[\protect\citeauthoryear{Wood}{Wood}{2000}]{wood2000}
Wood, S.~N. (2000).
\newblock Modelling and smoothing parameter estimation with multiple quadratic
  penalties.
\newblock {\em Journal of the Royal Statistical Society, Series B\/}~{\em 62},
  413--428.

\bibitem[\protect\citeauthoryear{Wood}{Wood}{2003}]{tprs}
Wood, S.~N. (2003).
\newblock Thin plate regression splines.
\newblock {\em Journal of the Royal Statistical Society, Series B\/}~{\em 65},
  95--114.

\bibitem[\protect\citeauthoryear{Wood}{Wood}{2006a}]{wood2006igam}
Wood, S.~N. (2006a).
\newblock {\em Generalized Additive Models: An Introduction with R}.
\newblock Boca Raton, FL: CRC press.

\bibitem[\protect\citeauthoryear{Wood}{Wood}{2006b}]{wood2006tensor}
Wood, S.~N. (2006b).
\newblock Low-rank scale-invariant tensor product smooths for generalized
  additive mixed models.
\newblock {\em Biometrics\/}~{\em 62\/}(4), 1025--1036.

\bibitem[\protect\citeauthoryear{Wood}{Wood}{2011}]{wood2011}
Wood, S.~N. (2011).
\newblock Fast stable restricted maximum likelihood and marginal likelihood
  estimation of semiparametric generalized linear models.
\newblock {\em Journal of the Royal Statistical Society: Series B (Statistical
  Methodology)\/}~{\em 73\/}(1), 3--36.

\bibitem[\protect\citeauthoryear{Yee and Wild}{Yee and Wild}{1996}]{yee1996}
Yee, T.~W. and C.~Wild (1996).
\newblock Vector generalized additive models.
\newblock {\em Journal of the Royal Statistical Society. Series B
  (Methodological)\/}, 481--493.

\bibitem[\protect\citeauthoryear{Yoshida and Naito}{Yoshida and
  Naito}{2014}]{yoshida2014asymptotics}
Yoshida, T. and K.~Naito (2014).
\newblock Asymptotics for penalised splines in generalised additive models.
\newblock {\em Journal of Nonparametric Statistics\/}~{\em 26\/}(2), 269--289.

\bibitem[\protect\citeauthoryear{Yu and Yau}{Yu and Yau}{2012}]{yu.yau2012}
Yu, D. and K.~K. Yau (2012).
\newblock Conditional {A}kaike information criterion for generalized linear
  mixed models.
\newblock {\em Computational Statistics \& Data Analysis\/}~{\em 56\/}(3),
  629--644.

\bibitem[\protect\citeauthoryear{Zhou and Wolfe}{Zhou and
  Wolfe}{2000}]{zhou2000}
Zhou, S. and D.~A. Wolfe (2000).
\newblock On derivative estimation in spline regression.
\newblock {\em Statistica Sinica\/}~{\em 10\/}(1), 93--108.

\end{thebibliography}
\bibliographystyle{chicago}

\end{document}